\documentclass{aa}  

\usepackage{graphicx}
\usepackage{txfonts}
\usepackage{caption}
\usepackage{subcaption}
\usepackage{lscape}
\usepackage[breaklinks=true]{hyperref} %% to avoid \citeads line fills
\usepackage{orcidlink}

\hypersetup{
  colorlinks=true,   %% links colored instead of frames
  urlcolor=blue,     %% external hyperlinks
  linkcolor=blue,     %% internal latex links (eg Fig)
  citecolor=blue
}

\newcommand{\msun}{\mbox{M$_{\odot}$}}

\newcommand{\lsun}{\mbox{L$_{\odot}$}}

\makeatletter
\renewcommand*\aa@pageof{, page \thepage{} of \pageref*{LastPage}}
\makeatother

\begin{document} 

  \title{Investigating episodic mass loss in evolved massive stars}

\subtitle{III. Spectroscopy of dusty massive stars in three northern galaxies}
% Order of co-authors not final 
   \author{S. de Wit\inst{\ref{noa} \thanks{These authors contributed equally and should be considered joint first authors.}}\orcidlink{0000-0002-9818-4877} 
        \and
        G. Mu\~noz-Sanchez\inst{\ref{noa}, \ref{nkua} ^\star}\orcidlink{0000-0002-9179-6918}
        \and
        G. Maravelias\inst{\ref{noa}, \ref{crete}}\orcidlink{0000-0002-0891-7564}
        \and 
        A.Z. Bonanos\inst{\ref{noa}}\orcidlink{0000-0003-2851-1905}
        \and
        K. Antoniadis\inst{\ref{noa}, \ref{nkua}}\orcidlink{0000-0002-3454-7958} 
        \and \\
        D. Garc\'ia-\'Alvarez \inst{\ref{tenerife}, \ref{palma}}
        \and
        N. Britavskiy \inst{\ref{brussels}, \ref{liege}}\orcidlink{0000-0003-3996-0175}      \and
        A. Ruiz \inst{\ref{noa}}\orcidlink{0000-0002-3352-4383}
        \and
        A. Philippopoulou \inst{\ref{oxford}}
        }
   \institute{
        IAASARS, National Observatory of Athens, I. Metaxa \& Vas. Pavlou St., 15236, Penteli, Athens, Greece\label{noa}
        \and
        Department of Physics, National and Kapodistrian University of Athens, Panepistimiopolis, Zografos, 15784, Greece\label{nkua}
        \and
        Institute of Astrophysics, FORTH, 71110, Heraklion, Greece\label{crete}
        \and
        Instituto de Astrofísica de Canarias, Avenida Vía Láctea, 38205 La Laguna, Tenerife, Spain \label{tenerife}
        \and
        Grantecan S. A., Centro de Astrofísica de La Palma, Cuesta de San José, 38712 Breña Baja, La Palma, Spain \label{palma}
        \and Royal Observatory of Belgium, Avenue Circulaire, Ringlaan 3, B-1180 Brussels, Belgium \label{brussels}
        \and
        University of Li\`ege, All\'ee du 6 Ao\^ut 19c (B5C), B-4000 Sart Tilman, Li\`ege, Belgium \label{liege}
        \and
        Department of Physics, University of Oxford, Oxford, United Kingdom \label{oxford}}

   \date{}

  \abstract
   {Mass loss in massive stars is crucial to understanding how these stars evolve and explode. Despite increasing evidence indicating its importance, episodic mass loss remains poorly understood. Here we report the results of an optical spectroscopic survey of evolved massive stars in NGC~6822, IC~10, and IC~1613, conducted by the ASSESS project (Episodic Mass Loss in Evolved Massive Stars: Key to Understanding the Explosive Early Universe), which aimed to investigate the role of episodic mass loss, by targeting stars with infrared excesses indicating a dusty circumstellar environment. We assigned a spectral class to 122 unique sources, the majority of which are dusty. The rate of evolved massive stars was over 60\% for the highest-priority targets. We discovered 2 blue supergiants, 1 yellow supergiant, 1 emission-line object, and confirmed 2 supernova remnant candidates, a Wolf-Rayet star, and 2 H~\textsc{ii} regions. Twenty-eight unique sources were classified as red supergiants, 21 of which are new discoveries. In IC~10, we increased the sample of spectroscopically confirmed RSGs from 1 to 17. We used the \textsc{marcs} models to obtain their surface properties, most importantly the effective temperature, and spectral energy distribution fitting to obtain the stellar luminosity for 17 of them. The dusty RSGs are cooler, more luminous, more extinguished, and more evolved than the non-dusty ones, in agreement with previous findings. Investigating the optical photometric variability of the RSGs from light curves covering a period over a decade revealed that the dusty RSGs are more variable.
   We further highlight the very extinguished emission-line object, two RSGs that display a significant change in spectral type between two observed epochs, and four dusty K-type RSGs as candidates for having undergone episodic mass loss.}


   \titlerunning{Spectroscopy of dusty massive stars in three northern galaxies}
   \authorrunning{de Wit et al.}

   \maketitle
%
%-------------------------------------------------------------------

\section{Introduction}\label{sec:intro}
Massive stars lose a significant fraction of their mass throughout their evolution. For stars between 8 and 30~\msun, mass is gradually lost through stationary, line-driven winds in the OB phase \citep[e.g.,][]{Castor1975, Vink2000, Vink2021} and through enhanced winds during the red supergiant (RSG) phase \citep[e.g.,][]{Jager1988, Loon2005, Beasor2020, Yang2023, Antoniadis2024}. The evolution of such stars \citep[e.g., whether they transition to yellow hypergiants;][]{Yoon2010, Koumpia2020, Humphreys2023}, their supernova explosion characteristics \citep{Smartt2009, Smartt2015, Forster+2018}, and lastly, the properties of the resulting compact object \citep{Sukhbold2016, Vink2021} are all affected by the degree of mass loss.

Reductions in the observed mass-loss rates in both the OB phase \citep{Fullerton2006, Cohen2014, Telford2024} and the RSG phase \citep{Beasor2020, Antoniadis2024, Decin2024} suggest that mass lost through stationary winds is inadequate to strip stars of their hydrogen-rich envelopes. Evidence for episodic mass loss \citep{Smith2014} is mounting from theory \citep[e.g.,][]{Cheng2024} and direct observations of massive stars \citep[e.g.,][]{Ohnaka2008, Gvaramadze2010, Montarges2021, MunozSanchez2024}, superluminous supernovae \citep{Gal-Yam2012, Neill2011, Gal-Yam2019} and `SN imposters' \citep[e.g., SN2009ip;][]{Mauerhan2013,Smith2022}.

Clues of ongoing or recent epochs of enhanced mass loss in RSGs may lie in their circumstellar environments \citep{Ohnaka2008, Montarges2021, Dupree2022, Decin2024}. The presence of infrared (IR) excesses due to recent dust formation provides a laboratory to study mass-losing RSGs with increased surface activity \citep{Ma2024, Drevon2024}. Pulsations \citep{Yoon2010}, convective cycles \citep{Josselin2007, Kravchenko2019}, radiation pressure on molecules \citep{ArroyoTorres2015}, and turbulence \citep{Josselin2007, Kee2021} are all thought to contribute to the surface conditions enabling enhanced mass loss. Therefore, the large-scale characterization of RSG surfaces is crucial to understanding which RSGs exhibit the conditions for episodic mass loss and how frequent they are. The latter has implications for the duration of such a phase, whereas the surface properties may be linked to the stage of the RSG evolution where this occurs.

The ASSESS project\footnote{\url{http://assess.astro.noa.gr/}} so far has observed and classified 185 evolved massive stars \citep{Bonanos2024}, and further modeled 127 RSGs \citep{deWit2024} to obtain their properties. Additionally, the project has presented spectroscopic and photometric variability studies of extreme RSGs such as [W60]~B90 and WOH~G64 \citep[][respectively]{MunozSanchez2024, MunozSanchez2024b}, to understand their variable nature and the connection to episodic mass loss. This paper aims to increase the number of classified evolved massive stars, from which the classified RSGs are modeled, by presenting and analyzing spectra from 3 galaxies in our Northern survey: NGC~6822, IC~10, and IC~1613. Sect.~\ref{sec:observations} presents the target selection, observations, and data reduction. Sect.~\ref{sec:classification} provides the spectral classification for each object and the catalog. Sect.~\ref{sec:properties} presents RSG luminosities, optical light curves, and the RSG properties resulting from modeling, and we discuss them in Sect.~\ref{sec:discussion}. Sect.~\ref{sec:conclusions} summarizes our findings.

\section{Observations and data reduction}
\label{sec:observations}

\subsection{Target selection}\label{sec:targets}

\begin{table*}
\centering
\small
\caption{Properties of target galaxies.\label{tab:galaxies}}
\begin{tabular}{l c c c c c c}
\hline\hline
Galaxy		& R.A.		& Dec.	 	& Distance & Diameter	& 12$+\log(\rm O/H)$\tablefootmark{a} & $Z$ \\
			& \small{(J2000)}	& \small{(J2000)}	& \small{(Mpc)}	& \small{(\arcmin)} 	& 	& \small{($Z_{\sun}$)} \\
\hline \\[-9pt]
IC 10		& 00:20:23.11   & $+$59:17:35.2	& 0.77$^{+0.01}_{-0.02}$ $^1$	&  7 & 8.37$^2$ & 0.30$^3$ \\
IC 1613		& 01:04:47.78   & $+$02:07:03.7	& 0.74$^{+0.03}_{-0.05}$ $^4$	& 18 & 7.62$^5$ & 0.20$^6$ \\
NGC 6822    & 19:44:56.98   & $-$14:48:01.1 & 0.48$\pm0.03$$^7$, 0.45$^{+0.01}_{-0.02}$ $^8$	& 19 & 8.14$^9$ & 0.30$^{10}$ \\

\hline
\end{tabular}
\tablefoot{
\tablefoottext{a}{12$+\log(\rm O/H)_{\odot} = 8.75$ \citep{Bergemann2021}.}
}
\tablebib{(1) \citet{McQuinn2017}, (2)  \citet{Cosens2024}, (3) \citet{Polles2019}, (4) \citet{Gorski2011}, (5) \citet{Lee2003},  (6) \citet{Bresolin2007}, (7) \citet{Tully2013}, (8) \citet{Zgirski2021}, (9) \citet{Moustakas2010}, (10) \citet{Patrick2015}. 
}
\end{table*}

Table~\ref{tab:galaxies} presents the properties of the target galaxies (i.e., IC~10, IC~1613, and NGC~6822), such as coordinates, distance, diameter (i.e., the size of the galaxy based on visual inspection) used to match the catalogs, the oxygen abundance $\log(\rm O/H)$, and metallicity $Z$. For NGC~6822, we built the catalog and selected the targets using the CosmicFlows-2 distance of 0.48~Mpc \citep{Tully2013}. An updated distance \citep[$0.45 \pm 0.01$~Mpc,][]{Zgirski2021} was published after our observations, hence, we used it in the analysis (see Sect.~\ref{sec:properties}). We selected targets using multiwavelength photometric catalogs. We began with \textit{Spitzer} IR catalogs from \citet{Khan2015} for NGC 6822 and \citet{Boyer2015} for IC 10 and IC 1613. We cross-matched them using a 1\arcsec\ radius with \textit{Gaia} astrometry \citep{Gaia2016, gaia2018}, optical photometry \citep[Pan-STARRS1;][]{Chambers2016} and near-IR photometry. For the latter, we used the UK Infra-Red Telescope Hemisphere Survey \citep{Dye2018, Irwin2013} for IC 10 and the VISTA Hemisphere Survey \citep{McMahon2012} for NGC 6822. No data was available for IC 1613. We used the photometric datasets for the target selection and \textit{Gaia} to identify and remove foreground sources. 

We applied the criteria presented by \citet{Bonanos2024}, which rely on early roadmaps from \citet{Bonanos2009, Bonanos2010} showing the separation of classes of evolved massive stars based on their mid-IR colors. First, we considered sources with $[3.6]-[4.5]\geq0.1$ mag, corresponding to dusty sources and excluding foreground stars \citep[see e.g.][]{Britavskiy2015}. Then, cuts in the absolute magnitude $M_{3.6}\leq-9.0$ mag and the apparent magnitude of $[4.5]\leq15.5$ mag were applied to exclude contamination from the asymptotic giant branch (AGB) stars and background sources (such as galaxies and quasars), respectively. We assigned a priority scheme (from P1 to P6) depending on the mid-IR color, absolute magnitude criteria, and the presence of optical or near-IR counterparts (see \citealt{Bonanos2024} for details). The highest priority candidates, P1 and P2, correspond to the brightest IR sources with strong and moderate IR excess, respectively. 
Table \ref{tab:fields} presents the distribution of priorities for all the available targets, as well as for the observed targets in each field of multi-object spectroscopy (MOS) and long-slit observation. Overall, we observed 97 priority targets: 42 targets in IC~10 (34\% of the total available priority targets), 3 targets in IC~1613 (18\% of the total), and 52 targets in NGC~6822 (54\% of the total). We had a notable coverage of the highest-priority classes, with 9 P1 and 5 P2 targets (60\% and 71\% of the total, respectively).

\subsection{Pre-imaging and mask design (MOS)}

We obtained the spectra for a large number of targets with the MOS mode of OSIRIS \citep[Optical System for Imaging and low-Intermediate-Resolution Integrated Spectroscopy;][]{Cepa2000, Cepa2003} at the 10.4~m Gran Telescopio Canarias (GTC), at La Palma, Canary Islands. 

During our first observing campaign (proposal ID 115-GTC83/20A, PI: N. Britavskiy) MOS observations were carried out for IC 10 and NGC 6822. We first acquired short exposures (10--15 s in the $r$ filter) of the fields to be observed. This pre-imaging, combined with the Mask Designer tool (v. 3.26\footnote{\url{https://www.gtc.iac.es/instruments/osiris/osirisMOS\_Cass.php\#MDtool}}), helped to securely identify targets in the field of view and design the mask (i.e., the positions of the slits). We used the 1000R grism and a $1.2\arcsec$ slit width, which yielded a spectral resolving power ($R$ = $\lambda$/$\Delta$$\lambda$) of $\sim$600--1000, depending on the seeing (see Table~\ref{tab:obslog}), over the wavelength range of $\sim$5300--9000~\r{A}. The resolution and wavelength coverage provide access to a large number of spectral features critical for spectral classification (see Section \ref{sec:classification}). Targets at the edges of the detector yield a more limited spectral range. The pointings were chosen to maximize the number of P1 and P2 targets to be observed. Additional space on each mask was filled with lower-priority targets.

For the second campaign (proposal ID 29-GTC34/22A, PI: D. Garc\'ia-\'Alvarez) the MOS observing mode was not available. However, we were able to secure several long-slit observations for a selection of targets in IC 10, IC 1613, and NGC 6822. For these, we used the 1000R grism and a 0.8$\arcsec$ slit width, achieving $R\sim$1000. Whenever possible, we placed two targets in the slit. In all cases, we used the standard bias, flat-field, and arc exposures provided by the observatory. 
 
Table~\ref{tab:obslog} presents the observing log for all OSIRIS pre-imaging, MOS, and long-slit observations, providing the coordinates of the field center for MOS spectra or of the target for long-slit spectra, the UT date, MJD, mode, exposure time, airmass, and seeing for each observation.

\subsection{Data reduction} \label{sec:mos}

We performed the data reduction of the MOS observations in two steps. First, we used the \textsc{gtcmos} package \citep[IRAF\footnote{IRAF is distributed by the National Optical Astronomy Observatory, operated by the Association of Universities for Research in Astronomy (AURA) under agreement with the National Science Foundation.} based pipeline,][]{GomezGonzalez2016} to merge the images from each CCD\footnote{For observations taken in 2020, while the 2022 upgrade implemented a monolithic CCD.} and subtract the bias. Due to an offset between the flat field and the science images, which compromised the sky subtraction process in the narrow slits by creating artifacts close to the edges of the slits, we discarded the flat correction. We therefore proceeded to manually reduce the spectra using the IRAF tasks \textsc{identify}, \textsc{reidentify}, \textsc{fitcoords}, and \textsc{transform} to get a 2D-wavelength calibrated image, and \textsc{apall} to extract the spectra. We flux-calibrated the extracted spectra with the flux standard by using the tasks \textsc{standard}, \textsc{sensfunc}, and \textsc{calibrate}. We followed the same manual approach to reduce the long-slit spectra. For both the MOS and long-slit observations, serendipitous extracted stars, which do not fall into any priority category, are labeled as `R' (i.e., random) in Table~\ref{tab:fields} (e.g., IC10-R1). We computed and applied the heliocentric correction with \textsc{rvcorrect}.  

\begin{table*}[h]
\centering
\small
\caption{Distribution of classified targets per galaxy and spectral class.}\label{tab:classified}
\begin{tabular}{l | c | l l l r l l l l l l l}
\hline\hline
Galaxy	& Class. & RSG & BSG & YSG & C-stars & H~\textsc{ii} & Em. & Fgd & Other\tablefootmark{a} & Clusters & Galaxies & SNR\\
Name	&  & & & & &  & Obj. &  &  stars & & & \\
\hline \\[-9pt]
IC~10	& 46 (12) & 6 (10) & 1     & -     & 6 (2) & - & - & 17 & 11     & 2 & - & 3 \\
IC~1613 & 3 (3)  & 1 (1)  & - (1) & - (1) & -     & - & - & 1  & -      & - & 1 & - \\
NGC~6822& 36 (7) & 7 (3)  & -     & -     & 7 (2) & 2 & 1 & 3  & 15 (2) & 1 & - & - \\
\hline
{Total}	&  {85 (22)} &  {14 (14)} & {1 (1)} & {0 (1)} & {13 (4)} & {2} & {1}  & {21} & {26 (2)} & {3} & {1} & {3} \\
\hline
\end{tabular}

\tablefoot{
\small The parentheses indicate the number of candidate sources per spectral class.
\tablefoottext{a}{Other A, F, G, M stars, hot, warm, cool stars, and the WC star are grouped into ``Other stars''.}
}

\end{table*}

\begin{table*}[h]
\centering
\caption{Catalog of spectral classifications.}
\label{tab:grand}
\small
\begin{tabular}{l l l c c l c}
\hline
\hline      
ID    & R.A.    & Dec.   & ... & Spectral Class.   & Classification Notes \\
\hline
\small
IC10-5545 & 5.09071	& 59.30489 & ... & K0-K5 I	& Weak TiO abs. Strong Ca~\textsc{ii} triplet. \\
IC10-5660 & 5.10571	& 59.29033 & ... & M4-M6 I: $\&$ M0-M2 I: & Strong TiO abs (TiO $\lambda$8430 present). Strong Ca~\textsc{ii} triplet. Neb em lines. \\
IC10-6884 & 5.09643	& 59.29376 & ... & AF-star	& Lack of spectral lines. Ca~\textsc{ii} $\lambda$8542 abs. \\
IC10-7387 & 5.10304	& 59.26844 & ... & C-star & CN band abs. \\
IC10-8523 & 5.10129	& 59.33703 & ... & C-star & CN band abs. \\
IC10-9165 & 5.09258	& 59.29533 & ... & M4-M6 I: $\&$ M0-M2 I &	Strong TiO abs (TiO $\lambda$8430 present). Strong Ca~\textsc{ii} triplet. \\
IC10-10128 & 5.05371 & 59.29040 & ... & K V &  Strong D1/D2 abs. Weak Ca~\textsc{ii} triplet.  \\
IC10-11499 & 4.93359 & 59.32721 & ... & AFG V &	Strong H$\alpha$ abs. Strong metal abs. \\
IC10-16046 & 5.05946  & 59.28856 & ... & Red &	\\
IC10-16200 & 5.05233  & 59.34828 & ... & Red &	Unidentified Molecular abs around $\lambda$7000 \\
IC10-17332 & 5.03544  & 59.29600 & ... & M V & TiO abs. \\
IC10-17517 & 5.02783  & 59.32278 & ... & Composite: & Multiple stellar contributions \\
IC10-18665 & 5.07206 &	59.30387 & ... & AFG V & Weak H$\alpha$. Ca~\textsc{ii} triplet present. \\
IC10-19636 & 5.03346  & 59.25064 & ... & M-star & TiO abs. \\
IC10-20273 & 5.07797 &	59.31839 & ... & AFG V & Strong H$\alpha$ abs. Strong metal abs. \\
\hline
\end{tabular}
\vspace{0.3cm}

\tablefoot{
\small This table is available in its entirety at the CDS. A portion is shown here for guidance regarding its form and content.
}
\end{table*}

\section{Spectral classification and stellar content} \label{sec:classification}

\begin{figure*}[h]
\begin{subfigure}[t]{0.5\textwidth}
\centering
    \includegraphics[width=1\columnwidth]{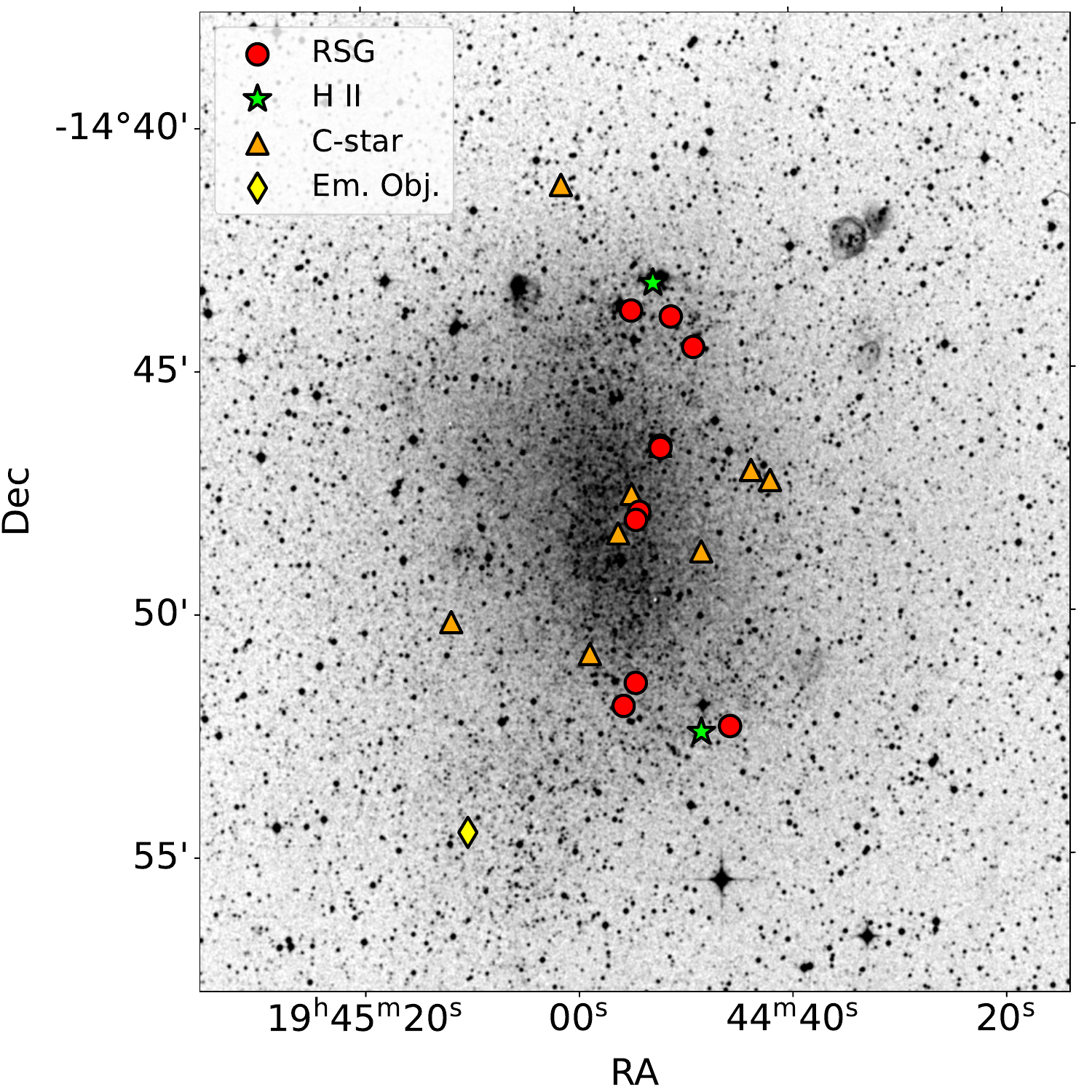}
\end{subfigure}
\begin{subfigure}[t]{0.5\textwidth}
    \includegraphics[width=1\columnwidth]{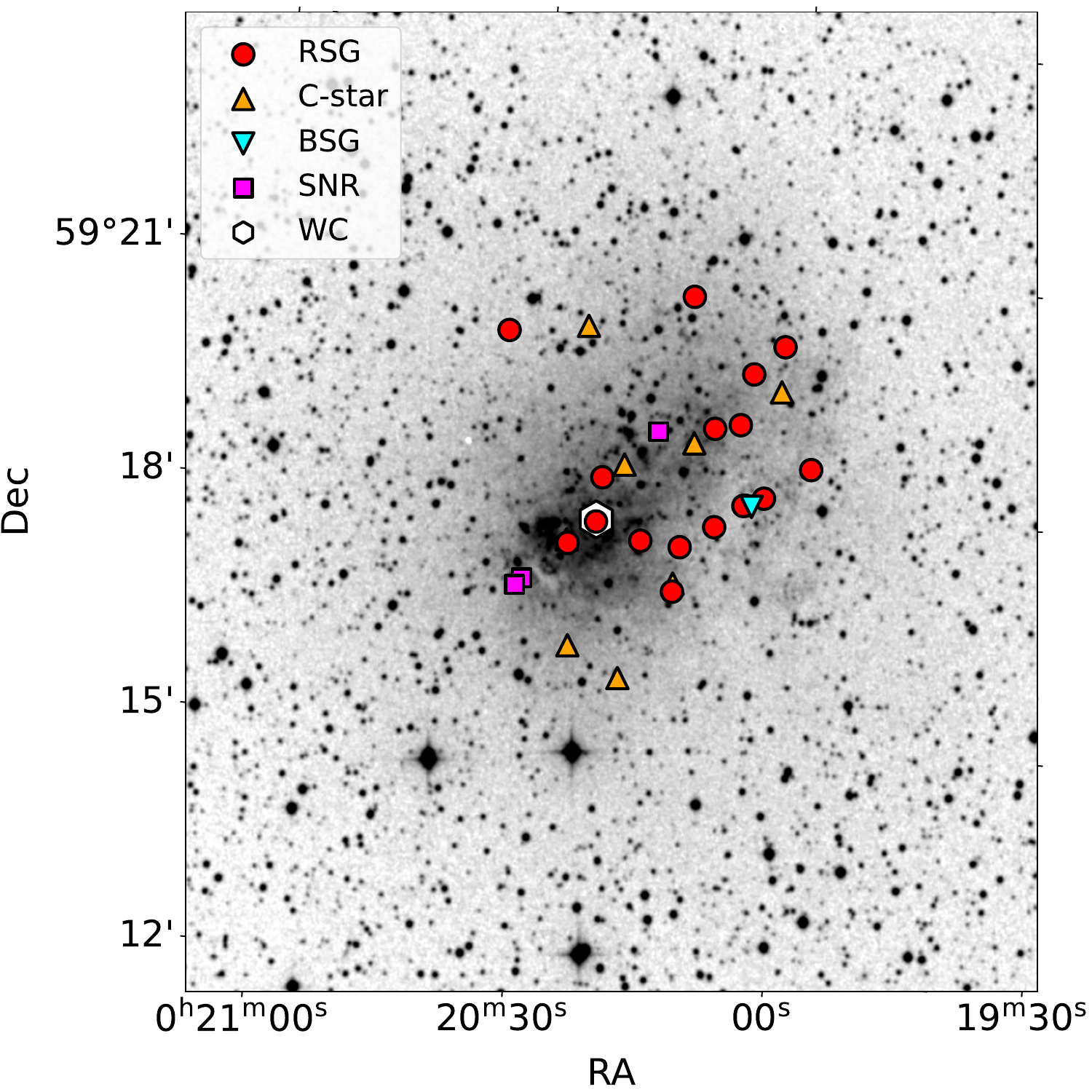}
\end{subfigure}
\caption {Spatial distribution of classified sources in NGC~6822 (\textit{left}) and IC~10 (\textit{right}); background images are from DSS2-red.}
\label{fig:spatial}
\end{figure*}

\begin{figure*}[h]
\begin{subfigure}[t]{0.5\textwidth}
\centering
    \includegraphics[width=1\columnwidth]{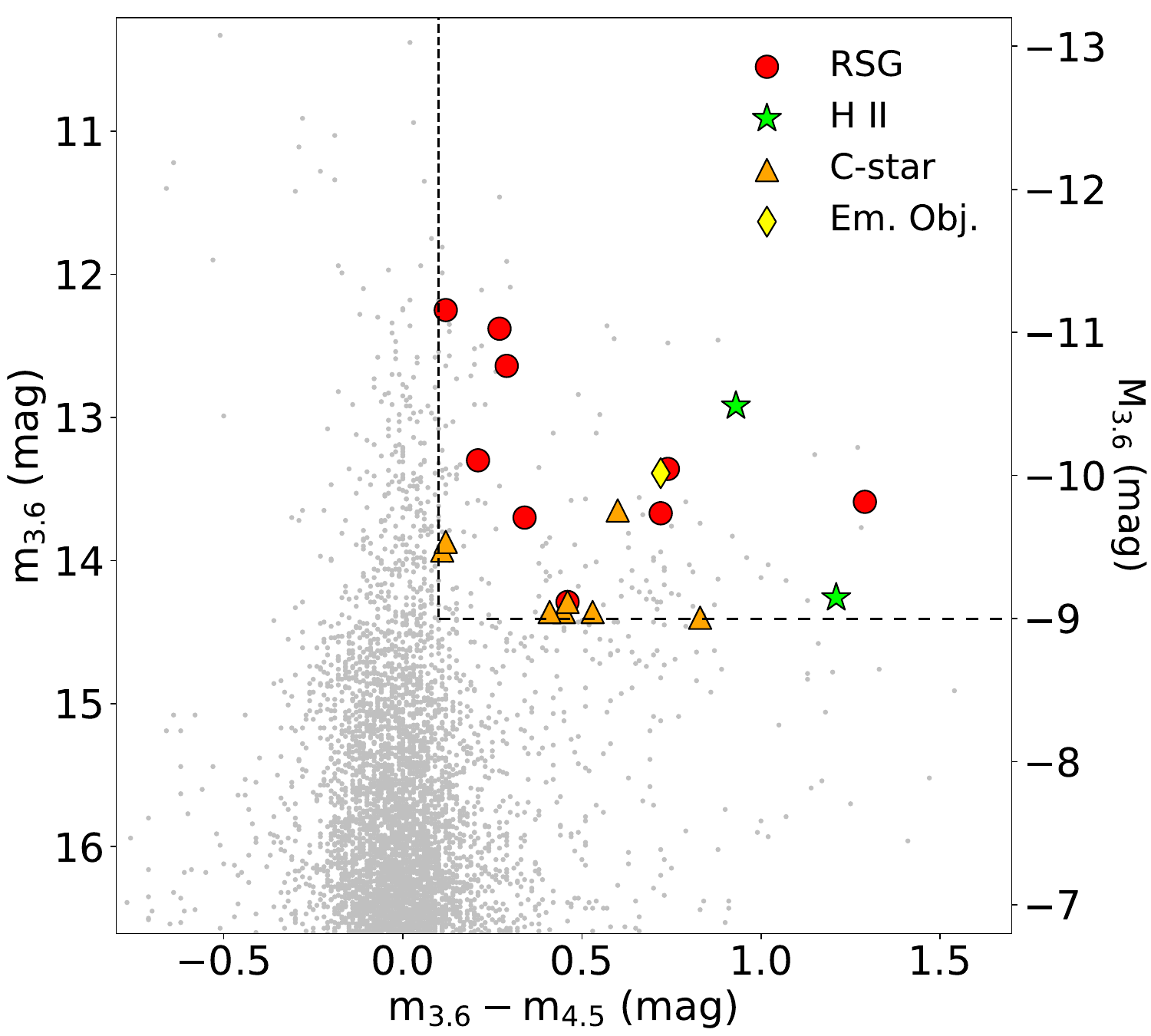}
\end{subfigure}
\begin{subfigure}[t]{0.5\textwidth}
    \includegraphics[width=1\columnwidth]{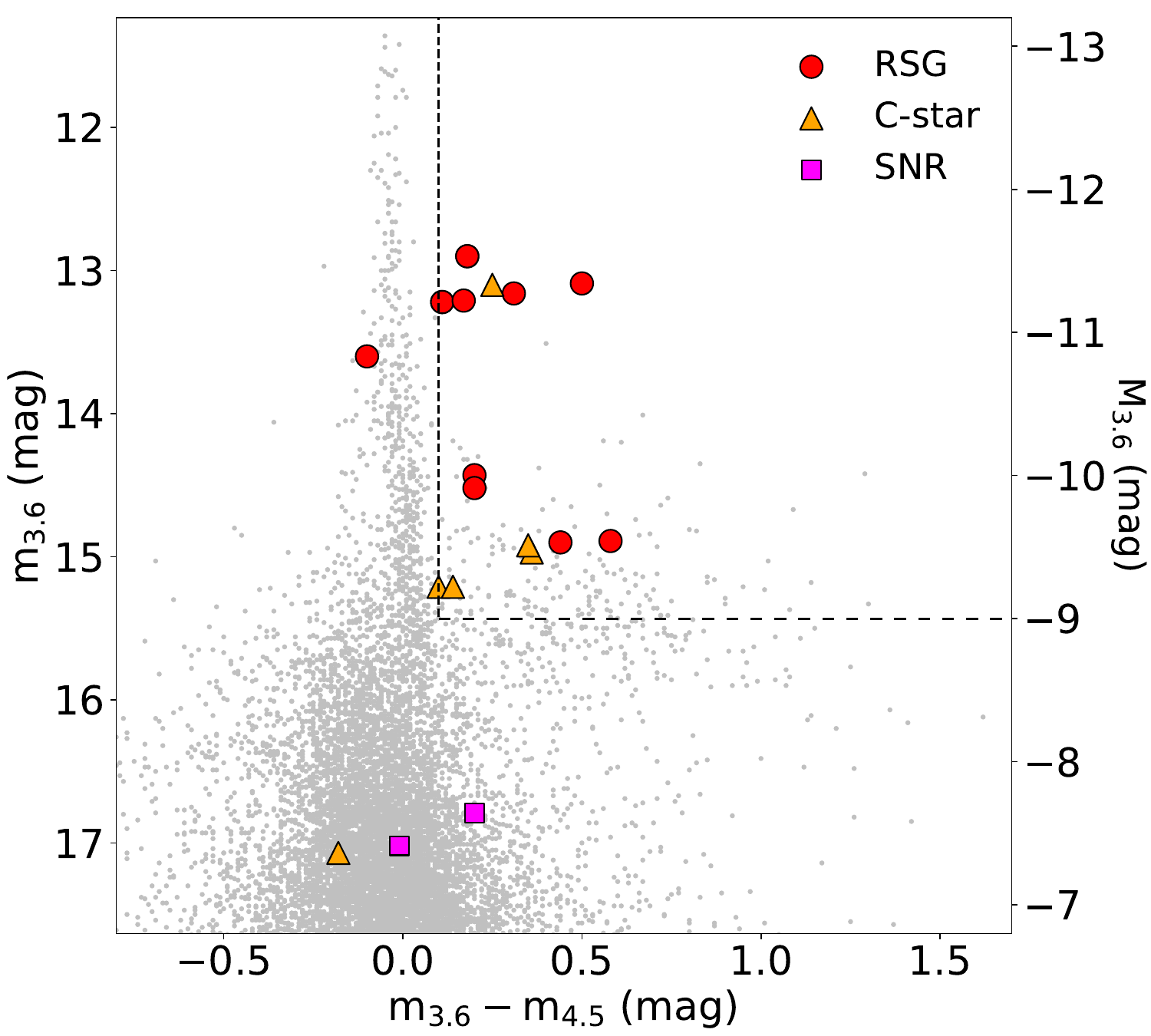}
\end{subfigure}
\begin{subfigure}[t]{0.5\textwidth}
    \includegraphics[width=1\columnwidth]{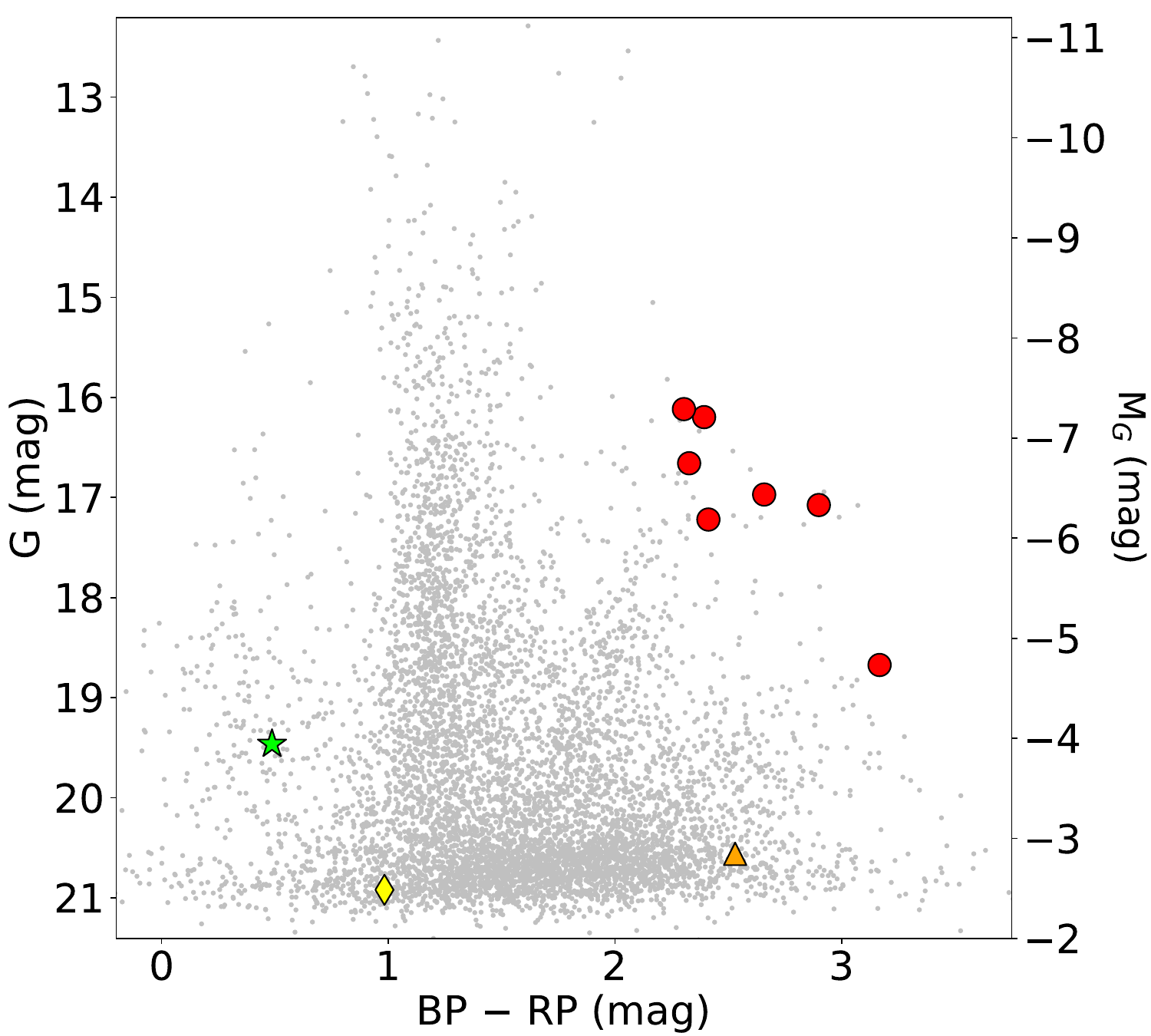}
\end{subfigure}
\begin{subfigure}[t]{0.5\textwidth}
    \includegraphics[width=1\columnwidth]{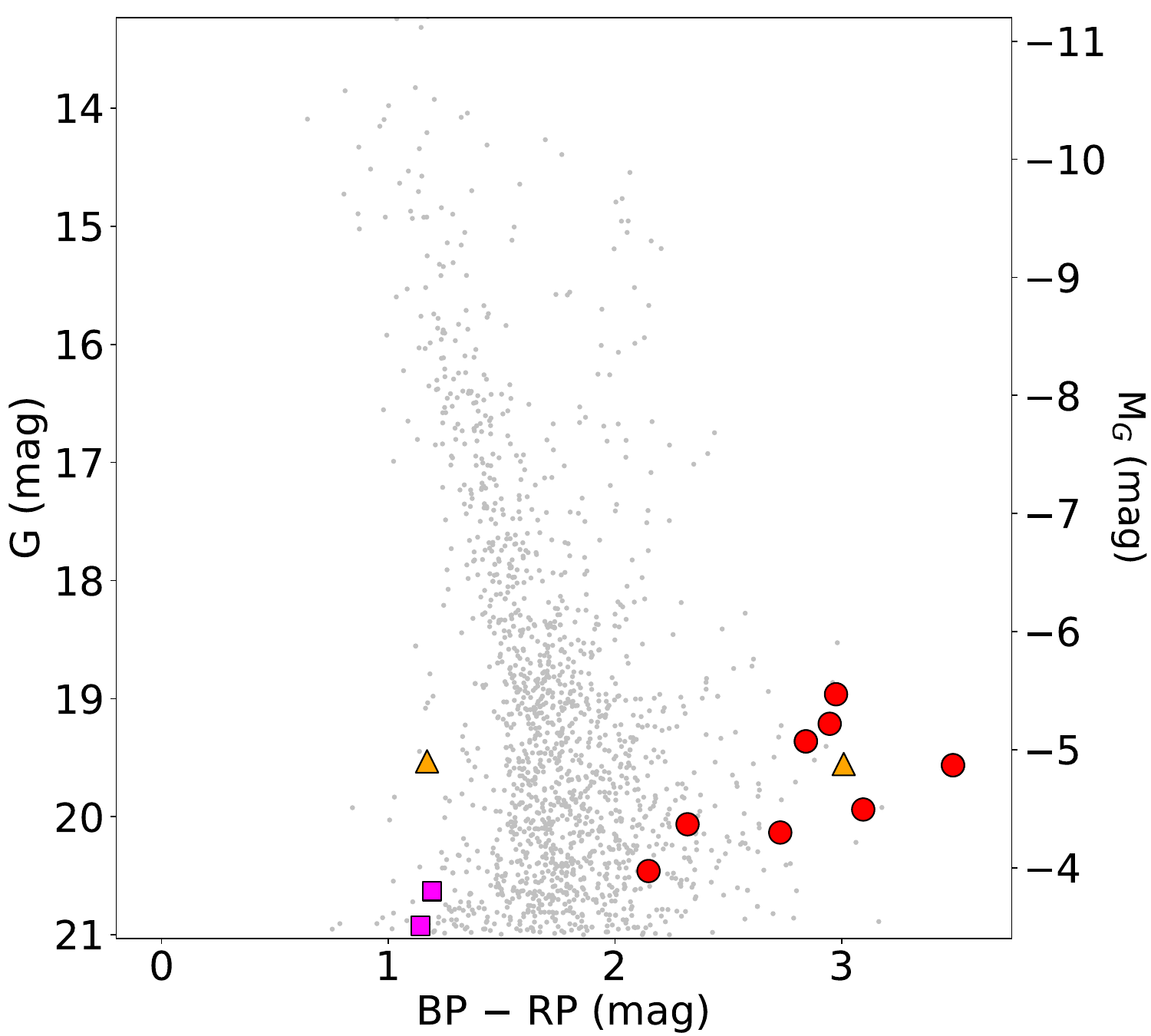}
\end{subfigure}
\caption{Near-infrared and optical color-magnitude diagrams of classified sources. We show the locations of classified targets with \textit{Spitzer} IRAC photometry for NGC~6822 (\textit{top left}) and IC~10 (\textit{top right}) and with \textit{Gaia} photometry for NGC~6822 (\textit{bottom left}) and IC~10 (\textit{bottom right}). The dashed lines indicate the priority criteria from \citet{Bonanos2024}. IC10-52406 is the only RSG that shows no IR excess, as it was a serendipitous observation.}
\label{fig:cmd}
\end{figure*}

Out of the 163 spectra, we classified 124 into broad classes, following the procedure and criteria described by \citet{Bonanos2024}. The other 39 spectra could not be classified due to their low S/N. Two sources were re-observed (IC10-5660 and IC10-9165) with the OSIRIS long-slit mode due to their apparent brightness and late RSG type (see Sect.~\ref{sec:RSGclass}), yielding 122 unique sources classified. 15 targets out of the 122 were classified more generally, like a nebula or a red source.  Table~\ref{tab:classified} presents the distribution of classified targets (107) for both robust classifications (85) and candidates (22) per galaxy and spectral class. Figure~\ref{fig:spatial} shows the spatial distribution of the H~\textsc{ii} regions, carbon stars, and evolved massive stars and Fig.~\ref{fig:cmd} their location in the color-magnitude diagrams. Note, particularly in IC~10, that the absolute magnitudes $M_G$ and $M_{3.6}$ are not corrected for foreground extinction \citep[A$_V=4.3$~mag,][]{Schlafly2011}.

Table~\ref{tab:grand} presents the catalog, including all 122 spectroscopically classified targets. The targets are sorted by galaxy name and the ID, which is based on [4.5] \citep[see][for a description]{Bonanos2024}. The columns of Table~\ref{tab:grand} indicate the J2000 coordinates (in degrees), the observed field, assigned priority, \textit{Gaia}~DR3 photometry, \textit{Gaia}~DR3 astrometry \citep{GaiaDR3}, \textit{Spitzer} photometry \citep{Fazio2004, Rieke2004}, Pan-STARRS1~DR2 photometry \citep{Chambers2016}, UKIRT Hemisphere Survey photometry \citep{Dye2018}, the VISTA Hemisphere Survey photometry \citep[][NGC~6822 only]{McMahon2012}, Median Absolute Deviations (MAD, see Sect.~\ref{sec:lcs}), spectral type, previous classifications from the literature and comments on the spectral classification of each source. Previous classifications from the literature were obtained by cross-matching our sources using the VizieR catalog access tool and a 1\arcsec radius. The catalog includes 21 newly discovered RSGs and 4 other evolved massive stars.

\subsection{RSG classification} \label{sec:RSGclass}

We classified the RSGs more precisely than the other classes following the classification criteria by \citet{deWit2024}. In addition to those criteria, if a strong Ca~\textsc{ii} triplet was observed and its radial velocity corresponded to that of the host galaxy, the sources were robustly classified as RSGs. We securely classified 14 targets as RSGs and verified that their modeled properties agreed with the expected range of properties for RSGs in these galaxies. However, we added a colon after the luminosity indicator (e.g., K5--M0~I:) when the S/N was low at the Ca~\textsc{ii} triplet or at least one modeled property was outside the expected range (e.g., their surface gravity was too high, or the recovered E($B-V$) too low, in the case of IC~10). We considered the other 14 RSGs as candidates, given that oxygen-rich AGB stars are likely contaminants \citep[reaching up to log~$(L/\rm L_{\odot}) \sim 5.0$, but with weaker Ca~\textsc{ii} lines;][]{Groenewegen2018}.

\begin{figure*}[h]
\begin{center}
    \centerline{\includegraphics[width=1.8\columnwidth]{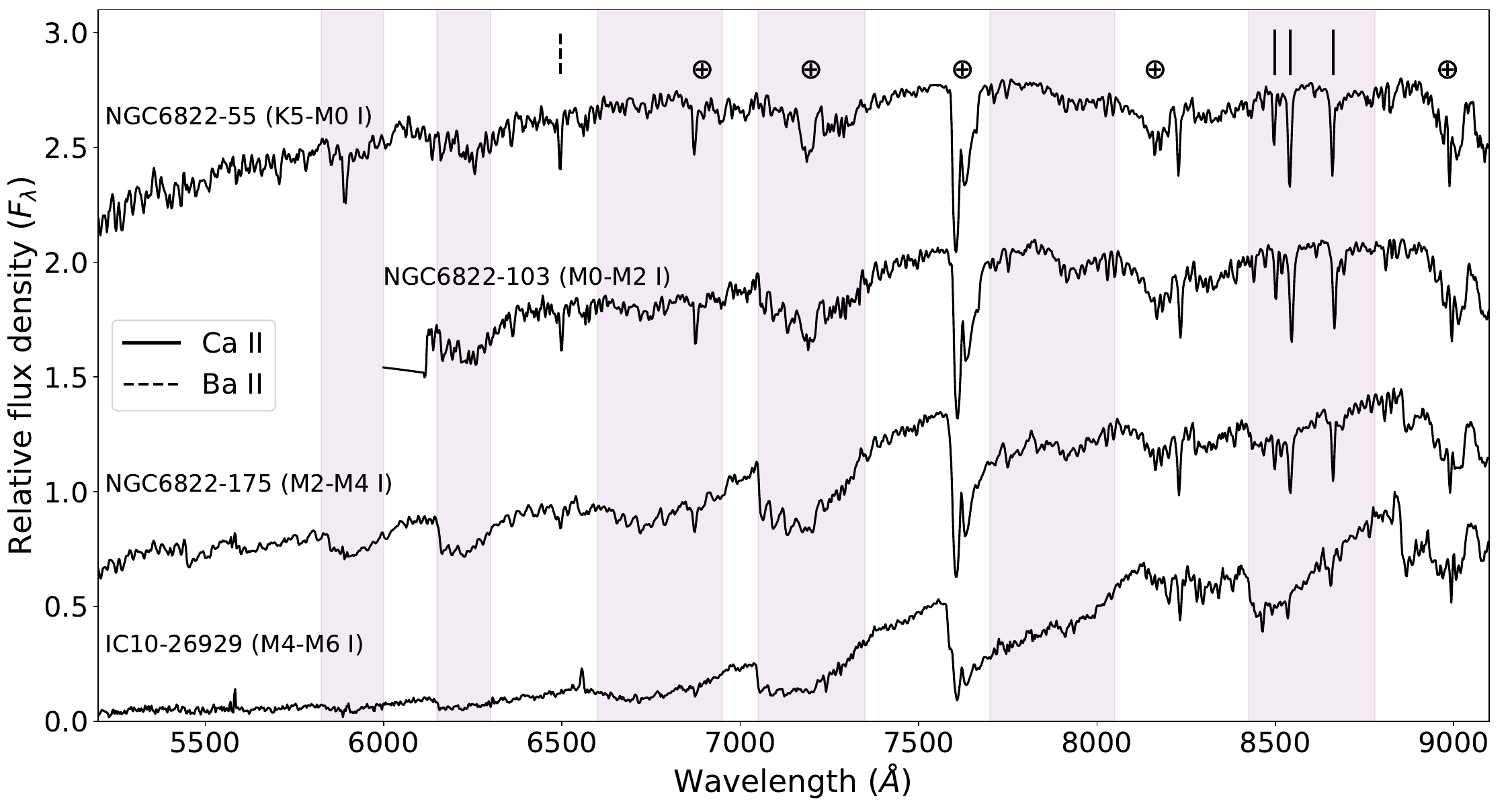}} 
    \caption{OSIRIS spectra of classified RSGs. We show four RSGs with increasing TiO absorption strengths, highlighting the spectral type sequence of RSGs. Purple-shaded areas indicate regions with TiO absorption. A $\oplus$ marks telluric absorption.}
    
    \label{fig:RSGexamples}
\end{center}
\end{figure*}

The 28 verified and candidate RSGs include 16 in IC~10, 2 in IC~1613, and 10 in NGC~6822. We classified them as: one GK~I, two K0--K5~I, three K5--M0~I, fourteen M0-M2~I, six M2--M4~I, and four M4--M6~I (see some examples in Fig.~\ref{fig:RSGexamples}). The criteria for the latter class were not presented by \citet{deWit2024} as their spectra did not cover the TiO band starting around $\lambda$8430. Typically, when this TiO band dominates the absorption in the $I$~band, stars are classified with spectral types ranging from M4 to M6 \citep{Solf1978}. The targets classified twice (IC10-5660 and IC10-9165) shifted from M4--M6~I in their MOS spectrum (August 2020) to M0--M2 in their long-slit spectrum (August 2022), significantly changing their TiO band strengths. The GK~I supergiant (IC1613-28306) lacks TiO bands, and the `forest' of metallic lines from $\lambda\lambda$5500--6500 indicated a slightly hotter source, bordering the Yellow Supergiant (YSG) class. However, some individual spectral lines (e.g., H$\alpha$) are too strong for such a star, hinting at a potential binary component. These spectra are modeled for their stellar properties in Sect.~\ref{subsec:grid}.

We find the average spectral type of the sample to be M0--M2~I, which agrees with that expected at $Z$$\sim$0.3 \citep{Levesque2012} and increases the sample at this metallicity. The metallicity dependence of the average spectral type, and, subsequently, of the effective temperature, was first reported by \citet{Elias1985} and is due to the shift of the Hayashi limit to higher temperatures at low metallicities. The small fraction of K-type RSGs (21$\%$) can also be attributed to the absolute magnitude selection criteria, as we mainly targeted bright RSGs. These are more likely to exhibit higher mass loss rates, increasing the TiO band strengths and shifting the RSGs to later types \citep{Davies2021}. 

Seven out of the 28~RSGs were previously reported in the literature: IC1613-28306, NGC6822-52, NGC6822-55, NGC6822-66, NGC6822-70, NGC6822-103, and NGC6822-175 \citep[][]{Massey1998a, Massey2007, Levesque2012, Patrick2015, Chun2022}. IC10-66828 is located in a known cluster \citep{Tikhonov2009}. \citet{Kacharov2012} classified NGC6822-248 as an M6~III star, but they only distinguished between foreground (V) and giant (III) stars, hence, our classification supersedes it. Finally, only one spectroscopically confirmed RSG was previously known in IC~10 \citep{Britavskiy2019a}, therefore, our work has significantly increased the RSG sample in this galaxy by 16 RSGs.

\subsection{Other sources}

We identified two blue supergiants (BSG) and a YSG in the sample (see Fig.~\ref{fig:BYSGexamples}): IC1613-8006 (A~I--III), IC1613-6612 (AFG~I--III), and IC10-21061 (BA~I). The Paschen absorption series in the spectra of IC1613-8006 and IC10-21061 indicated a spectral type A. In contrast, metal lines in the spectrum of IC1613-6612 suggested a cooler, YSG candidate. The slope of IC10-21061 was severely reddened due to the foreground extinction, but its significantly blue-shifted H$\alpha$ emission line profile supported the extragalactic supergiant classification. Additionally, we classified IC10-R15 as a Wolf-Rayet star based on the strong C~\textsc{iv} feature at 5800~\r{A}, which agrees with the previous classification from \citep{Crowther2003}.

We classified 17 carbon stars in our sample, of which 13 are robust classifications. Each of these exhibited strong CN absorption bands in the $R$ and $I$-bands, occasionally displaying the Swan C$_2$ bands if the signal in the blue was high enough (see Fig.~\ref{fig:Cstarexamples}; NGC6822-170 and NGC6822-387).

We classified three spectra as supernova remnants (SNRs) in IC~10. These spectra have emission lines (e.g., H$\alpha$, [S~\textsc{ii}], and [N~\textsc{ii}]) of nebular origin. The emission line ratios indicate a shock-ionization mechanism ([S~\textsc{ii}]/H$\alpha$ $\geq 0.4$), most commonly observed in supernova remnants \citep{Kopsacheili2020}. IC10-R23 matches the location of the radio source [YS93]~HL20b \citep{Yang1993}, a known SNR. The other two are located where other SNR candidates have been reported from radio observations \citep{Yang1993}; hence, our optical spectra confirm their nature.

Figure~\ref{fig:Hiiexamples} plots the H~\textsc{ii} regions and Fig.~\ref{fig:EmObj} the emission line object NGC6822-106. The latter shows strong emission features but differs from those observed in the H~\textsc{ii} regions (e.g., [Ar~\textsc{iii}] and He~\textsc{i} lines are absent, Ca~\textsc{ii} triplet in emission). NGC6822-48 is a well-studied extended H~\textsc{ii} region \citep[Hubble V;][]{Hubble1925, Rubin2016}, hosting an O-star in the center with visible He~\textsc{ii} absorption lines (i.e., He~\textsc{ii} $\lambda$5412), while NGC6822-169 is a known compact H~\textsc{ii} region \citep[Region~$\gamma$;][]{Kinman1979}. 

\subsection{High-priority sources}

We highlight the highest-priority classified targets (P1 and P2), potentially very dusty evolved massive stars, which occupy the reddest region of Fig.~\ref{fig:cmd}. 

Three of the nine P1 targets could not be classified due to their low S/N. We found three RSGs, one as K0--K5~I (IC10-26089) and two as M0--M2~I (NGC6822-66 and NGC6822-103), an H~\textsc{ii} region (NGC6822-48), an emission line object (NGC6822-106), and a star of spectral type M5--M9 (NGC6822-77). The lack of a visible Ca~\textsc{ii} triplet and its late spectral type suggest a foreground dwarf for the latter. We could not verify this with \textit{Gaia} due to the absence of proper motion data.

From the five P2 targets, three were classified as cool RSGs (NGC6822-70; M2--M4~I and NGC6822-52 and IC10-26929; M4--M6~I), suggesting these are evolved, mass-losing RSGs \citep{deWit2024}. The remaining two targets were classified as a carbon star (IC10-50206) and a cluster (NGC6822-234), known as Hubble~VII \citep{Hubble1925}, which may contain a large amount of dust causing the apparent IR excess.

\section{Properties of RSGs} \label{sec:properties}

\subsection{SED fitting}

We used the photometry presented in Table~\ref{tab:grand} to construct the spectral energy distribution (SED) of each RSG. We also added the median value of the time-series photometry from the Zwicky Transient Facility \citep[ZTF;][]{ZTF2019}. Following the procedure described by \citet{deWit2024}, the magnitudes were corrected for foreground Galactic extinction from \citet{Schlafly2011} in their respective bands. We fitted the observed SED with a Planck function to obtain the total observed flux. We adopted a distance of $0.77 \pm 0.02$~Mpc for IC~10 \citep{McQuinn2017}, $0.74 \pm 0.03$~Mpc for IC~1613 \citep{Gorski2011}, and $0.45 \pm 0.01$~Mpc for NGC~6822 \citep{Zgirski2021} to convert the observed flux into an intrinsic flux, yielding the bolometric luminosity. Table~\ref{tab:params} presents the obtained luminosity, log($L/\rm L_{\odot}$), of 17 RSGs through this SED fitting, and Fig.~\ref{fig:SEDAll} shows their fits. We could not obtain the luminosity for 11 RSGs because of the poor photometric coverage of their SED or contaminated photometry due to low spatial resolution. 

None of the upper luminosity limits exceed or are even close to the observed upper luminosity limit of $\log(L/\rm L_{\odot}$)~$\sim 5.5$ for cool supergiants \citep{Davies2018, McDonald2022}. The two RSGs that show a large shift in their TiO band strengths (IC10-5660 and IC10-9165) have high log($L/\rm L_{\odot}$)~$\sim 5.3$, which indicates a very evolved state \citep{Davies2013}.

\subsection{Light curves} \label{sec:lcs}
\begin{figure*}[h]
\begin{center}
\centerline{\includegraphics[width=2\columnwidth]{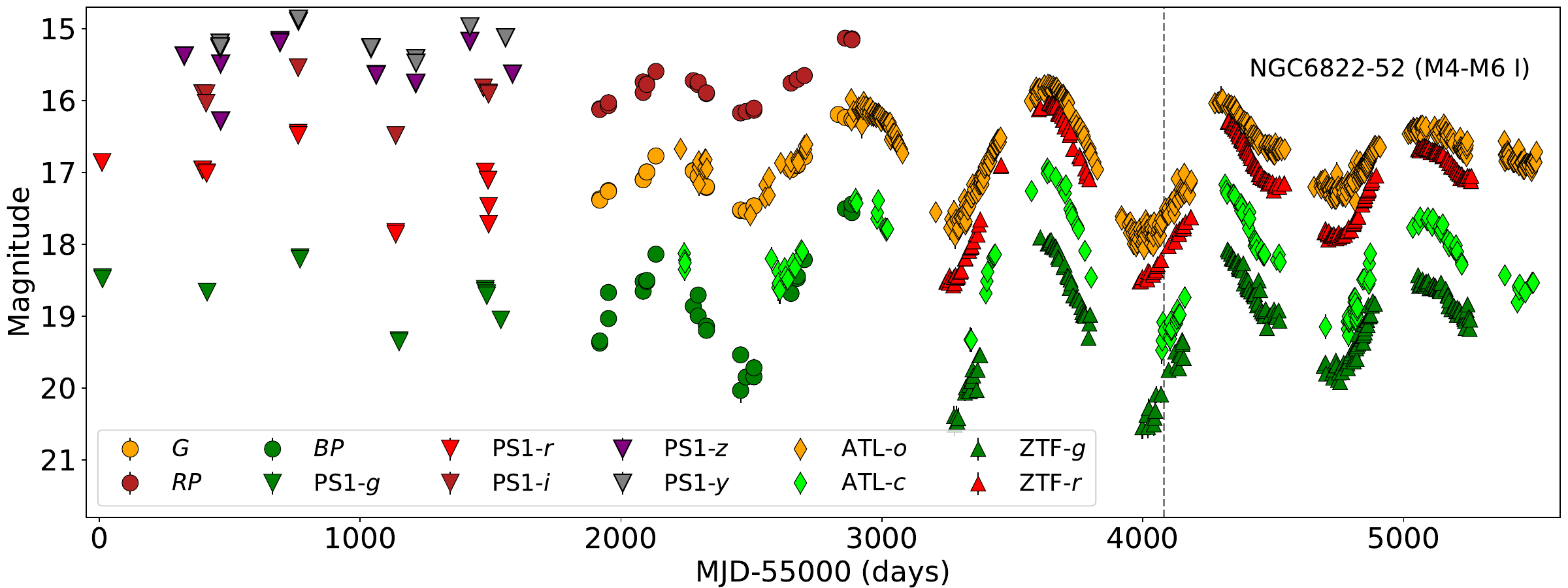}} 
    \caption{Multi-survey optical light curve of NGC6822-52. Apparent magnitudes from Pan-STARRS1 DR2 (triangles: $gri$), \textit{Gaia}-DR3 (circles: $G, BP, RP$), ZTF (upper triangles: $g, r$), and ATLAS (diamonds: $c, o$) are shown. The vertical dashed line indicates the epoch of the OSIRIS spectroscopy.}
    \label{fig:LC52}
\end{center}
\end{figure*}

\begin{figure}[h!]
\begin{subfigure}[t]{0.49\textwidth}
\centering
    \includegraphics[width=1\columnwidth]{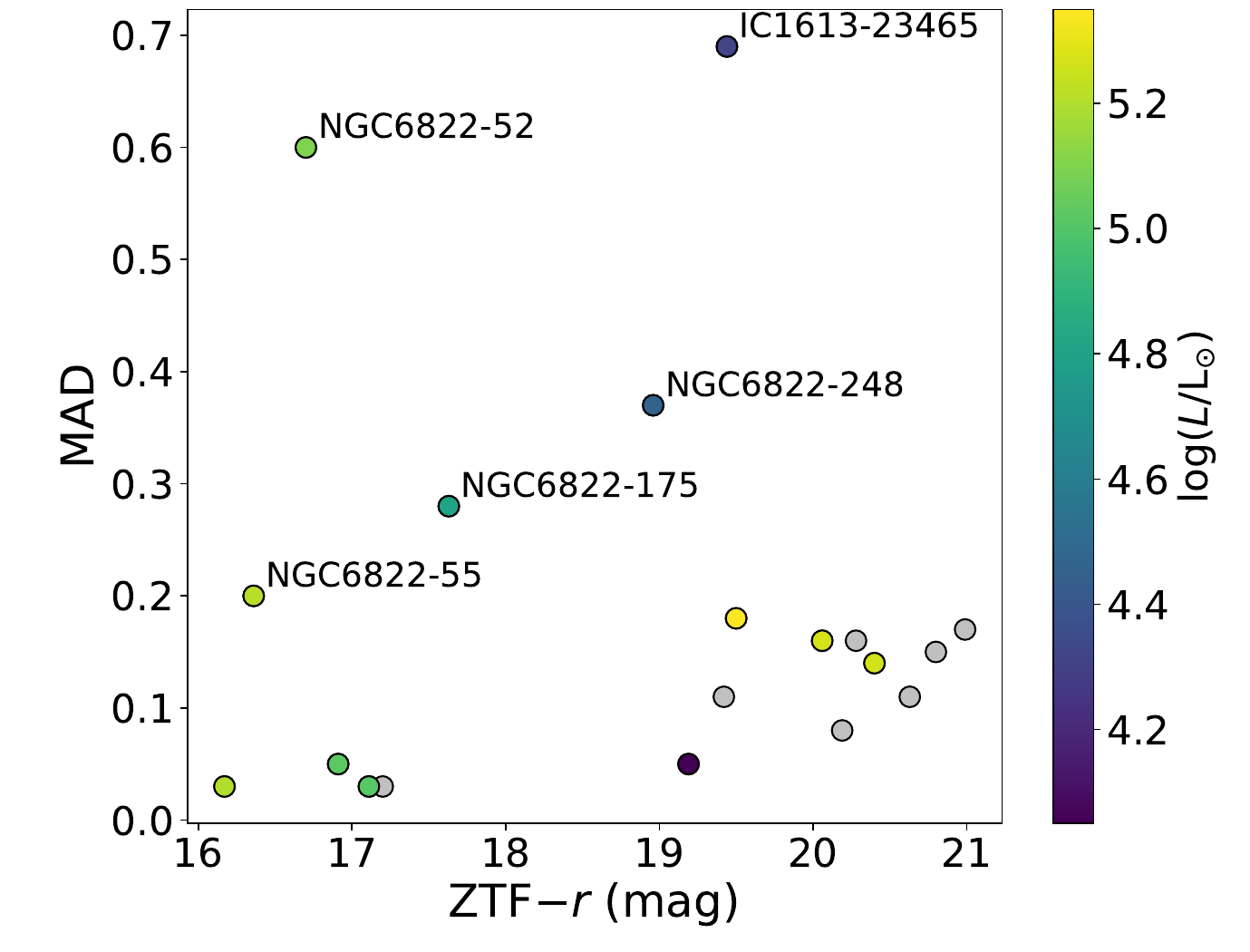}
\end{subfigure}
\begin{subfigure}[t]{0.49\textwidth}
    \includegraphics[width=1\columnwidth]{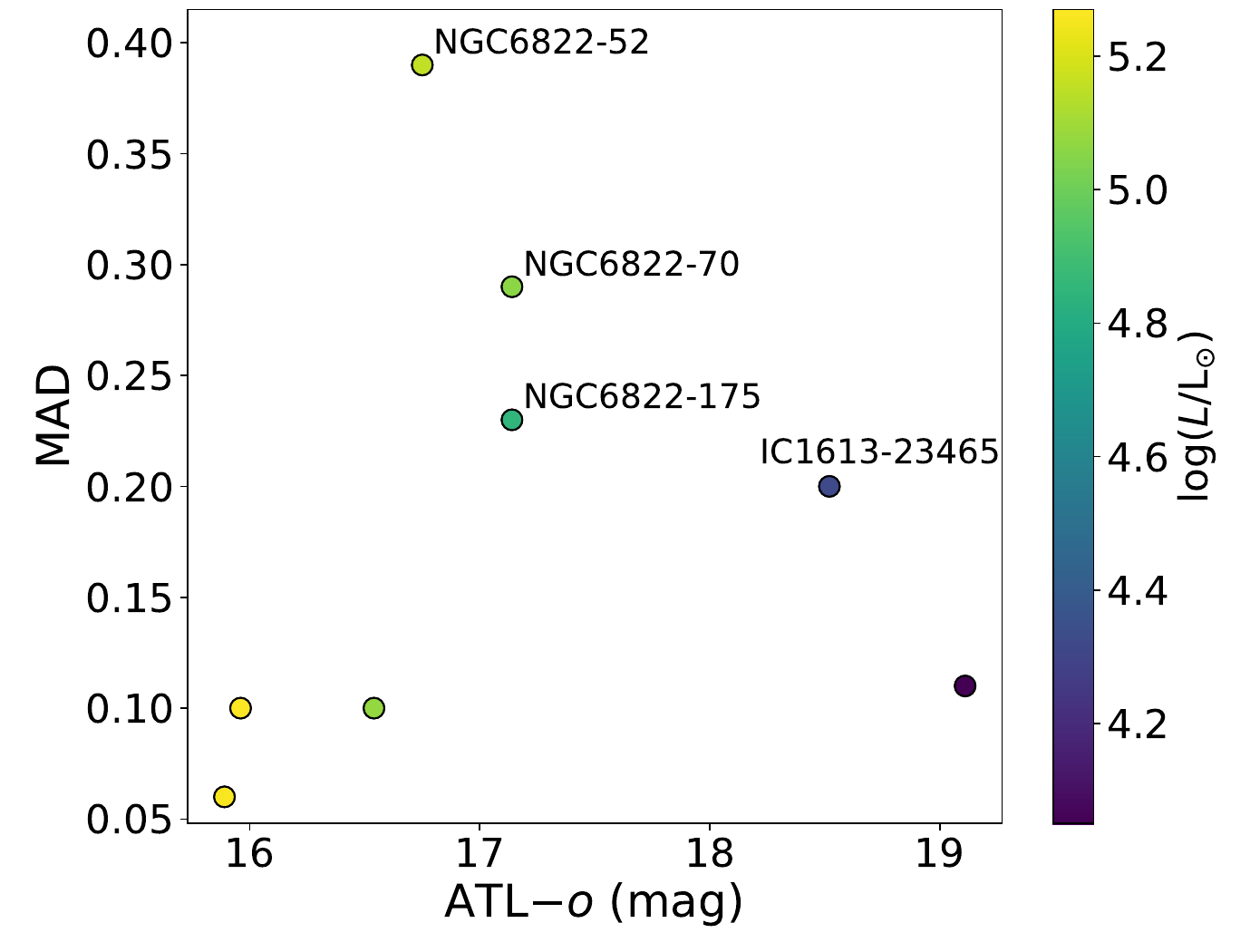}
\end{subfigure}
\begin{subfigure}[t]{0.49\textwidth}
    \includegraphics[width=1\columnwidth]{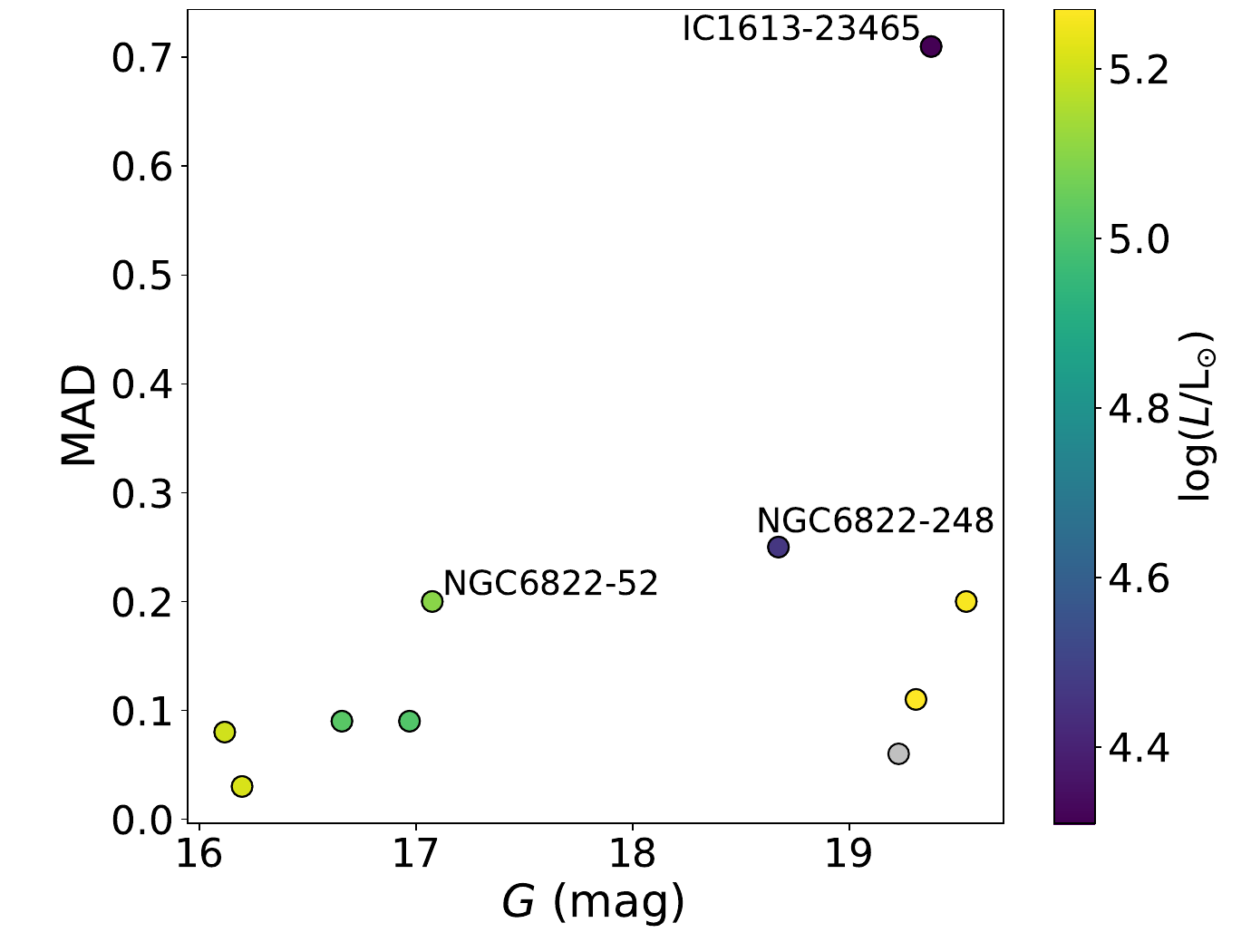}
\end{subfigure}
\caption{Median Absolute Deviations vs. magnitude diagram of RSGs in the ZTF$-r$ (\textit{top}), ATLAS$-o$ (\textit{middle}), and \textit{Gaia}$-G$ (\textit{bottom}) bands. Variable RSGs with firm classifications are labeled.}
\label{fig:MADs}
\end{figure}

We constructed light curves of the RSGs to search for correlations of variability with mass loss \citep[e.g.,][]{Yang2018, Antoniadis2024}. Optical multi-epoch photometry from \textit{Gaia}~DR3, ZTF, the Asteroid Terrestrial-impact Last Alert System \citep[ATLAS;][]{ATLAS2018}, and Pan-STARRS1~DR2 were collected; the ZTF and ATLAS data from the same night were binned by taking the median magnitude per night. Fig.~\ref{fig:LC52} presents the light curve of the well-sampled NGC6822-55, while Fig.~\ref{fig:LC1} plots the light curves of the other 19 RSGs. Four targets in IC~10 are only well-sampled in the $izy$ bands due to the high foreground extinction. The cadence provided by the ZTF light curves is sufficient to study their periodicity \citep[see the analysis of 7 targets by][]{Christodoulou2025}. Fig~\ref{fig:filters} shows the transmission curves of each filter as a reference.

We computed the Median Absolute Deviations (MAD) of each RSG in the ZTF-$r$, ATLAS-$o$, and $G$ bands (Table~\ref{tab:grand}) and present the MAD vs. magnitude in Fig.~\ref{fig:MADs}. To calculate the MAD, we require a target to have at least two measurements in one band, with an uncertainty of less than 0.1~mag. We consider a RSG as a certain variable when the MAD value is at least 1$\sigma$ above the general distribution for a given magnitude. The most variable RSGs among our sample (NGC6822-52, NGC6822-55, and IC10-63627) are also among the most luminous ($\log(L/\lsun)>5.10$) and, therefore, are expected to have the highest mass-loss rates. The two most luminous (IC10-5660 and IC10-9165) RSGs either do not have good coverage or high-quality photometry to properly asses their variability. Additionally, NGC6822-175 exhibits high variability despite having lower luminosity ($\log(L/\lsun)=4.80 \pm 0.05$). The low-mass RSGs (i.e., 10~\msun) only reach $\log(L/\lsun)\approx5.0$, hence, NGC6822-175 might be an evolved low-mass RSG, explaining its large variability. On the other hand, we found extreme short-timescale variability --commonly seen in AGB stars-- in two RSG candidates (NGC6822-151 and IC1613-23465) and the confirmed NGC6822-248. We could not determine the luminosity of NGC6822-151, however, the other two RSGs have low luminosities $\log(L/\lsun)<4.50$, at the boundary where RSGs and massive AGBs overlap (see Sect.~\ref{sec:hrds} for more discussion).

\subsection{Grid of models and spectral fitting method} \label{subsec:grid}

To obtain the properties of the RSGs, we fitted \textsc{marcs} model atmospheres \citep{Gustafsson2008} to our spectra. The wavelength range covers the Ca~\textsc{ii} (i.e., $\lambda$8498, $\lambda$8542, and $\lambda$8662) triplet and multiple TiO bands (i.e., $\lambda$6150, $\lambda$6650, $\lambda$7050, $\lambda$7700, and $\lambda$8420), constraining on the surface gravity (log~$g$) and effective temperature ($T_{\rm eff}$), respectively. We used the spherical \textsc{marcs} model grid presented by \citet{deWit2024}, ranging from 3300--4500~K and $-0.5$ to $1.0$ for the $T_{\rm eff}$ and log~$g$, respectively. For the metallicity, we approximated the observed values for NGC~6822 \citep[$Z = 0.3 \,\ \rm Z_{\odot}$;][]{Patrick2015}, IC~10 \citep[$Z = 0.3 \,\ \rm Z_{\odot}$;][]{Polles2019} and IC~1613 \citep[$Z = 0.2 \,\ \rm Z_{\odot}$;][]{Bresolin2007} to the closest metallicity of our model grid (log~$(Z/Z_{\odot}) = -0.5$~dex). We modeled $T_{\rm eff}$, log~$g$, the color excess E$(B-V)$, and the radial velocity of the system $v_{\rm rad}$ as free parameters. We fixed $Z$ to break its degeneracy with the $T_{\rm eff}$ when using the TiO bands.

Each model was downsampled to match the resolving power of the data: $R$$\sim$800 for the MOS data and $R$$\sim$1000 for the long-slit spectra. Similar to \citet{deWit2023}, we first fitted the Ca~\textsc{ii} triplet to constrain log~$g$ separately from the other parameters. The inclusion of the log~$g$ in the main fit, containing all regions of the spectrum, tends to overestimate its value. We then fixed the log~$g$ when fitting the TiO bands for $T_{\rm eff}$ and the spectral slope for E$(B-V)$. Lastly, we obtained $v_{\rm rad}$ from a cross-correlation of the model to the spectrum.

To fit the models to each star, we used \textsc{ultranest} \citep{Ultranest}, a Bayesian nested sampling approach \citep{Skilling2004} to establish the maximum likelihood model for each star. For each parameter, we assume a flat prior. \textsc{ultranest} then computes the posterior distributions for each free parameter ($T_{\rm eff}$, log~$g$, $v_{\rm rad}$ and E$(B-V)$) and finds physical correlations between each parameter pair. Specifically for $T_{\rm eff}$, \textsc{ultranest} obtained unrealistically low uncertainties, therefore we adopted a minimal uncertainty of 50~K.

\subsection{Modeled properties} \label{subsec:properties}

\begin{table*}[h]
\centering
\small
\caption{Stellar parameters of the RSG sample.}
\label{tab:params}
\renewcommand{\arraystretch}{1.2}
\begin{tabular}{l | c | r c c c r r c r}
\hline\hline
ID & Spectral & S/N & Flags\tablefootmark{a} & $T_{\rm eff,TiO}$ & E$(B-V)$ & log~$g$ & $v_{\rm rad}$ & log($L/\rm L_{\odot}$) & $R/\rm R_{\odot}$ \\ 
& Class. & & & (K) & (mag) & (dex) & (km s$^{-1}$) & (dex) &  \\  
\hline                
NGC6822-52 & M4--M6 I: & 61 & 1,2 & 3350$\pm 50$ & 0.30$\pm0.01$ & 0.98$\pm 0.01$ & $-$55$\pm6$ & 5.10$\pm$0.06 & 1050$^{+110}_{-100}$ \\
NGC6822-55 & K5--M0 I & 23 & - & 3960$\pm 50$ & 0.20$\pm 0.02$ & 0.2$ \pm 0.2$ & $-$79$\pm6$ & 5.21$\pm$0.05 & 850$\pm 70$ \\
NGC6822-66 & M0--M2 I & 54 & - & 3740$\pm 50$ & 0.14$\pm 0.01$ & 0.05$\pm 0.08$ & $-$79$\pm3$ & 5.19$\pm$0.05 & 940$\pm 80$ \\
NGC6822-70 & M2--M4 I & 51 & - & 3510$\pm 50$ & 0.24$\pm 0.01$ & 0.06$\pm 0.09$ & $-$32$\pm4$ & 5.01$\pm$0.05 & 860$^{+80}_{-70}$ \\
NGC6822-103 & M0--M2 I & 41 & - & 3680$\pm 50$ & 0.16$\pm 0.01$ & $-$0.07$\pm 0.09$ & 85$\pm1$ & 5.01$\pm$0.05 & 790$^{+70}_{-60}$ \\
NGC6822-151 & K5--M0 I:& 69 & 4 & 3750$\pm 50$ & 0.10$\pm 0.01$ & $-$0.44$\pm 0.04$ & $-$169$\pm3$ & - & - \\
NGC6822-175 & M2--M4 I: & 52 & 2 & 3530$\pm 50$ & 0.19$\pm 0.01$ & 0.8$\pm 0.1$ & $-$20$\pm4$ & 4.81$\pm$0.05 & 670$\pm 60$ \\
NGC6822-248 & M2--M4 I& 46 & - & 3420$\pm 50$ & 0.39$\pm 0.01$ & 0.70$\pm 0.13$ & 0$\pm 6$ & 4.46$\pm$0.04 & 480$\pm 40$ \\
NGC6822-407 & M2--M4 I: & 7 & 4 & 3510$\pm 50$ & 0.12$ \pm 0.05$ & 0.7$^{+0.2}_{-0.4}$ & 56$\pm 23$ & - & - \\
NGC6822-R10 & M0--M2 I & 14 & - & 3610$\pm 50$ & 0.43$\pm 0.02$ & 0.3$\pm 0.3$ & 61$\pm13$ & - & - \\
IC10-5545 & K0--K5 I & 14 & - & 4250$\pm 100$ & 0.74$\pm 0.05$ & $-$0.3$^{+0.2}_{-0.1}$ & $-$520$\pm15$ & 4.61$\pm$0.06 & 370$^{+50}_{-40}$ \\
IC10-5660A & M4--M6 I:& 48 & 3 & 3450$\pm 50$ & 0.45$\pm 0.01$ & 0.4$\pm 0.1$ & $-$170$\pm5$ & 5.27$\pm$0.06 & $1210\pm120$ \\
IC10-5660B$^b$ & M0--M2 I:& 25 & - & 3530$\pm 50$ & 0.44$\pm 0.02$ & 0.5$\pm 0.2$ & $-$302$\pm9$ & 5.27$\pm$0.06 & $1160\pm90$ \\
IC10-9165A & M4--M6 I:& 53 & 3 & 3360$\pm 50$ & 0.94$\pm 0.01$ & 0.7$\pm 0.1$ & $-$140$\pm6$ & 5.26$\pm$0.03 & $1260\pm80$ \\
IC10-9165B$^b$ & M0--M2 I& 16 & - & 3510$\pm 50$ & 1.31$\pm 0.04$ & 0.5$\pm 0.3$ & $-$336$\pm12$ & 5.26$\pm$0.03 & $1150\pm70$ \\
IC10-20296 & M2--M4 I:& 16 & 2,4 & 3400$\pm 50$ & 0.84$\pm 0.06$ & 0.92$^{+0.06}_{-0.14}$ & $-$455$\pm21$ & 4.13$\pm$0.07 & 330$^{+40}_{-30}$ \\
IC10-26089 & K0--K5 I& 36 & - & 4250$\pm 50$ & 1.04$\pm 0.01$ & $-$0.44$^{+0.07}_{-0.04}$ & $-$240$\pm5$ & 5.28$\pm$0.04 & 800$\pm 60$ \\
IC10-26929 & M4--M6 I:& 47 & 2,3 & 3330$^{+50}_{-30}$ & 1.05$\pm 0.01$ & 0.97$^{+0.02}_{-0.05}$ & $-$181$\pm8$ & - & - \\
IC10-35694$^b$ & M0--M2 I& 22 & - & 3530$\pm 50$ & 0.67$\pm 0.02$ & 0.2$\pm 0.2$ & $-$317$\pm 9$ & - & - \\
IC10-52406$^b$ & M0--M2 I& 18 & - & 3420$\pm 50$ & 1.71$\pm 0.04$ & $-$0.1$\pm 0.2$ & $-$435$\pm11$ & 5.07$\pm$0.03 & 980$^{+70}_{-60}$ \\
IC10-63627$^b$ & M0--M2 I & 25 & - & 3650$\pm 50$ & 0.83$\pm 0.02$ & 0.2$\pm 0.2$ & $-$384$\pm8$ & 5.35$\pm$0.03 & 1180$^{+80}_{-70}$ \\
IC10-66828$^b$ & M0--M2 I:& 38 & 1,2 & 3670$\pm 50$ & 0.10$\pm 0.01$ & 0.96$^{+0.03}_{-0.06}$ & $-$355$\pm6$ & - & - \\
IC10-66981$^b$ & M0--M2 I:& 10 & 4 & 3550$\pm 50$ & 0.69$\pm0.04$ & 0.4$\pm 0.4$ & $-$353$\pm25$ & - & - \\
IC10-R3$^c$ & M0--M2 I:& 69 & 2 & 3630$\pm 50$ & 0.23$\pm 0.01$ & 1.00$\pm 0.01$ & $-$167$\pm3$ & 4.05$\pm$0.08 & 270$\pm 30$ \\
IC10-R10 & M0--M2 I:& 11 & 4 & 3480$\pm 50$ & 0.85$\pm 0.05$ & 0.4$\pm 0.4$ & $-$402$^{+27}_{-28}$ & - & - \\
IC10-R16 & K5--M0 I:& 9 & 4 & 3600$\pm 50$ & 0.88$\pm 0.07$ & 0.1$\pm 0.5$ & $-$390$\pm 40$ & - & - \\
IC10-R17 & M2--M4 I:& 18 & 4 & 3420$\pm 50$ & 0.36$\pm 0.02$ & $-$0.43$^{+0.10}_{-0.05}$ & $-$216$\pm 15$ & - & - \\
IC10-R20 & M0--M2 I:& 12 & 2,4 & 3620$\pm 50$ & 0.82$\pm 0.05$ & 0.70$\pm 0.3$ & $-$480$\pm 30$ & - & - \\
IC1613-23465$^b$ & M0--M2 I:& 44 & 2,4 & 3540$\pm 50$ & 0.17$\pm 0.01$ & 1.00$\pm 0.01$ & $-$244$\pm5$ & 4.35$\pm$0.09 & 400$^{+60}_{-50}$ \\
IC1613-28306$^b$ & GK I& 41 & - & 4380$\pm 50$ & 0.09$\pm 0.01$ & 0.23$\pm 0.09$ & $-$278$\pm4$ & 4.06$\pm$0.09 & 190$^{+30}_{-20}$ \\
\hline
\end{tabular}

\tablefoot{Reported uncertainties are 68.3\% confidence level errors. The v$_{\rm rad}$ of the MOS spectra have systematic uncertainties of $\sim$60 km~s$^{-1}$. The ending A/B in the ID indicates different observations of the same source.\\
\tablefoottext{a}{Flags: (1) bad fit to spectrum, star too cool, parameters unreliable.
(2) log~$g$ near the edge of the grid, lower limit.
(3) acceptable fit, reliable parameters, but to be interpreted with caution.
(4) bad fit to Ca II triplet, surface gravity unreliable.} \\
\tablefoottext{b}{Long-slit spectra}\\
\tablefoottext{c}{The high S/N, high log~$g$, and low luminosity suggest a foreground nature for this star.}
}

\end{table*}

\begin{figure}[h]
\begin{subfigure}[t]{0.5\textwidth}
\centering
    \includegraphics[width=1.0\columnwidth]{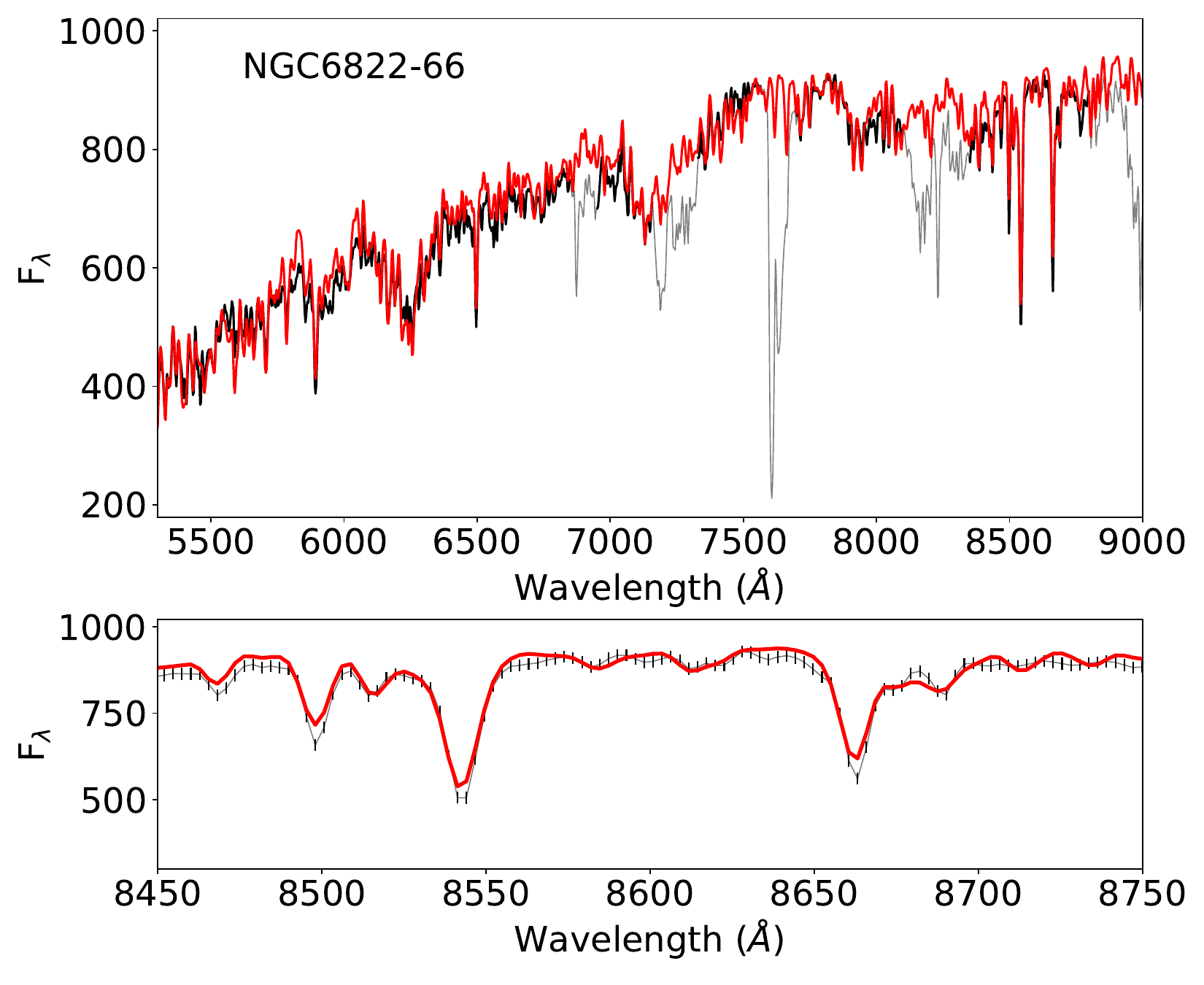}
\end{subfigure}
\caption{\textsc{marcs} model (red) fitted to the spectrum of NGC6822-66 (black). Gray regions are excluded from the fit due to contamination by telluric absorption. F$_\lambda$ has units $\rm 10^{-18}\,\ erg \,\ cm^{-2}s^{-1}A^{-1} $. \textit{Top}: The overall OSIRIS spectrum. \textit{Bottom}: Zoom in on the Ca~\textsc{ii} triplet.}
\label{fig:fit66}
\end{figure}

\begin{figure}[h]
\begin{subfigure}[t]{0.5\textwidth}
\centering
    \includegraphics[width=1.0\columnwidth]{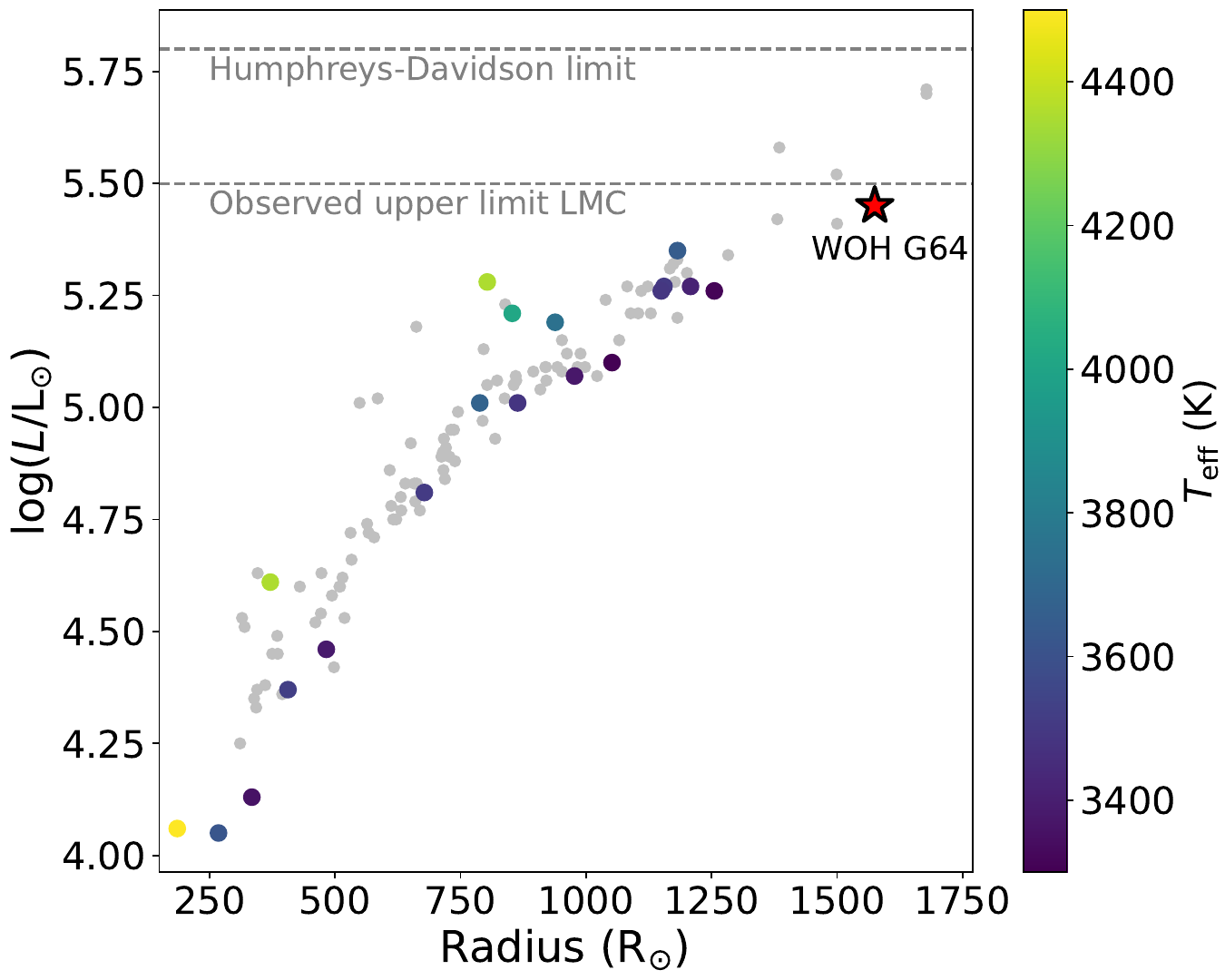}
\end{subfigure}
\caption{Radius vs. luminosity of RSGs in the sample. Gray sources are from the ASSESS southern sample \citep{deWit2024}. The color indicates the effective temperature. We compare it to the extreme WOH~G64 in its RSG state.}
\label{fig:RadvsL}
\end{figure}

Table~\ref{tab:params} presents the best-fit properties of the 30 RSG spectra. Fig.~\ref{fig:fit66} shows an example fit for NGC6822-66, a dusty P1 target classified as M0--M2~I. The top panel shows the quality of the best-fit \textsc{marcs} model to our spectrum, illustrating the reliability of the derived extinction coefficients and effective temperature. The bottom panel zooms in on the Ca~\textsc{ii} triplet, showing the reliability of the derived surface gravity and radial velocity. Fig.~\ref{fig:fitAll} shows the other fitted spectra.

Stars with problems with the fitting were assigned one of four flags (see the third column of Table~\ref{tab:params}). Two spectra cannot be reproduced by the \textsc{marcs} models (flag = 1) because either the TiO bands were too extreme (NGC6822-52) or because we obtained an unreliable fit by eye (IC10-66828). Several stars have a weaker Ca~\textsc{ii} triplet, either due to a strong TiO band hiding the strength of the absorption line ($T_{\rm eff} \lesssim 3400$~K), or because their envelopes have not yet fully expanded, assuming that these are indeed RSGs. These then yielded values of log~$g$ near the upper limit of the grid at log~$g = 1.0$. Therefore, only the lower limits of their respective surface gravities could be derived (flag = 2). Some RSGs show TiO bands in minor disagreement with the \textsc{marcs} models, yielding trustworthy results (flag = 3). Finally, some spectra had bad Ca~\textsc{ii} triplet fits, yielding unreliable values for the surface gravity (flag = 4). Uncertain classifications with high surface gravities may be reclassified in the future as oxygen-rich AGB contaminants. Finally, IC10-R3 may be a foreground star, given that such a high S/N (the highest in our RSG sample) is not expected for a low-luminosity RSG.

MOS spectra suffered from wavelength calibration issues due to a reduced wavelength range for slits near the edges and the rotator angle, which can induce shifts and therefore systematic errors up to 60 km~s$^{-1}$. Due to this, the spread in radial velocities of sources in IC~10 is large, while for the long-slit spectra, the spread is more realistic. This issue explains the $v_{\rm rad}$ difference between the MOS and long-slit observations of IC10-5660 and IC10-9165.

Following \citet{deWit2024}, we computed the stellar radius and reported it in the last column of Table~\ref{tab:params}. Fig.~\ref{fig:RadvsL} compares the luminosity and radius for the present sample with the results from \citet{deWit2024}. The figure includes WOH G64 \citep{Ohnaka2008, Levesque2009}, the largest, most dusty RSG known in the Local Group, with a mass-loss rate of $\sim 10^{-4} \rm \,\ M_{\odot} yr^{-1}$, before its transition to a yellow hypergiant \citep{MunozSanchez2024b}. The smallest source (IC1613-28306) was classified as a GK supergiant, possibly indicating a star at the beginning of the RSG phase.

\section{Discussion}\label{sec:discussion}

The fraction of evolved massive stars among the classified P1 and P2 targets is 67\% and 60\%, respectively, confirming the success of our priority system. The fractions decrease down to 37\% for P5 and 17\% for random targets. In particular, the P5 class suffered from extra contamination of five carbon stars and an H~\textsc{ii} region in the NGC~6822 sample. We could have avoided this by using the newest distance available for this galaxy \citep[$0.45 \pm 0.01$~Mpc,][]{Zgirski2021}. All priority-selected RSGs were P1, P2, or P5. Finally, H~\textsc{ii} regions were found to contaminate the P1 and P5 targets \citep[as in][]{Bonanos2024}.

\subsection{Evolutionary diagrams}\label{sec:hrds}

\begin{figure}[h]
\begin{subfigure}[t]{0.5\textwidth}
\centering
    \includegraphics[width=0.98\columnwidth]{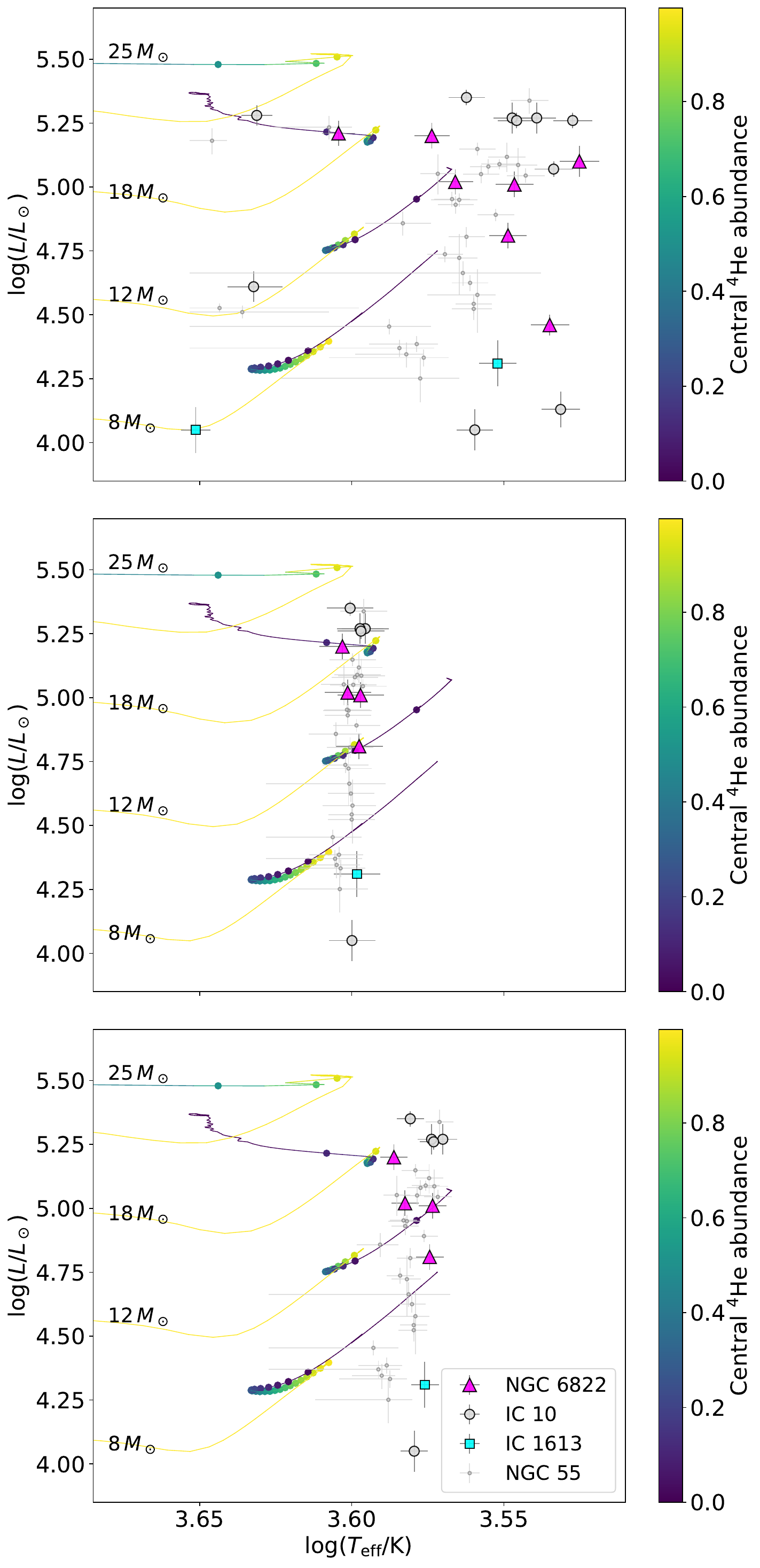}
\end{subfigure}
\caption{HR diagram of RSGs in NGC~6822, IC~10 and IC~1613, compared to RSGs in NGC~55 \citep{deWit2024}. Four \textsc{Posydon} evolutionary tracks for $Z/\rm Z_{\odot} = 0.2$ are overplotted; color indicates the central helium abundance, while dots indicate intervals of 10,000~yrs. Each panel includes effective temperature measurements from the following methods: TiO-band fits (\textit{top}), the $i$-band scaling relation applied to our TiO temperatures (\textit{middle}), and the $J$-band scaling relation (\textit{bottom}).}
\label{fig:HRDs}
\end{figure}

To compare the obtained properties to model predictions, we constructed Hertzsprung-Russell (HR) diagrams. We computed evolutionary tracks at $Z \sim 0.2 \,\ \rm Z_{\odot}$ using the grid of \textsc{Posydon} models \citep{Fragos2023, Andrews2024} to match the metallicities of the RSGs in the sample. Following \citet{deWit2024}, we assumed an exponential convective core overshoot, stellar winds that follow the "Dutch" recipe \citep[][]{Jager1988, Vink2001}, a mixing length fixed at $\alpha_{\rm MLT} = 1.93$, and no stellar rotation. The tracks evolve until the end of carbon-core burning.

Figure~\ref{fig:HRDs} compares the positions of the 17 RSGs for which we derived their luminosity to four illustrative tracks and to the RSGs in NGC~55 from \citet{deWit2024}. The top panel shows $T_{\rm eff}$ based on TiO-band fits.  We included IC10-5660 and IC10-9165 twice since we measured their temperature from two spectral epochs. Most sources occupy the forbidden zone, the cool region to which the evolutionary tracks do not extend. This result has been reproduced by many studies \citep[e.g.,][]{Davies2013, deWit2024} and is thought to be connected to the mass loss of RSGs affecting the optical temperature diagnostics \citep{Davies2021}. Moreover, if the metallicity of the RSGs differs significantly from the default assumed value for each galaxy, $T_{\rm eff,TiO}$ is affected. However, this cannot explain the systematic discrepancies, given that the effect of metallicity is symmetric (i.e., shifting temperatures to both slightly lower and higher values). 

We then applied the scaling relations from \citet{deWit2024} to ten RSGs to scale the $T_{\rm eff,TiO}$ up to more representative values. The remainder of the RSGs were dropped after this point, as the relations may only be applied to RSGs in the range 3450--3950~K. We show the $i$-band-scaled temperatures and $J$-band-scaled temperatures in the middle and bottom panels of Fig.~\ref{fig:HRDs}, respectively. In contrast to \citet{deWit2024}, the hotter $J$-band-scaled temperatures align better with the evolutionary predictions, strengthening the argued use of the $J$-band diagnostics to measure the effective temperature from RSG spectra. However, similar to \citet{deWit2024}, we refrain from a firm conclusion based on the evolutionary tracks alone, given the plethora of assumptions going into the \textsc{Posydon} models. \citet{Christodoulou2025} have performed a $J$-band spectral analysis for NGC6822-55, NGC6822-70, NGC6822-103, and NGC6822-175, determining more accurate temperatures of these RSGs. 

Four of our low-luminosity RSGs (IC10-20296, IC10-R3, IC1613-23465, and NGC6822-248) are too cool to reconcile with the 8~\msun\ track. Their low temperature cannot be explained by mass loss alone because the low-luminosity RSGs are expected to experience weak winds \citep[e.g.,][]{Antoniadis2024}. Optical photometry was not available for IC10-20296 and IC10-R3, but the large amplitude and short-timescale variability of IC1613-23465 and NGC6822-248 differ from those of the other RSGs. Among them, only NGC6822-248 was robustly classified as a RSG based on spectral features. RSGs and AGBs overlap in luminosity between $4.0<\log(L/\lsun)<4.5$, hence, these four targets might be evolved stars with masses $<8$~\msun. A more detailed study of these four objects is needed to determine their nature.

\subsection{Dusty RSGs}\label{sec:dusty}

Clues of episodic mass loss may lie in the circumstellar dust. We consider a RSG dusty when its infrared $[3.6]- [4.5]$ color exceeds 0.1~mag and its absolute magnitude M$_{3.6}\le-9.0$~mag. Out of the 28 RSGs, 20 satisfy both of these criteria, 7 random targets classified as RSGs did not have mid-IR photometry, and 1 random target had blue mid-IR colors. We compared the properties of the dusty RSGs to the statistics presented by \citet{deWit2024}. They reported a median temperature of 3570~K for 27 dusty RSGs compared to a median temperature of 3630~K for 84 non-dusty RSGs. Considering the 20 known dusty RSGs in our sample (with two duplicate measurements; 22 measured effective temperatures), we found a median temperature of 3533~K, which agrees with the findings of \citet{deWit2024}. This is approximately 100~K cooler than the non-dusty RSGs. This temperature difference could be due to the effect of mass loss, which simultaneously increases the molecular absorption (lowering $T_{\rm eff}$) and provides fresh dust to the circumstellar environment.

For the extinction, we use $R_V=3.1$, to convert the E$(B-V)$ to $A_V$, although the real value of the total-to-selective extinction may be higher for the circumstellar environments of massive stars \citep[e.g.,][]{Massey2005, MaizJesus2014, Brands2023}. Given the large foreground extinction of IC~10, we cannot reliably compare the median $A_V$ of our sample to those presented by \citet{deWit2024}. For this comparison, we only include RSGs from NGC~6822 and IC~1613. We calculate a median extinction of this sample of 11 dusty RSGs of $A_V = 0.53$~mag, compared to the $A_V = 1.08$~mag from the previous work for dusty RSGs and $A_V = 0.46$~mag for non-dusty RSGs. Despite the conservative criteria, we still find a marginally higher median extinction than the non-dusty sample. None of the 11 RSGs passes the criteria to be considered a dust-enshrouded RSG \citep{Beasor2022, deWit2024}.

Lastly, \citet{deWit2024} reported a median log$(L/$L$_{\odot}) = 5.09$ for dusty RSGs and log$(L/$L$_{\odot}) = 4.88$ for non-dusty RSGs. In this work, we derive log$(L/$L$_{\odot}) = 5.01$ from our sample of 20 dusty RSGs. We argue that the lower luminosity is a natural result of observing nearby galaxies and, as such, the ability to probe the lower end of the luminosity distribution. Each of these medians agrees with the conclusions of \citet{deWit2024} that dusty RSGs are, as a population, more evolved than dust-free RSGs. If dusty RSGs are more evolved, they should also be more variable. Each RSG indicated in Fig.~\ref{fig:MADs}, showing a high degree of variability, also shows IR excess.

For those without \textit{Spitzer} colors, for which we are unable to determine if they are dusty RSGs, and for those without dust, only one source was classified as a potential late M-type RSG (M2-M4~I:, ID: IC10-R17), whereas the other seven sources are either K or early M-types. This contrasts sharply with the sample of known dusty RSGs, which has nine late-type RSGs, around 40$\%$ of the sample. This can be explained as the TiO bands, which increase in strength with higher mass-loss rates, are used for classification. Higher mass-loss rates should, therefore, naturally lead to later-type classifications \citep{Davies2013, Davies2021}.

\subsection{Episodic mass loss}

We relied on mid-IR indicators to detect massive evolved stars \citep{Bonanos2009, Bonanos2010}. In our southern survey \citep[i.e.,][]{Bonanos2024}, we discovered massive stars in transitional phases where eruptive or episodic mass loss occurs, such as LBV candidates \citep{Maravelias2023}. In this work, we report an emission line object (NGC6822-106) that stands out as one of the brightest mid-IR sources in our sample and one of the faintest in the optical (Fig.~\ref{fig:cmd}). Its extreme $G-$[3.6] color (7.5 mag) closely resembles that of WOH~G64 ($G -$[3.6] = 9.2 mag), the most extreme RSG in the Large Magellanic Cloud, which recently transitioned into a Yellow Hypergiant \citep{MunozSanchez2024b}. NGC6822-106 has a redder color than Var~A ($G- \rm[3.6]=5.0$ mag), which is a Yellow Hypergiant from M33 that underwent an eruption for over 35 years \citep{Hubble1953, Humphreys2006}. The significant reddening and the strong Ca~\textsc{ii} triplet emission suggest that NGC6822-106 is undergoing substantial mass loss, making it a compelling target for future studies. Its optical faintness prevented us from constructing a light curve and analyzing the potential eruption.

Several studies have reported spectral-type variations in RSGs \citep[e.g.,][]{Levesque2007, Dorda2016} which are directly connected to changes in TiO strength. These variations can arise through different mechanisms. The rise and fall of convective cells at the stellar surface alter both the brightness and TiO absorption features of RSGs, creating hysteresis loops \citep{Kravchenko2019, Kravchenko2021}. Episodic mass ejections can also significantly impact the spectral type and photometric variability \citep[e.g.,][]{Massey2007, Humphreys2021, Dupree2022, Anugu2023, MunozSanchez2024}. Consequently, long-term light curves are essential for identifying disruptive events associated with episodic mass loss and disentangling them from convection effects.

Two of our sources, IC10-5660 and IC10-9165, exhibit spectral variability between our two epochs. Both are luminous ($\log(L/\lsun) > 5.2$), dusty RSGs that shifted from a late to an early M type over a span of two years. The \textit{Gaia} variability of IC10-9165 prior to our observations, along with the higher extinction observed in the second epoch, suggest a mass ejection event near our first epoch. However, due to the lack of photometric coverage during our observations, we cannot confirm it. In contrast, IC10-5660 maintains constant brightness and extinction between epochs, making its spectral change more puzzling and without a clear cause. 

We also classified four dusty RSGs as K-type supergiants: IC1613-28306, NGC6822-55, IC10-5545, and IC10-26089. For the targets in IC~10, we investigated whether the high foreground extinction could explain their reddened \textit{Spitzer} colors. Assuming a foreground extinction of $A_V =4.3$~mag \citep{Schlafly2011}, we estimate that foreground dust contributes approximately [3.6]$-$[4.5] $\sim 0.05$~mag. After correcting for this factor, a significant infrared excess remains, confirming these objects as genuinely dusty K-type RSGs. Following the recipe from \citet{Antoniadis2024, Antoniadis2025}, we estimated their mass-loss rate to be $\dot M < 10^{-7}$~$\rm M_{\odot}yr^{-1}$. These low rates imply an eruptive mass-loss mechanism causing enhanced dust production \citep[e.g.,][]{Beasor2020, Decin2024}. Their light curves do not show strong variability, indicating that they are currently in a stable state. However, the mid-IR photometry was taken 20 years before our observations. Besides the absence of photometric coverage from that period, we can only speculate that a mass ejection event may have occurred back then. A possible explanation for the GK~I supergiant IC1613-28306 is episodic mass loss during its approach to the Hayashi limit from the Yellow Supergiant phase \citep{Cheng2024}, forming dust around the supergiant.

\section{Summary}\label{sec:conclusions}

The ASSESS project \citep{Bonanos2024} has tackled the role of episodic mass loss in evolved massive stars. In this light, we have presented a new sample of observed, evolved massive stars in our northern survey, containing 3 low-metallicity galaxies: NGC~6822, IC~10, and IC~1613. We collected the photometry of 163 observed targets, obtained their spectra through multi-object and long-slit spectroscopy using GTC-OSIRIS, and spectroscopically classified 124 out of 163 spectra (122 unique objects). The majority show an infrared excess in their \textit{Spitzer} photometry. Among these, we confirmed 2 supernova remnant candidates, 2 H~\textsc{ii} regions and found 33 evolved massive stars: 28 RSGs, 2 BSGs, 1 YSG, 1 WR, and 1 emission-line object. Remarkably, the Ca~\textsc{ii} triplet emission and the obscuration of the emission-line object suggest an ongoing period of enhanced mass loss. All but 7 of the above RSGs and the WR are newly discovered evolved massive stars. This work increases the sample of RSGs in IC~10 from 1 to 17. 

We derived luminosities for 17 RSGs through SED fitting and the temperatures for 28 RSGs using the \textsc{marcs} radiative transfer models, adding significantly to the studied low-metallicity RSGs in northern galaxies. The properties of the dusty RSGs agree well with those by \citet{deWit2024}, with a median temperature of 3530~K and a median luminosity of log$(L/$L$_{\odot}) = 5.01$. The optical spectra of two of the RSGs varied over two years, changing by a few spectral types. In particular, we found larger extinction in the second epoch of one of IC10-9165, suggesting an episodic mass loss event. Lastly, we found four K-type RSGs with IR excess, reinforcing previous results by \citet{deWit2024}. We speculate that these experienced episodic mass loss, leading to excessive dust formation, giving rise to the infrared excess we observed.

Future studies can utilize photometry to optimize the process of finding and characterizing RSGs, specifically those with high mass loss that is potentially episodic. Using large surveys like ZTF, LSST, and NEOWISE, machine-learning algorithms such as those presented by \citet{Maravelias2022, Maravelias2025} can automatically predict the general spectral classes of the sources based on their colors and photometric variability. A combination of such algorithms can trigger low-risk, systematic large-scale studies of RSG populations in a variety of metallicity environments. Mapping the luminosities and temperatures of RSG populations can lead to luminosity functions of RSGs in galaxies, providing an indirect measurement of a time-averaged mass-loss rate of the RSG phase \citep{Massey2023, Zapartas2024}.

%--------------------------------------------------------------------

\begin{acknowledgements}
The authors acknowledge funding support from the European Research Council (ERC) under the European Union’s Horizon 2020 research and innovation program (ASSESS; Grant agreement No. 772086). We thank Emmanouil Zapartas and the POSYDON team for providing POSYDON evolutionary tracks prior to publication. Based on observations made with the Gran Telescopio Canarias (GTC), installed at the Spanish Observatorio del Roque de los Muchachos of the Instituto de Astrofísica de Canarias, on the island of La Palma. This work is based on data obtained with the instrument OSIRIS, built by a Consortium led by the Instituto de Astrofísica de Canarias in collaboration with the Instituto de Astronomía of the Universidad Autónoma de México. OSIRIS was funded by GRANTECAN and the National Plan of Astronomy and Astrophysics of the Spanish Government. The Digitized Sky Surveys were produced at the Space Telescope Science Institute under U.S. Government grant NAG W-2166. The images of these surveys are based on photographic data obtained using the Oschin Schmidt Telescope on Palomar Mountain and the UK Schmidt Telescope. The plates were processed into the present compressed digital form with the permission of these institutions. This research made use of Astropy, a community-developed core Python package for Astronomy \citep{astropy2013, astropy2018} and Photutils, an Astropy package for the detection and photometry of astronomical sources \citep{photutils}. "This research has made use of the VizieR catalogue access tool, CDS,
Strasbourg, France (\href{https://vizier.cds.unistra.fr/}{10.26093/cds/vizier}). The original description 
of the VizieR service was published in \citet{vizier2000}.

\end{acknowledgements}

\bibliographystyle{aa} % style aa.bst
\bibliography{assessGTC.bib} % your references Yourfile.bib
\clearpage

\begin{appendix}
\onecolumn

\clearpage

\section{Tables describing observations}

\begin{table*}[h!]
\centering
\small
\caption{Distribution of targets and target priorities per galaxy.}\label{tab:fields}
\begin{tabular}{lcccccccc}
\hline\hline
Galaxy				& P1	& P2	& P3	& P4	& P5	& P6  & R & { Total} \\
\hline \\[-9pt]
IC 10 	& {\it 2}	& {\it 3}	& {\it 6} & {\it 0} & {\it 39} & {\it 73} & - & { {123}}\\
\quad {\it Field A} 	& 1		& 2		& 2		& 0		& 16	& 16	& 18 &{ 55}	\\
\quad {\it long-slit} 	 & 0 & 0 & 0 & 0 & 5 & 0 & 25 & { 30}	\\
\ \\

IC 1613 	& {\it 2}	& {\it 0}	& {\it 0} & {\it 0} & {\it 7} & {\it 8} & - & { {17}} \\
 \quad {\it long-slit} 	 & 0 & 0 & 0 & 0 & 3 & 0 & 4 & { 7}	\\
 \ \\

NGC 6822 	& {\it 11}	& {\it 4}	& {\it 1} & {\it 0} & {\it 75} & {\it 6} & - & { {97}} \\
\quad {\it Field A} 	& 3		& 1		& 0		& 0		& 19		& 0	& 11 & { 34}	\\
 \quad {\it Field B} 	& 4		& 2		& 1		& 0		& 19		& 1	& 7 & { 35}	\\
\quad {\it long-slit} 	 & 1 & 0 & 0 & 0 & 1 & 0 & 0 & { 2}	\\

\hline
Total observed & 9 & 5 & 3 & 0 & 63 & 17 & 65 & { 163} \\
\hline
Total classified & 6 & 5 & 1 & 0 & 39 & 8 & 65 & { 124} \\
\hline
\end{tabular}

\tablefoot{Two sources in IC 10 (IC10-5660 and IC10-9165) were observed in both the MOS and long-slit modes.
}

\end{table*}

\begin{table*}[h!]
\centering
\small
\caption{Log of OSIRIS observations.}\label{tab:obslog}
\begin{tabular}{lccrrccccc}
\hline\hline
Galaxy	& Field or &  R.A.	  &  Dec. & UT Date\tablefootmark{a} & MJD\tablefootmark{b}	&   
Mode\tablefootmark{c} & $T_{\mathrm{exp}}$	& Airmass\tablefootmark{d}	& Seeing  \\
	& Target ID & \small(J2000) & \small(J2000) &	 &  &  & \small{(s)} &    & \small{(\arcsec)}\\
\hline \\[-9pt]
NGC 6822 & A & 19:44:54.07  & $-$14:50:36.3 & 29 Jun 2020 & 59029.17 & PRE &  1 $\times$ 15 & 1.55 & 1.4 \\
NGC 6822 & B & 19:44:53.67  & $-$14:44:55.1	& 29 Jun 2020 & 59029.18 &  PRE & 1 $\times$ 15 & 1.57 & 1.3 \\
IC 10 & A & 00:20:05.19	& 59:18:46.8 &	29 Jun 2020 & 59029.18 &  PRE & 1 $\times$ 15 & 1.32 & 1.4 \\
NGC 6822 & A & 19:44:54.07   & $-$14:50:36.3 & 15 Aug 2020 & 59076.94 & MOS &  2 $\times$ 1350 & 1.40 & 0.5 \\ 
NGC 6822 & A & 19:44:54.07   & $-$14:50:36.3 & 15 Aug 2020 & 59076.97 & MOS &  2 $\times$ 1350 & 1.38 & 0.6 \\ 
IC 10 & A & 00:20:05.19	& 59:18:46.8 & 16 Aug 2020 & 59077.10 &  MOS & 2 $\times$ 1350 & 1.20 & 0.7 \\ 
IC 10 & A & 00:20:05.19	& 59:18:46.8 & 16 Aug 2020 & 59077.13 &  MOS & 2 $\times$ 1350 & 1.17 & 0.7 \\ 
NGC 6822 & B & 19:44:53.67  & $-$14:44:55.1	& 16 Aug 2020 & 59077.93 &  MOS & 2 $\times$ 1350 & 1.41  & 0.8 \\ 
NGC 6822 & B & 19:44:53.67  & $-$14:44:55.1	& 16 Aug 2020 & 59077.96 &  MOS & 2 $\times$ 1350 & 1.38 & 0.8 \\
NGC 6822 & 106 &  19:45:11.69 &   $-$14:50:48.5 & 3 Aug 2022 & 59794.00 & LSS & 3 $\times$ 1200 & 1.38 & 0.8 \\ 
IC 1613 & 23465 & 01:04:17.83  &  02:13:24.7 & 6 Aug 2022 & 59797.15 & LSS & 3 $\times$ 1000 & 1.19 & 0.9 \\ 
IC 1613 & d2 & 01:04:22.33  &  02:07:28.5 & 7 Aug 2022 & 59798.18 & LSS & 3 $\times$ 1200 & 1.13 & 0.8 \\ 
IC 10 & d1 & 00:19:33.29   &  59:20:30.8 & 19 Aug 2022 & 59810.97 & LSS & 3 $\times$ 1200 & 1.56 & 0.9 \\ 
IC 10 & 35694 & 00:20:58.17  & 59:21:34.4 & 24 Aug 2022 & 59815.13 & LSS & 3 $\times$ 1200 & 1.16 & 0.9 \\
IC 10 & d2 & 00:19:58.81   &  59:19:53.9 & 24 Aug 2022 & 59815.02 & LSS & 3 $\times$ 1200 & 1.33 & 0.8 \\

\hline
\end{tabular}
\tablefoot{
\tablefoottext{a}{UT date at the start of the observations.}
\tablefoottext{b}{MJD at the start of the OB.}
\tablefoottext{c}{PRE, MOS, and LSS correspond to the pre-imaging, multi-object spectroscopy, and long-slit modes.}
\tablefoottext{d}{Mean value of the first and last exposure (except for PRE mode).}

}
\end{table*}

\clearpage

\section{Example spectra for highlighted classes} \label{sec:spectraAll}
\begin{figure*}[h]
\begin{center}
    \centerline{\includegraphics[width=0.95\columnwidth]{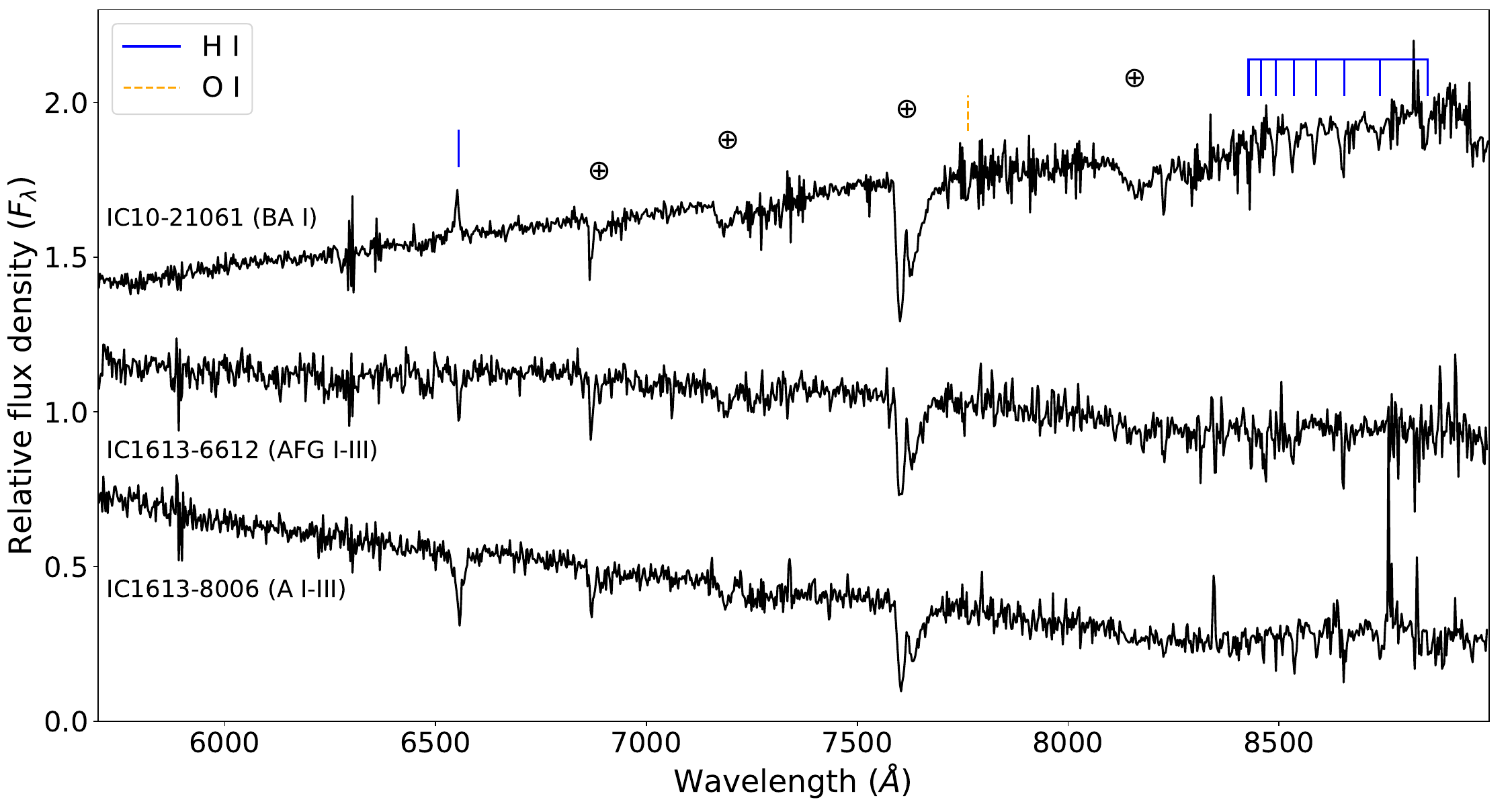}} 
    \caption{Similar to Fig.~\ref{fig:RSGexamples}, but for two BSGs and one YSG. The oxygen triplet, H$\alpha$, and the Paschen series are identified.}
    \label{fig:BYSGexamples}
\end{center}
\end{figure*}

\begin{figure*}[h]
\begin{center}
    \centerline{\includegraphics[width=0.95\columnwidth]{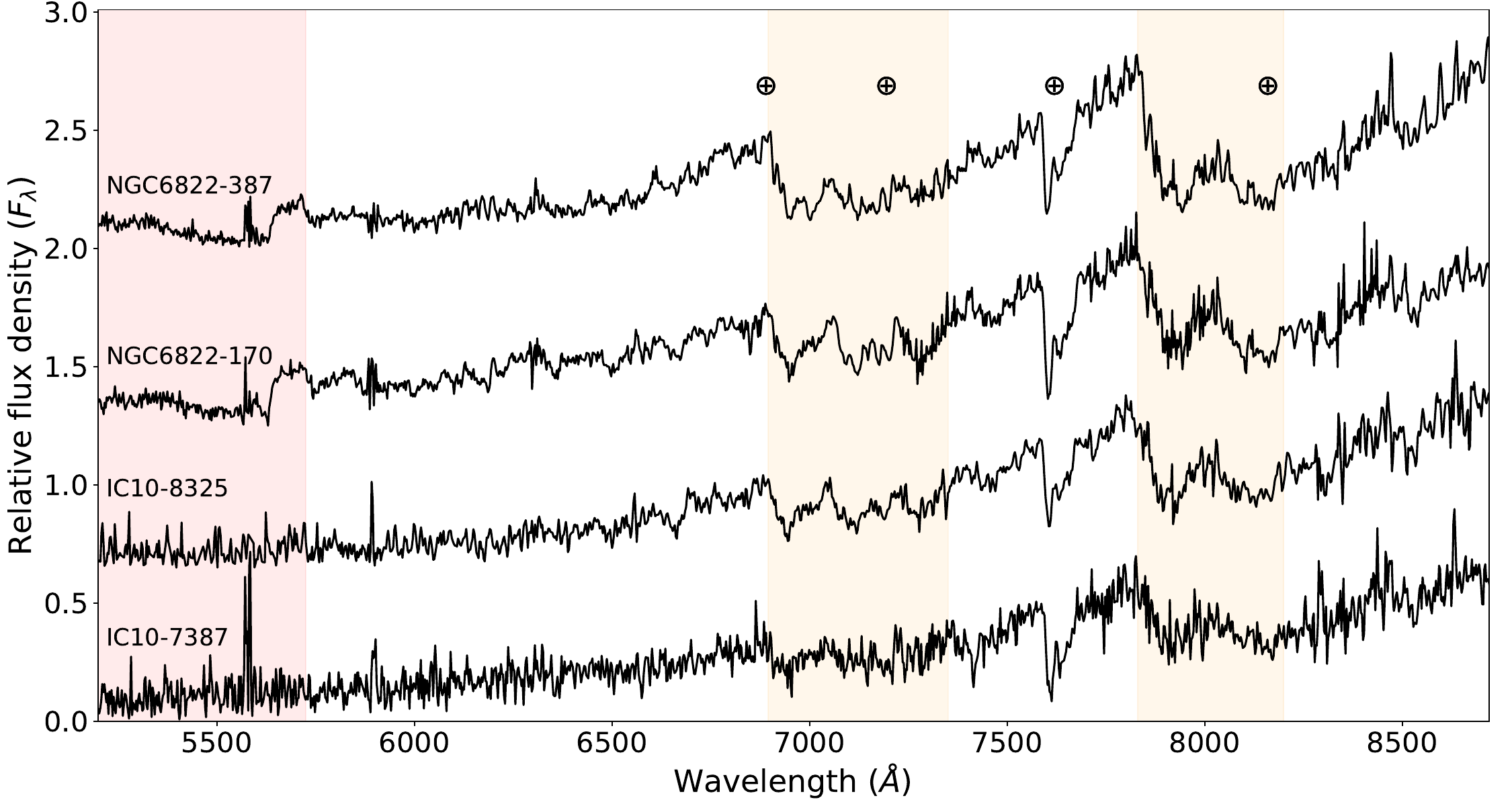}} 
    \caption{Similar to Fig.~\ref{fig:RSGexamples}, but for Carbon stars. Regions of significant carbon absorption are identified in orange (CN) and red (C$_2$, for NGC6822-387, NGC6822-170).}
    
    \label{fig:Cstarexamples}
\end{center}
\end{figure*}

\begin{figure*}[h]
\begin{center}
    \centerline{\includegraphics[width=0.95\columnwidth]{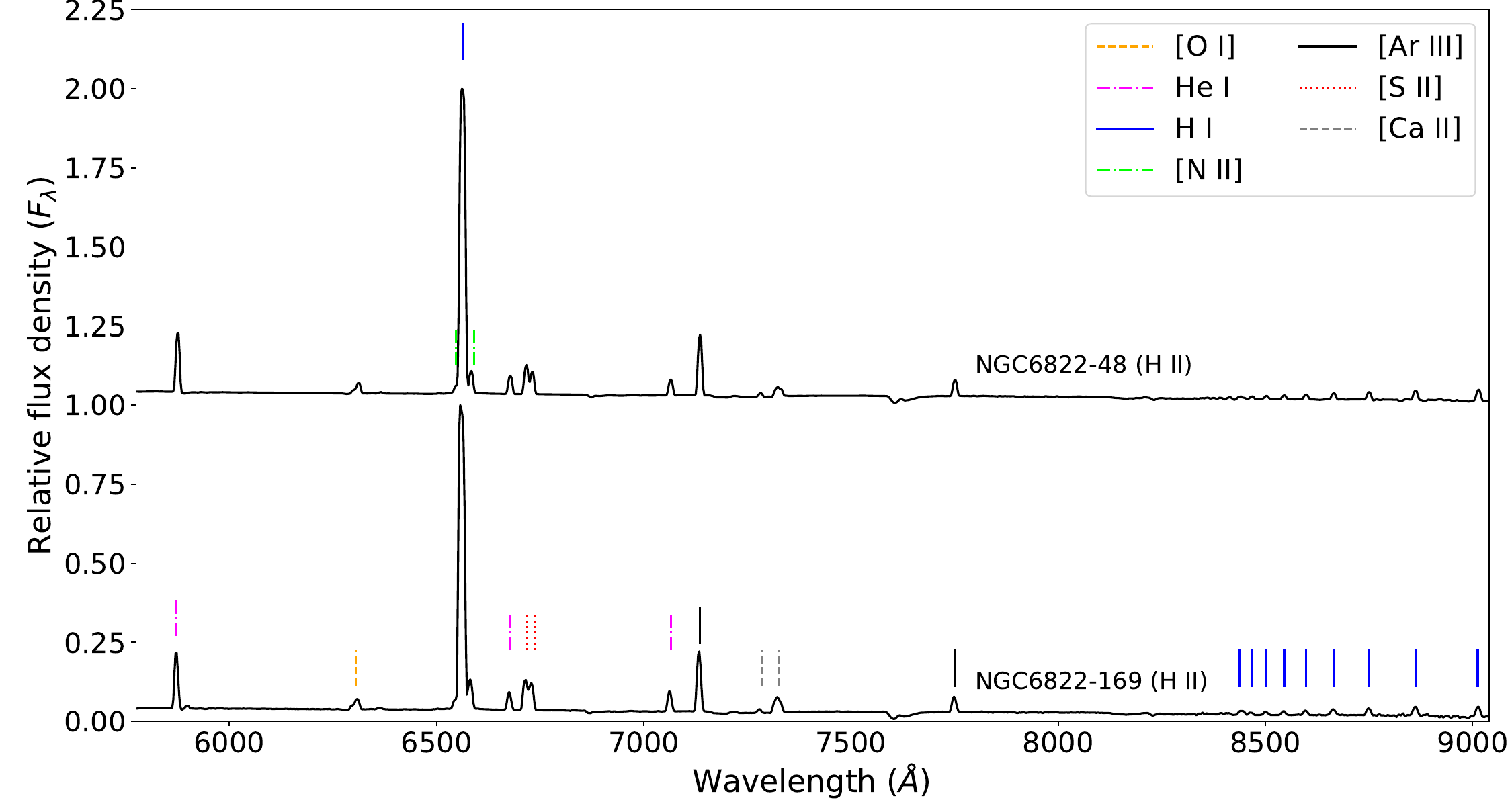}} %referee 1.1, aa 2 column width
    \caption{Similar to Fig.~\ref{fig:RSGexamples}, but for the H~\textsc{ii} regions. Primary emission lines are identified.}
    \label{fig:Hiiexamples}
\end{center}
\end{figure*}

\begin{figure*}[h]
\begin{center}
    \centerline{\includegraphics[width=0.95\columnwidth]{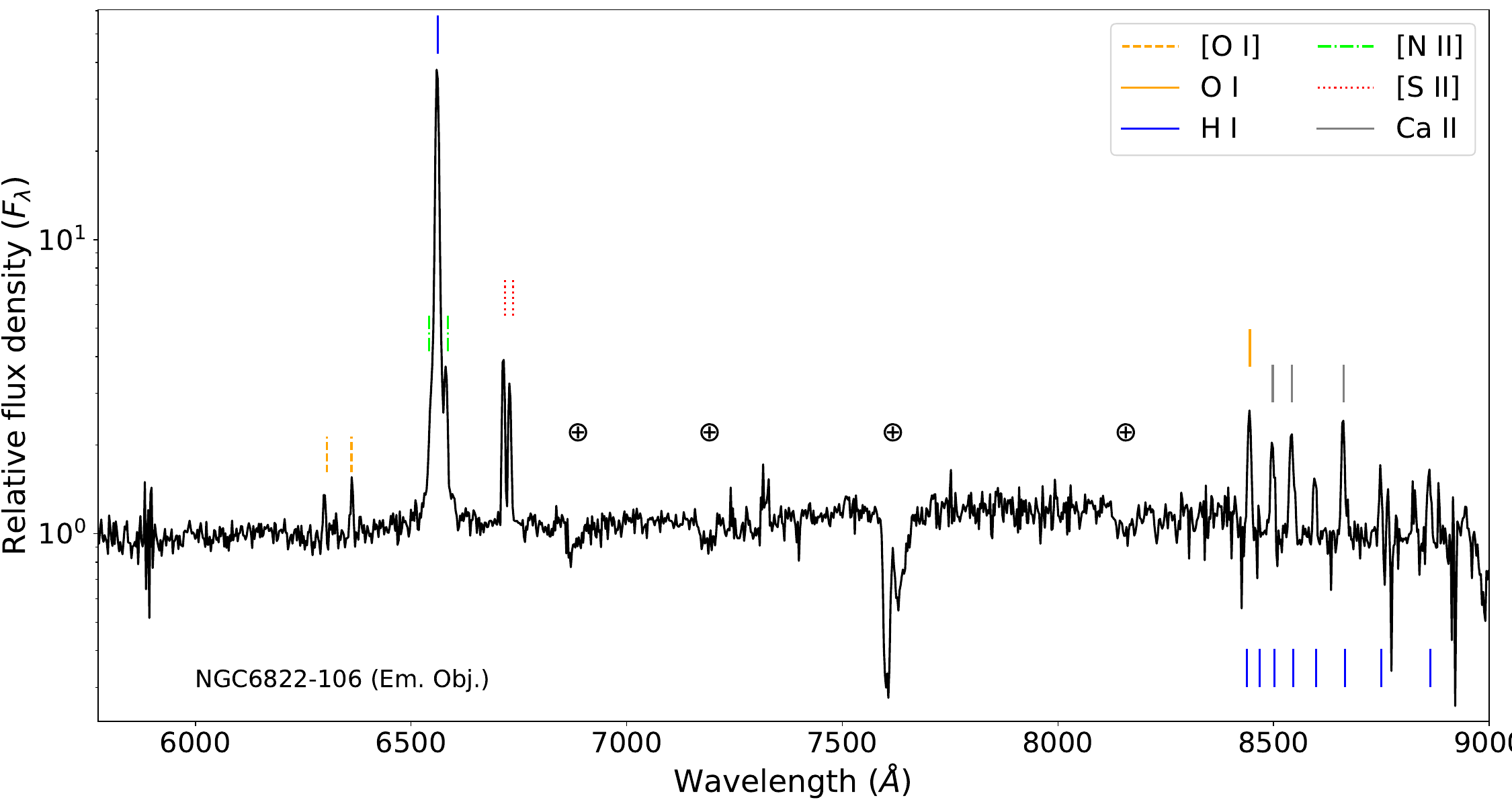}} %referee 1.1, aa 2 column width
    \caption{Similar to Fig.~\ref{fig:RSGexamples}, but for the emission line object. Primary emission lines are identified.}
    \label{fig:EmObj}
\end{center}
\end{figure*}

\clearpage
\section{SED fits} \label{sec:SEDall}

\begin{figure*}[h]
\begin{subfigure}[t]{0.247\textwidth}
    \includegraphics[width=1\columnwidth]{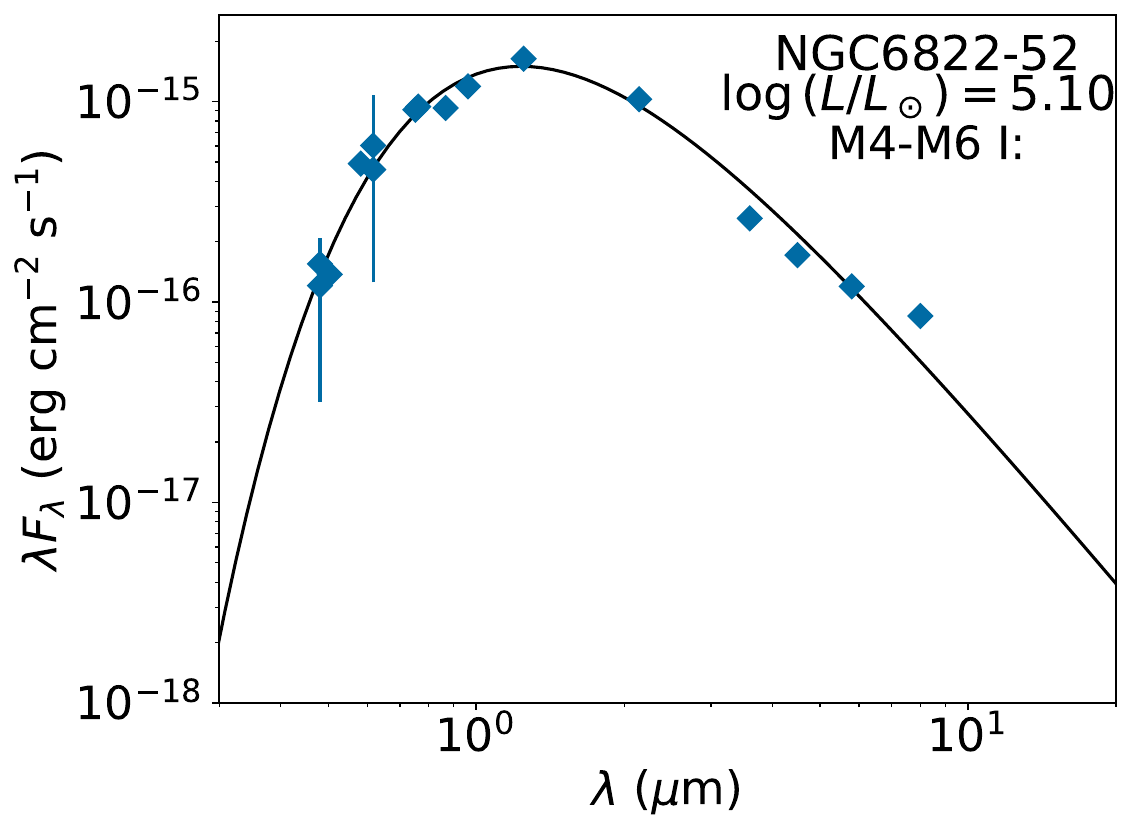}
\end{subfigure}
\begin{subfigure}[t]{0.247\textwidth}
    \includegraphics[width=1\columnwidth]{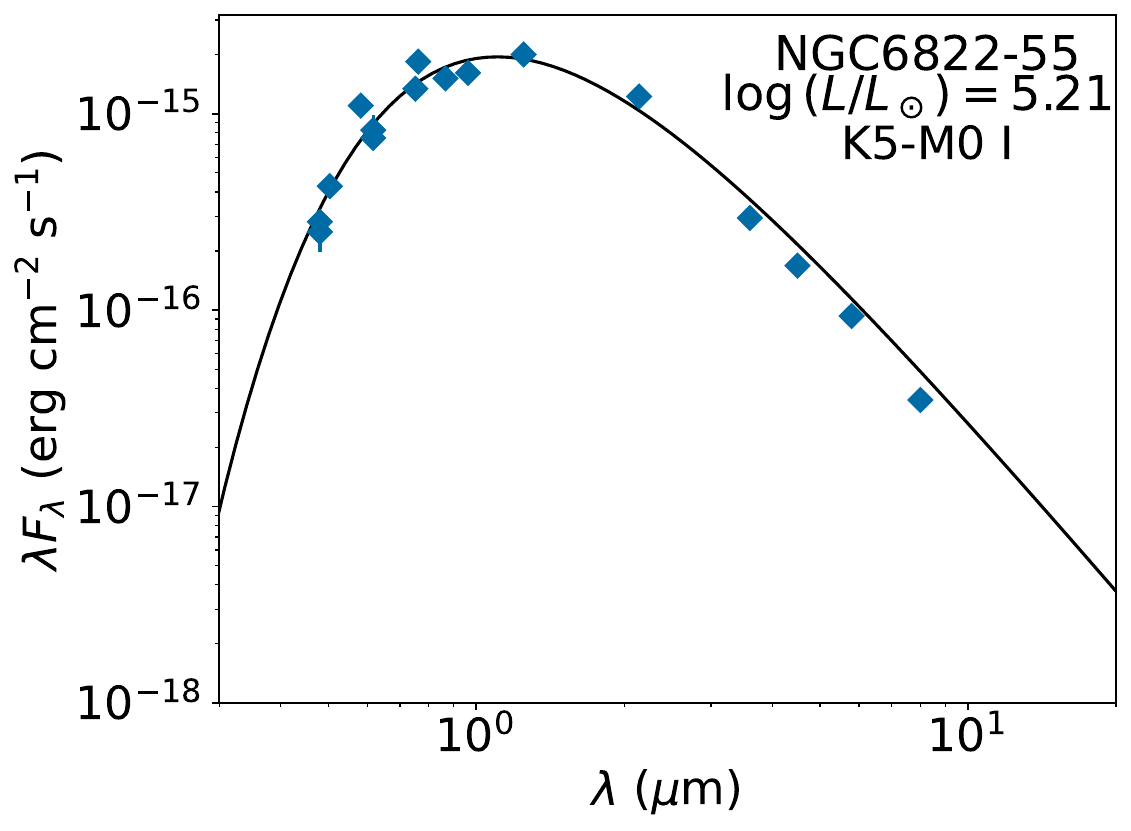}
\end{subfigure}
\begin{subfigure}[t]{0.247\textwidth}
    \includegraphics[width=1\columnwidth]{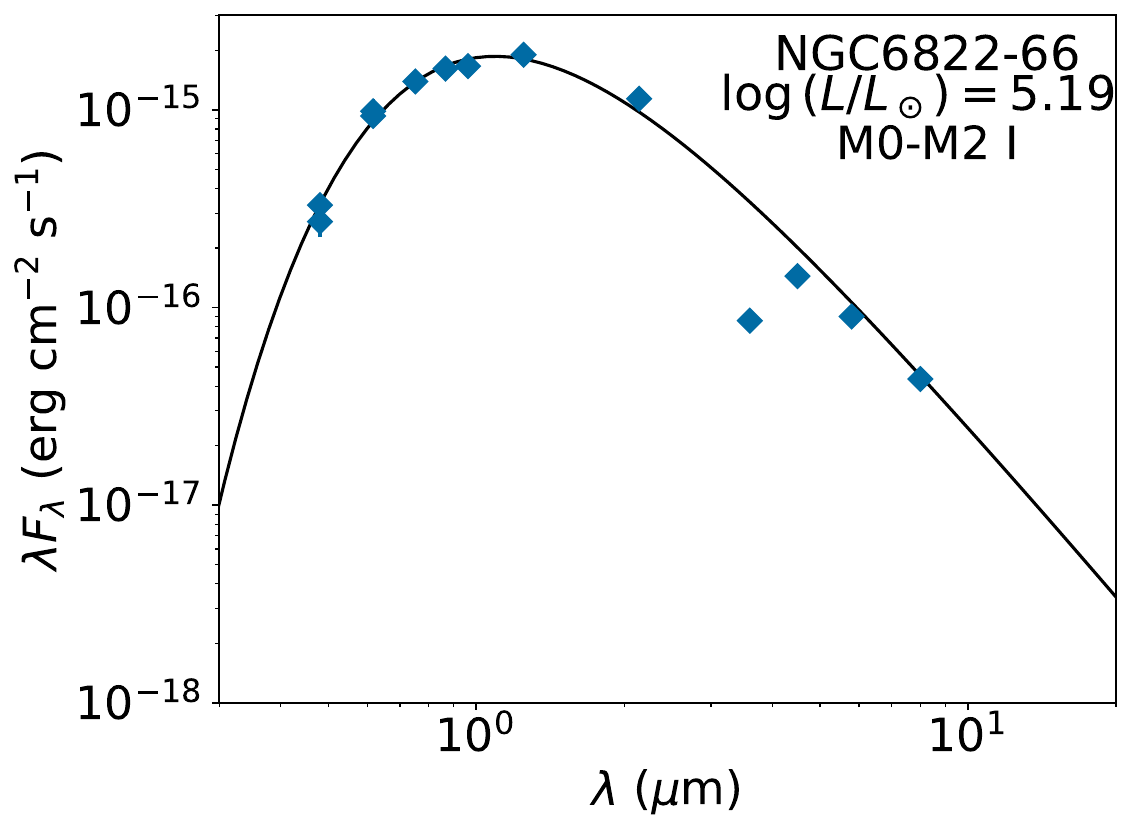}
\end{subfigure}
\begin{subfigure}[t]{0.247\textwidth}
    \includegraphics[width=1\columnwidth]{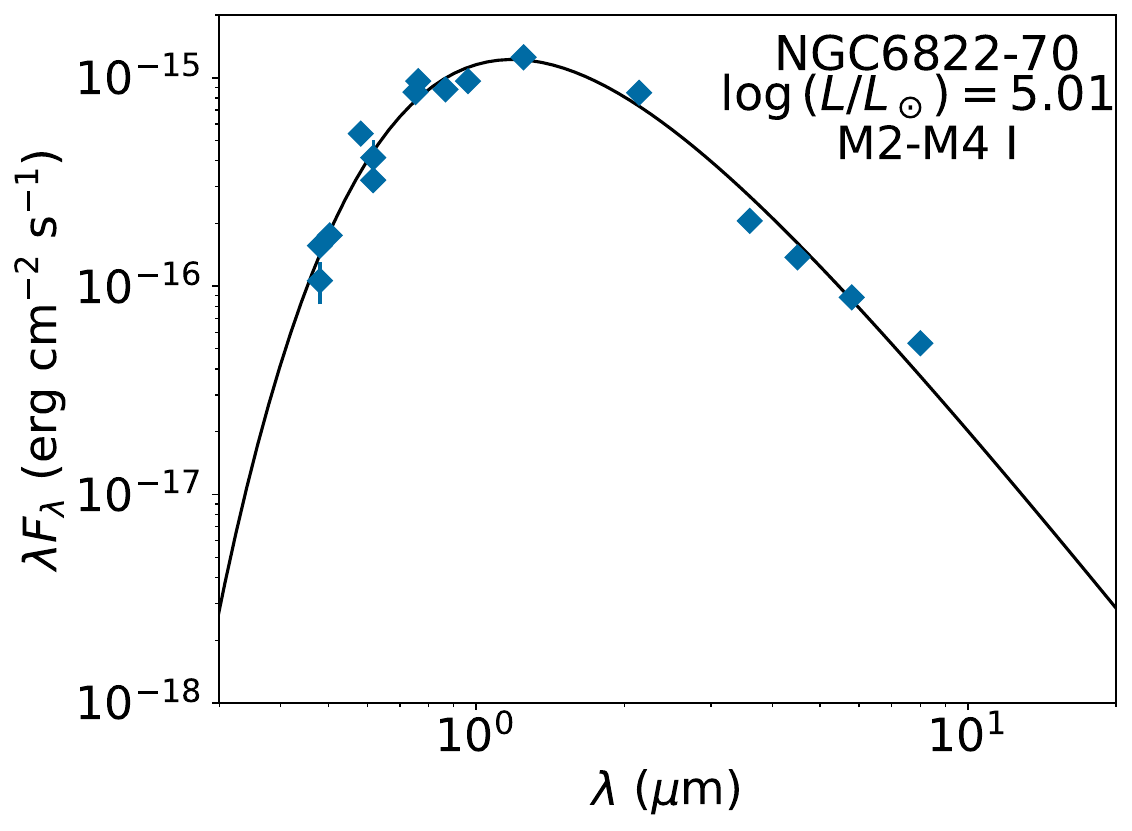}
\end{subfigure}
\begin{subfigure}[t]{0.247\textwidth}
    \includegraphics[width=1\columnwidth]{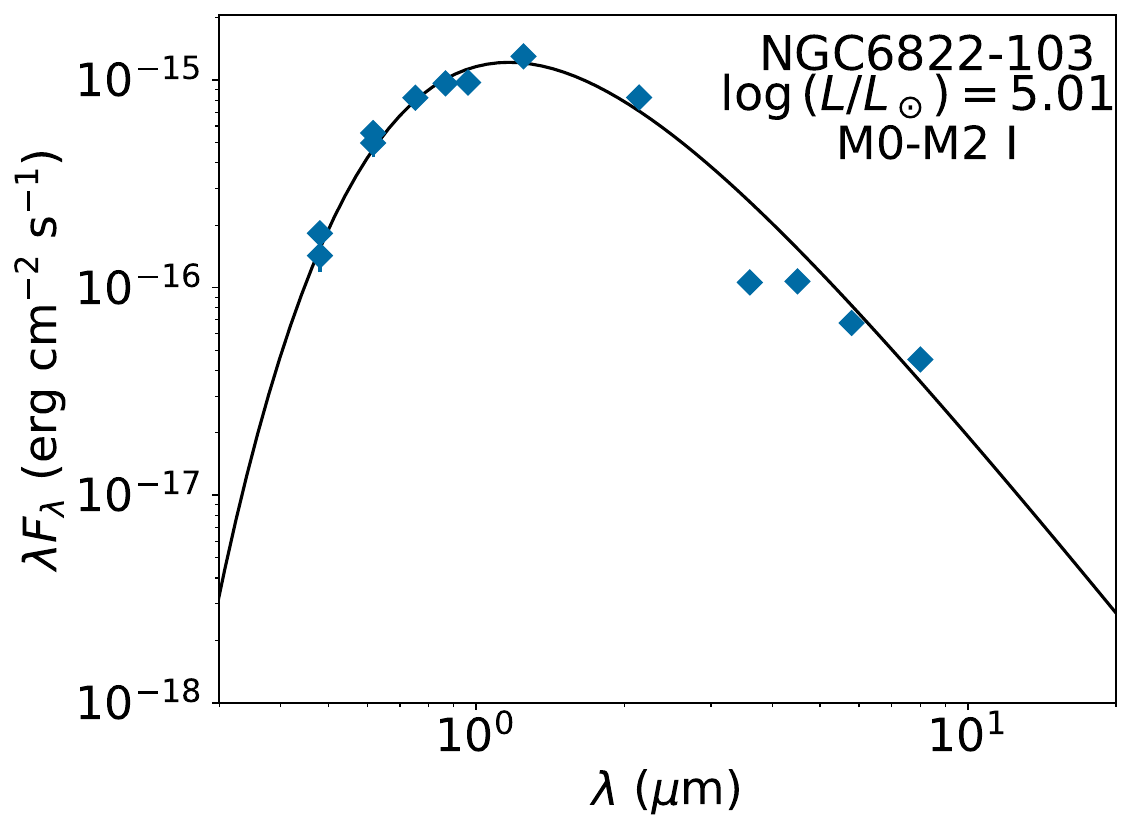}
\end{subfigure}
\begin{subfigure}[t]{0.247\textwidth}
    \includegraphics[width=1\columnwidth]{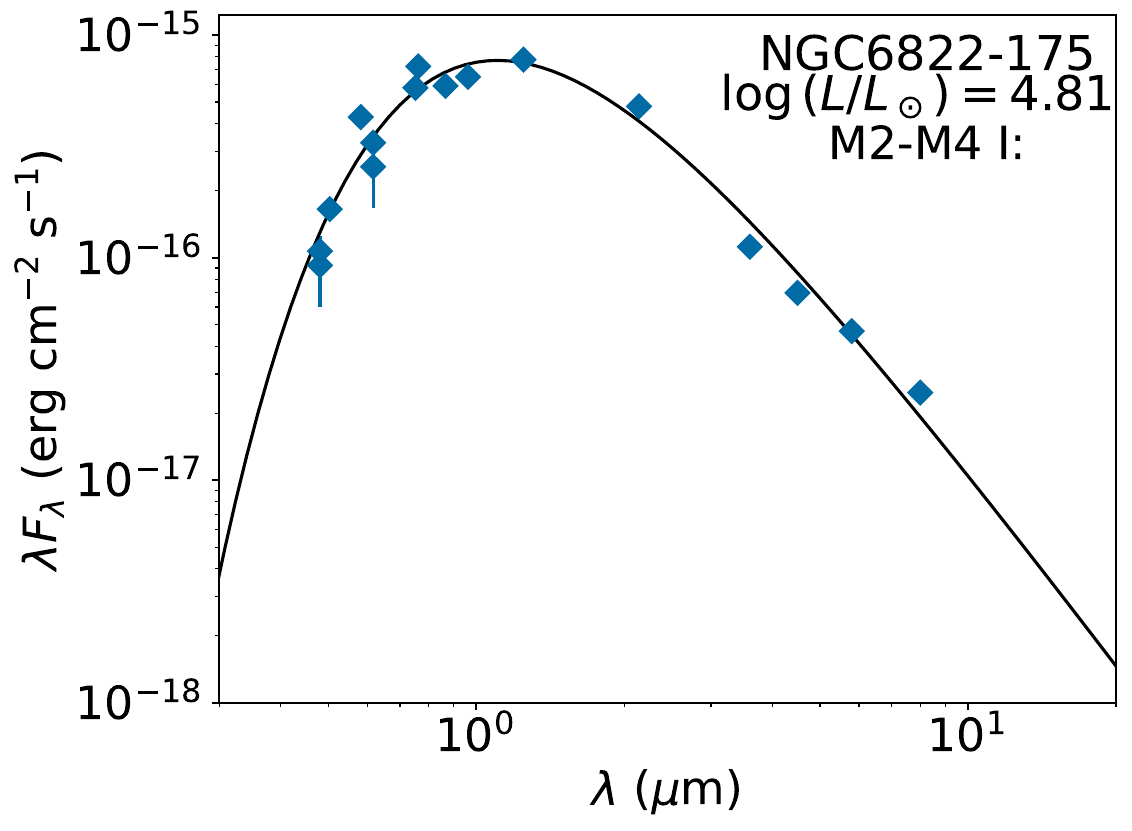}
\end{subfigure}
\begin{subfigure}[t]{0.247\textwidth}
    \includegraphics[width=1\columnwidth]{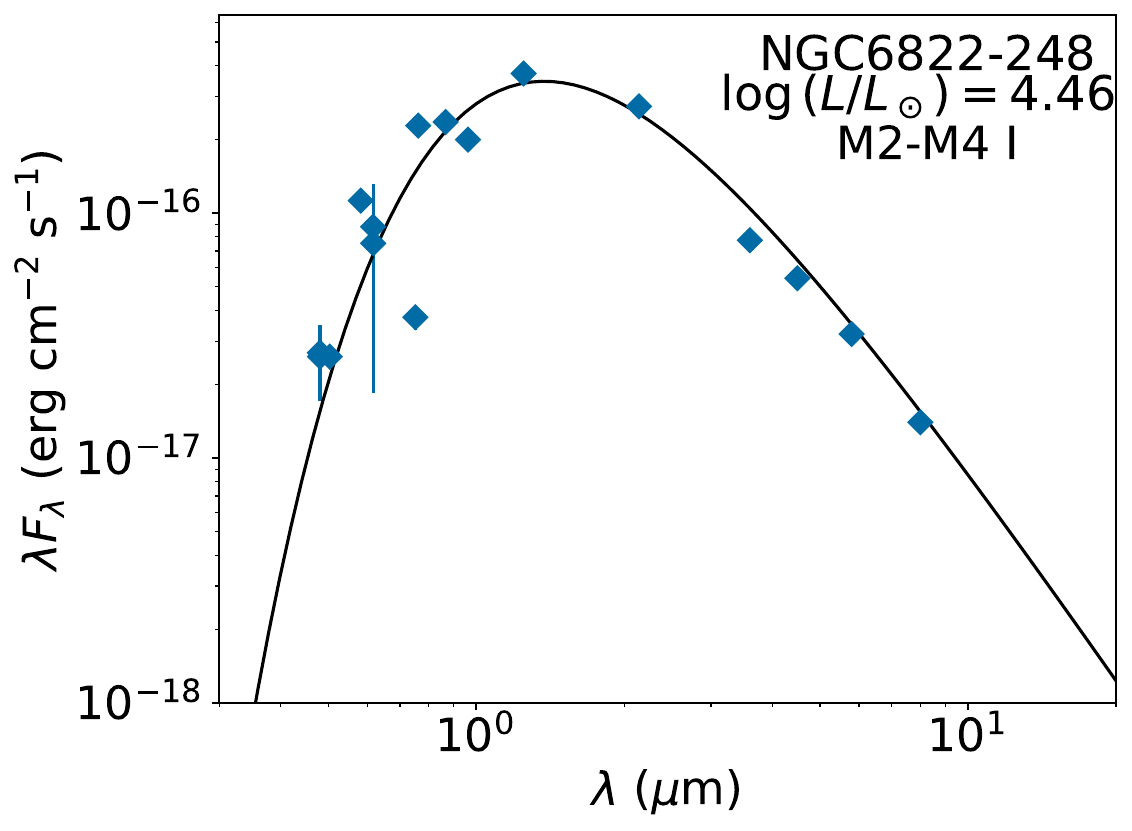}
\end{subfigure}
\begin{subfigure}[t]{0.247\textwidth}
    \includegraphics[width=1\columnwidth]{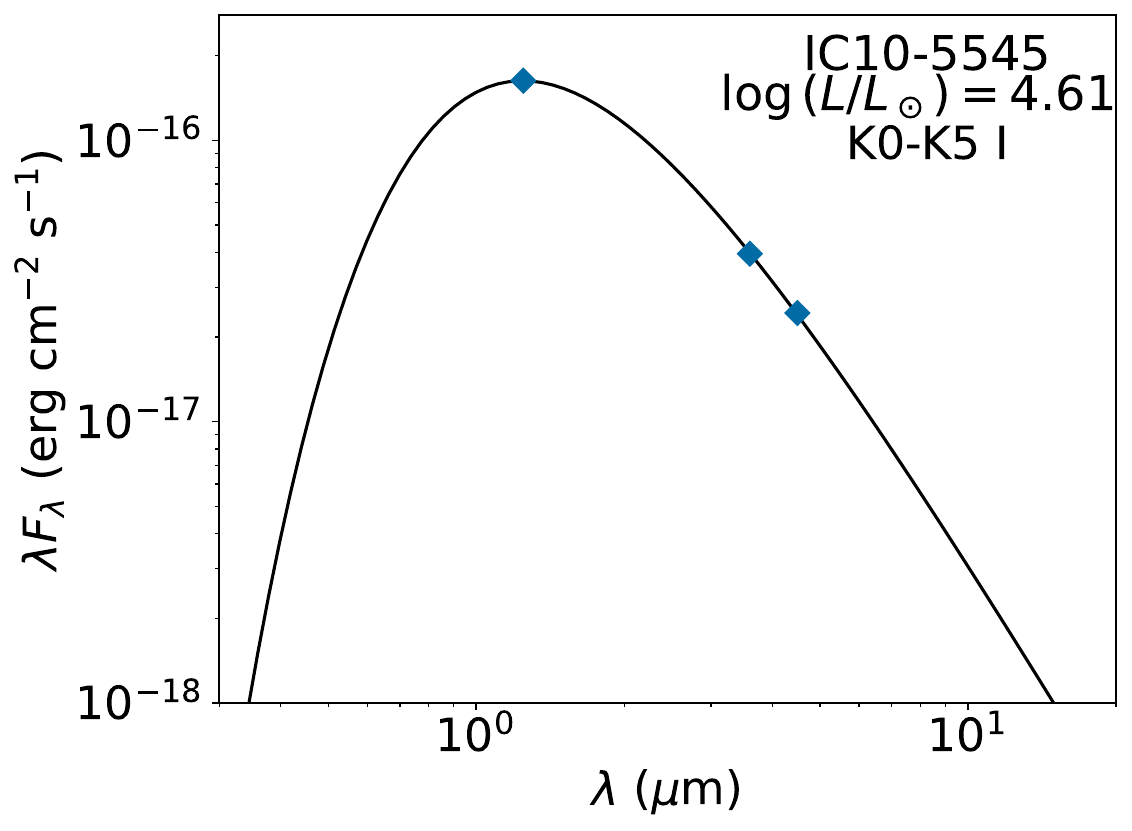}
\end{subfigure}
\begin{subfigure}[t]{0.247\textwidth}
    \includegraphics[width=1\columnwidth]{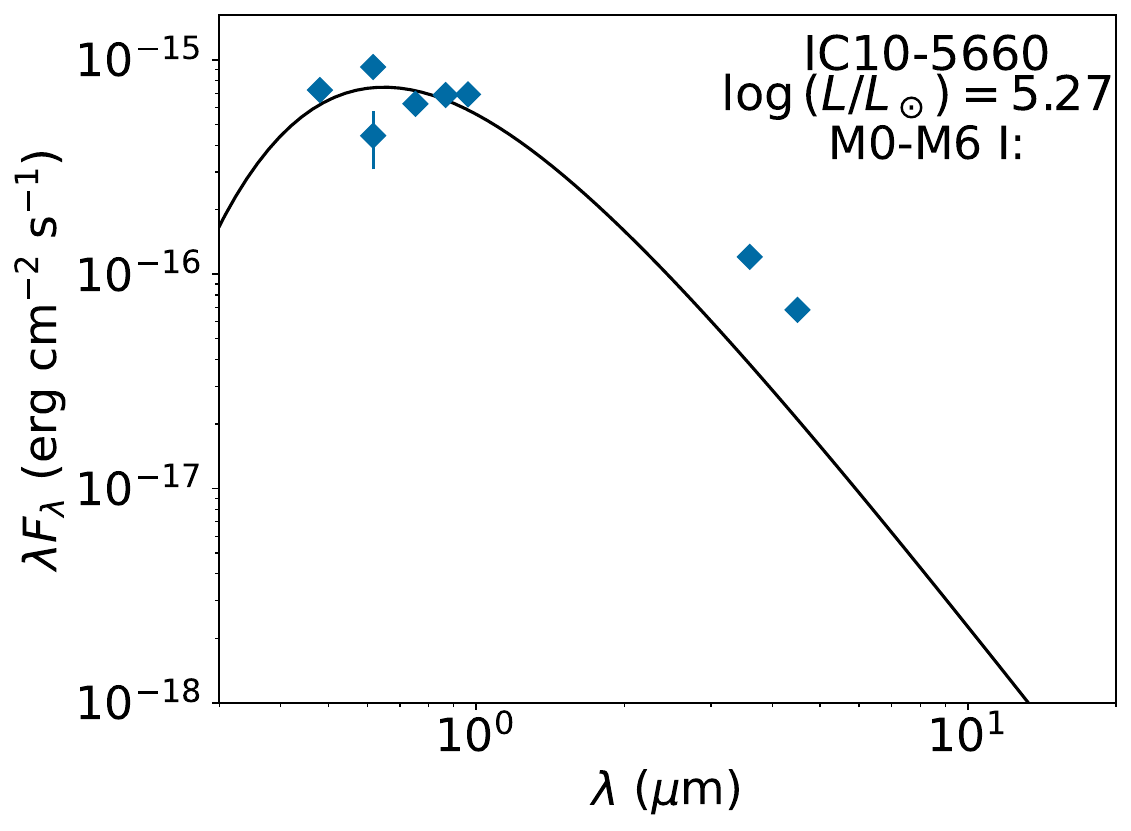}
\end{subfigure}
\begin{subfigure}[t]{0.247\textwidth}
    \includegraphics[width=1\columnwidth]{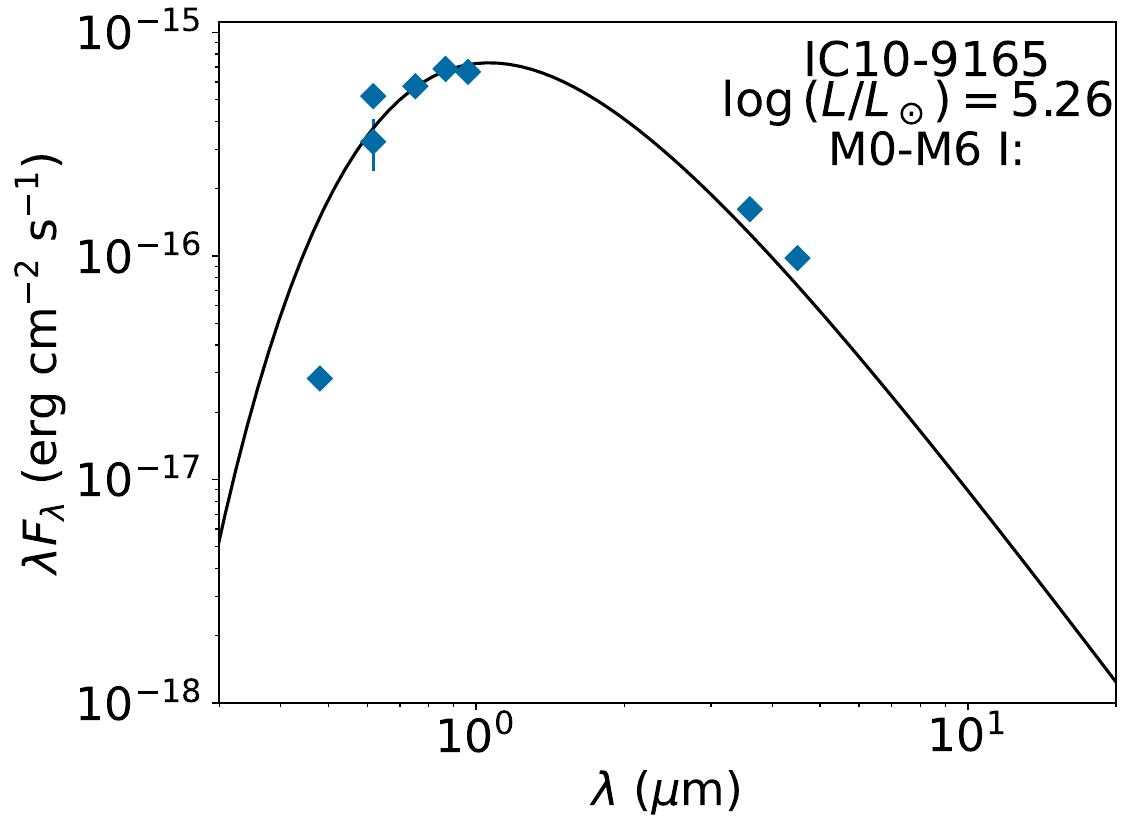}
\end{subfigure}
\begin{subfigure}[t]{0.247\textwidth}
    \includegraphics[width=1\columnwidth]{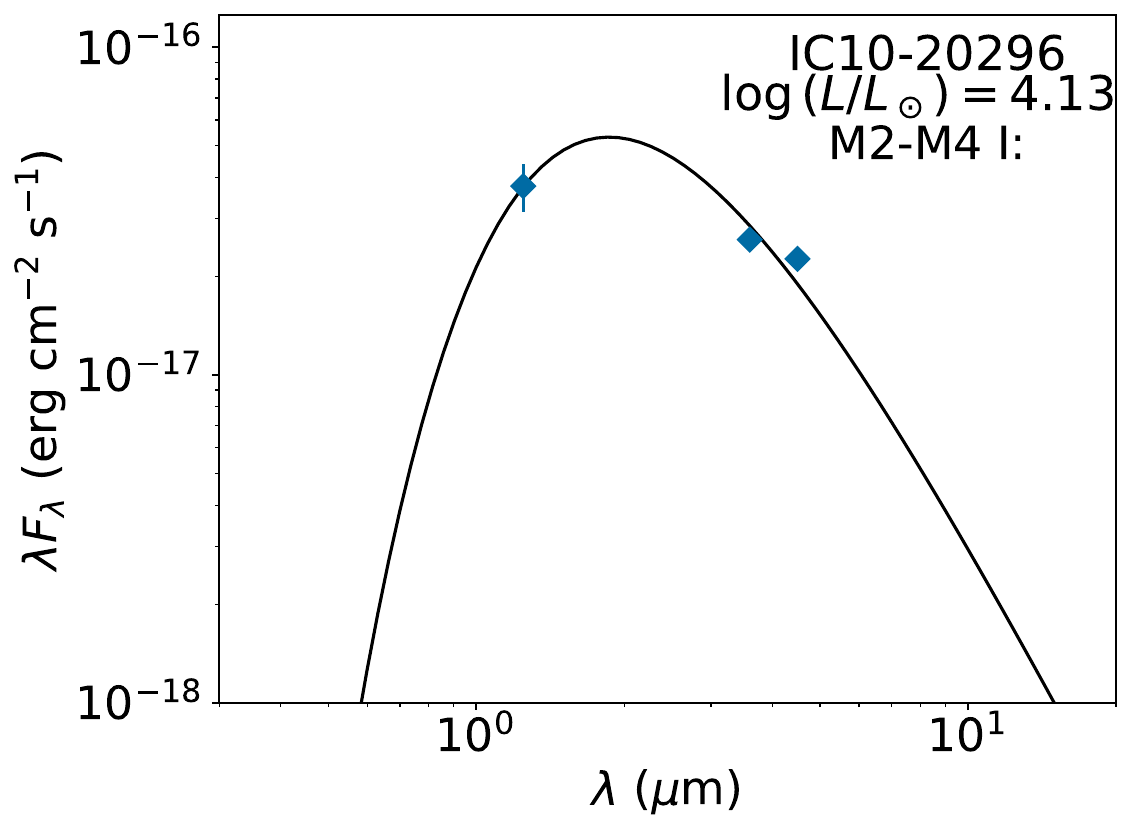}
\end{subfigure}
\begin{subfigure}[t]{0.247\textwidth}
    \includegraphics[width=1\columnwidth]{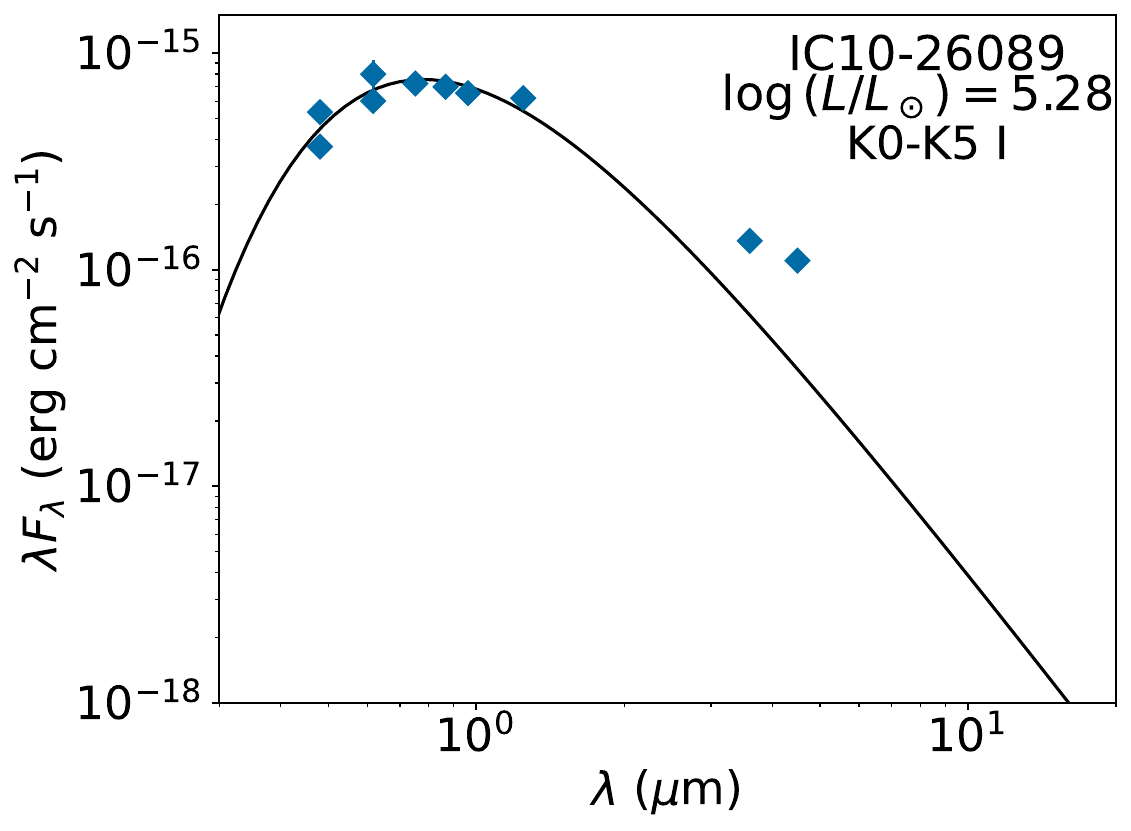}
\end{subfigure}
\begin{subfigure}[t]{0.247\textwidth}
    \includegraphics[width=1\columnwidth]{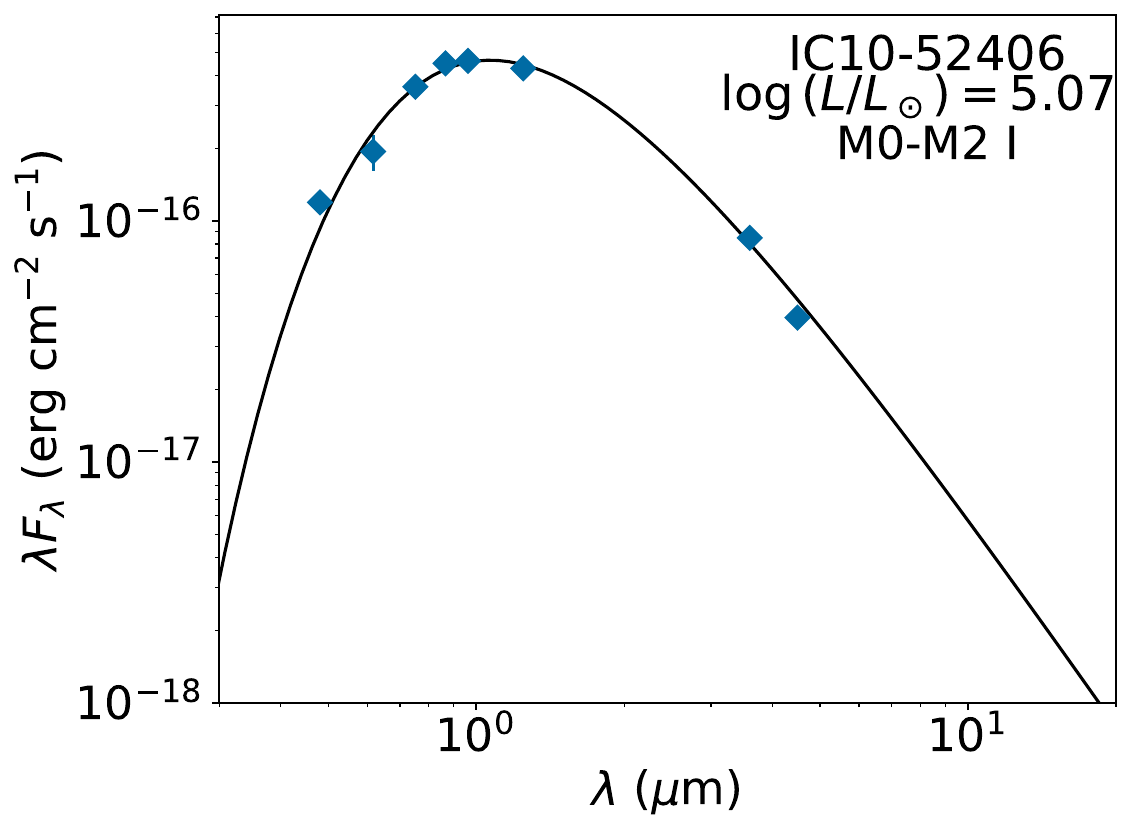}
\end{subfigure}
\begin{subfigure}[t]{0.247\textwidth}
    \includegraphics[width=1\columnwidth]{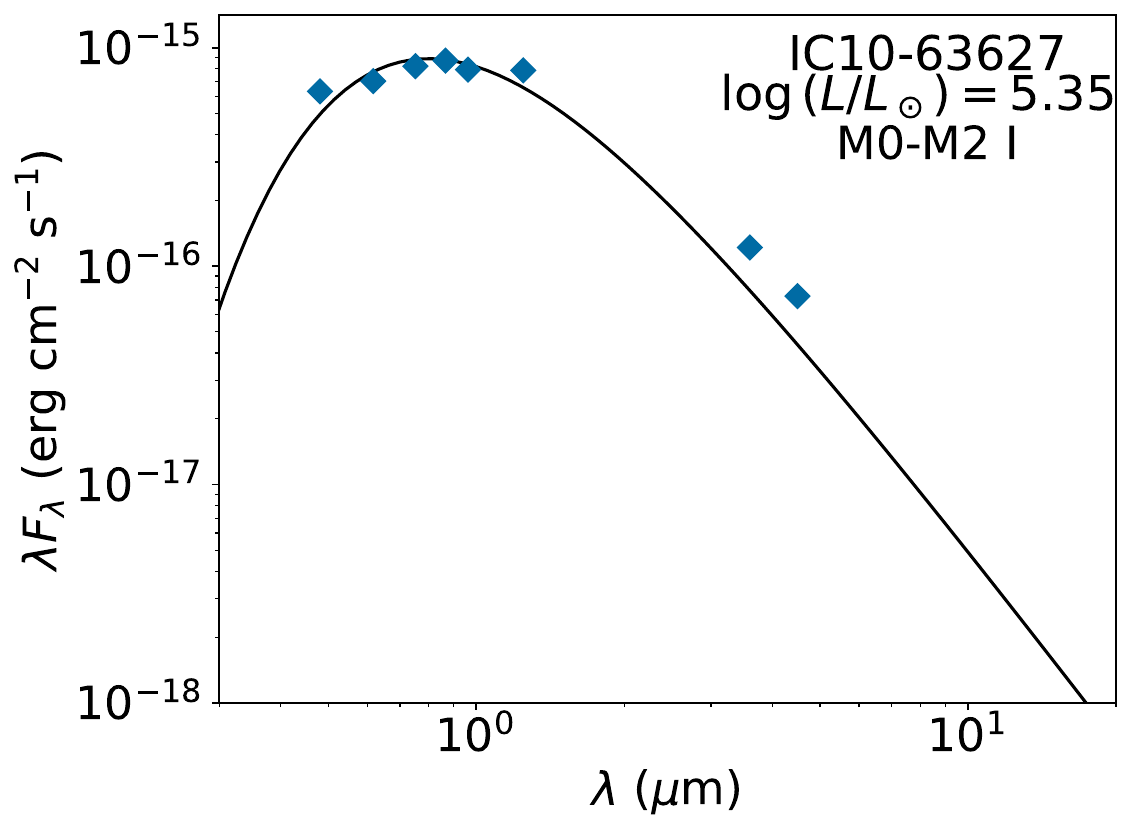}
\end{subfigure}
\begin{subfigure}[t]{0.247\textwidth}
    \includegraphics[width=1\columnwidth]{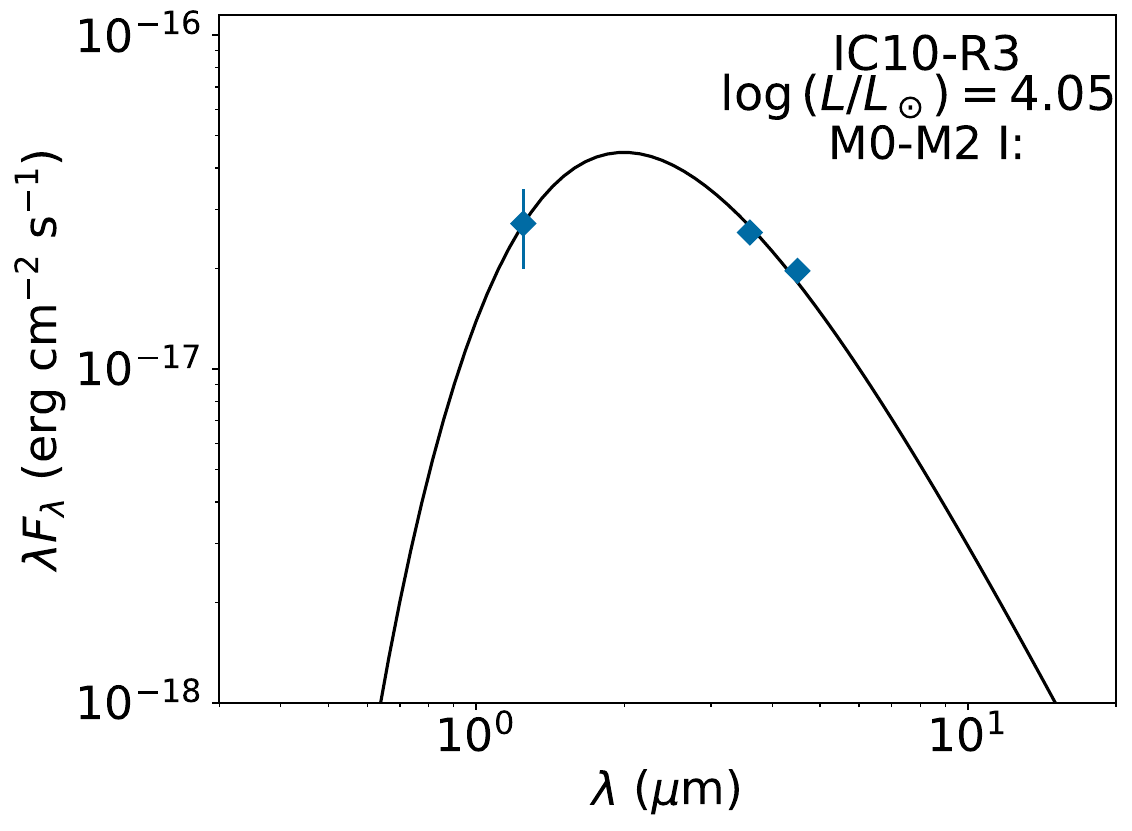}
\end{subfigure}
\begin{subfigure}[t]{0.247\textwidth}
    \includegraphics[width=1\columnwidth]{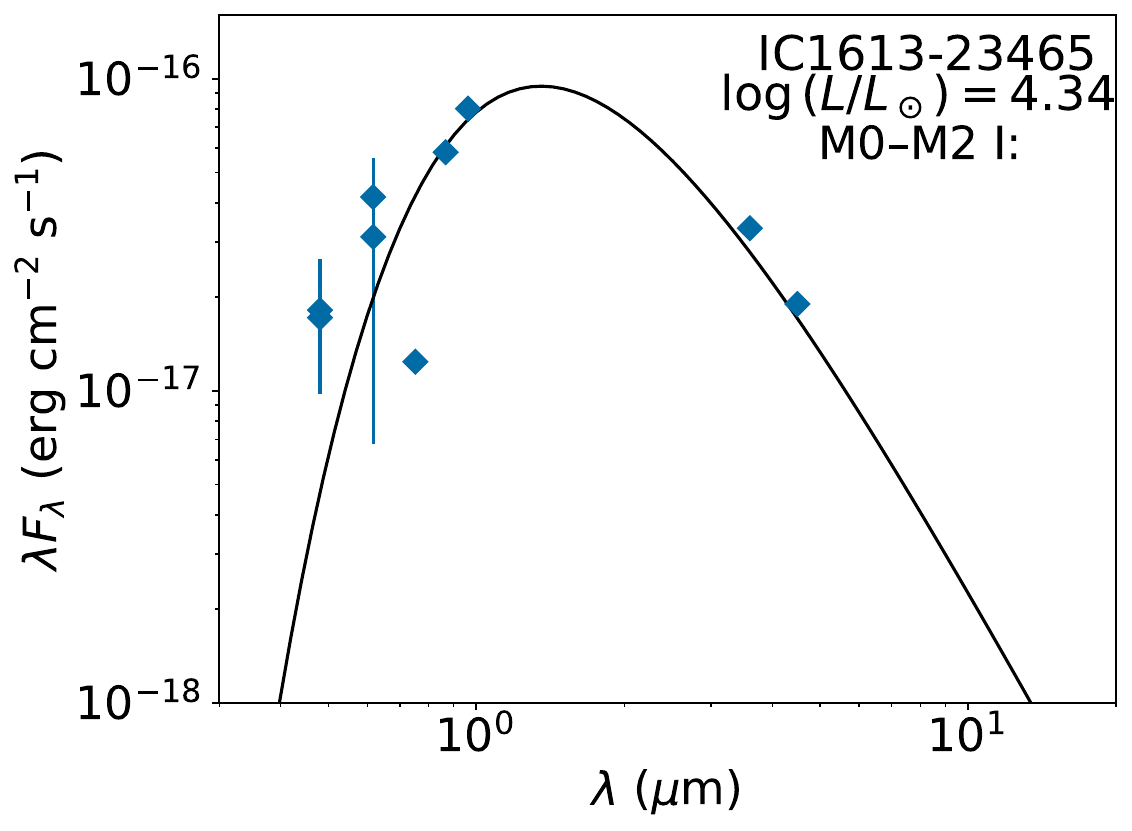}
    %referee 1.1, aa 2 column width
\end{subfigure}
\begin{subfigure}[t]{0.247\textwidth}
    \includegraphics[width=1\columnwidth]{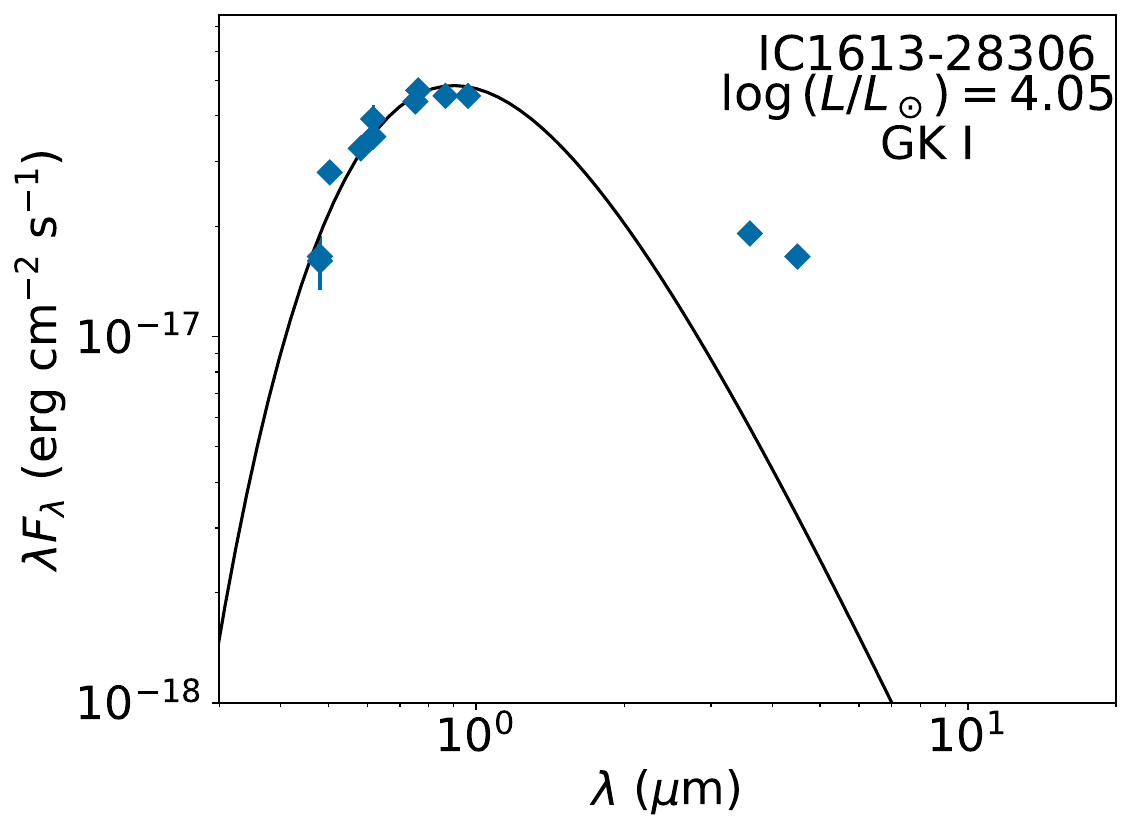}
\end{subfigure}
\caption{Blackbodies fitted to the photometry from \textit{Gaia}, Pan-STARRS1, VISTA, ZTF and \textit{Spitzer}. The target name, derived luminosity, and spectral type are indicated in each panel.}
\label{fig:SEDAll}
\end{figure*}

\clearpage

\section{Light curves} \label{sec:LCall}

\begin{figure*}[h]

\begin{subfigure}[t]{0.47\textwidth}
    \includegraphics[width=1\columnwidth]{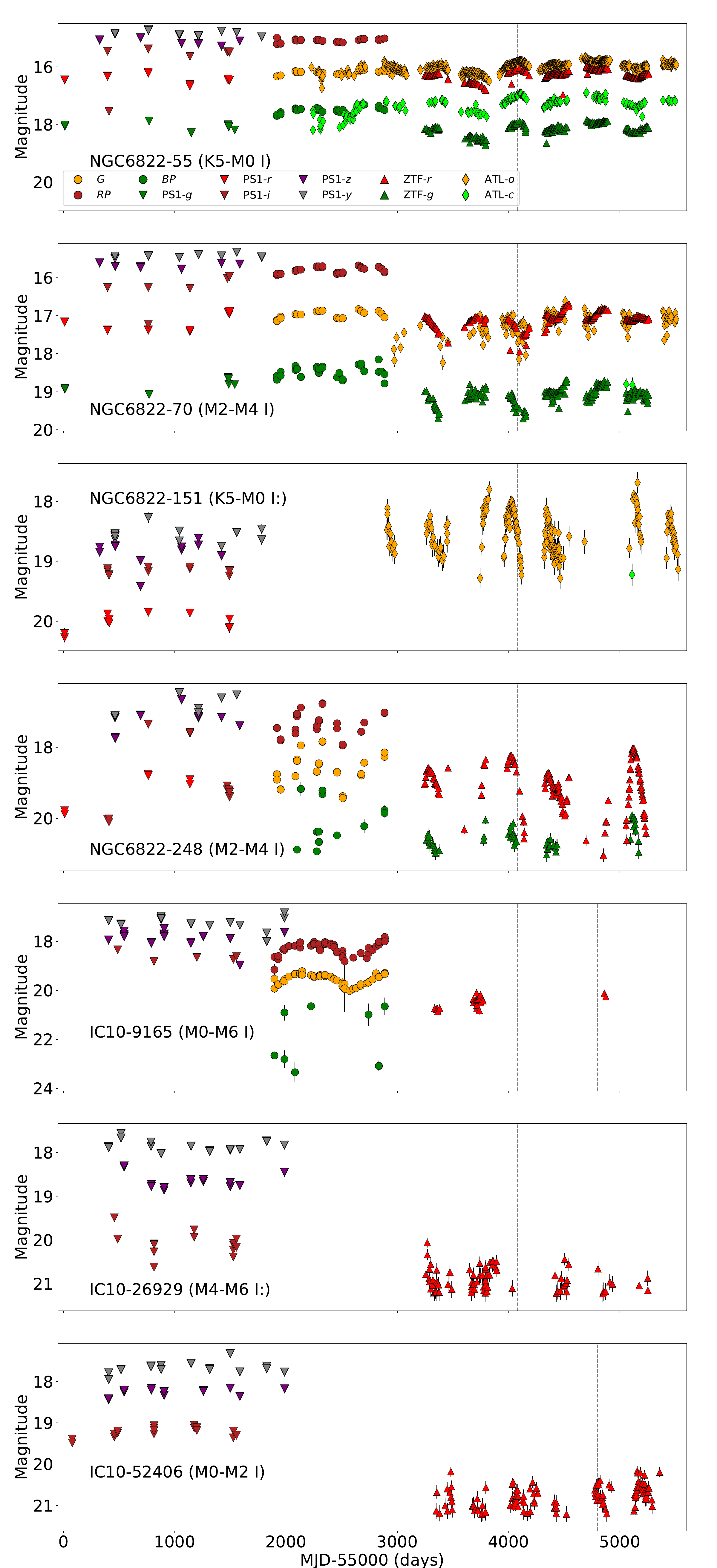}
\end{subfigure}
\begin{subfigure}[t]{0.47\textwidth}
    \includegraphics[width=1\columnwidth]{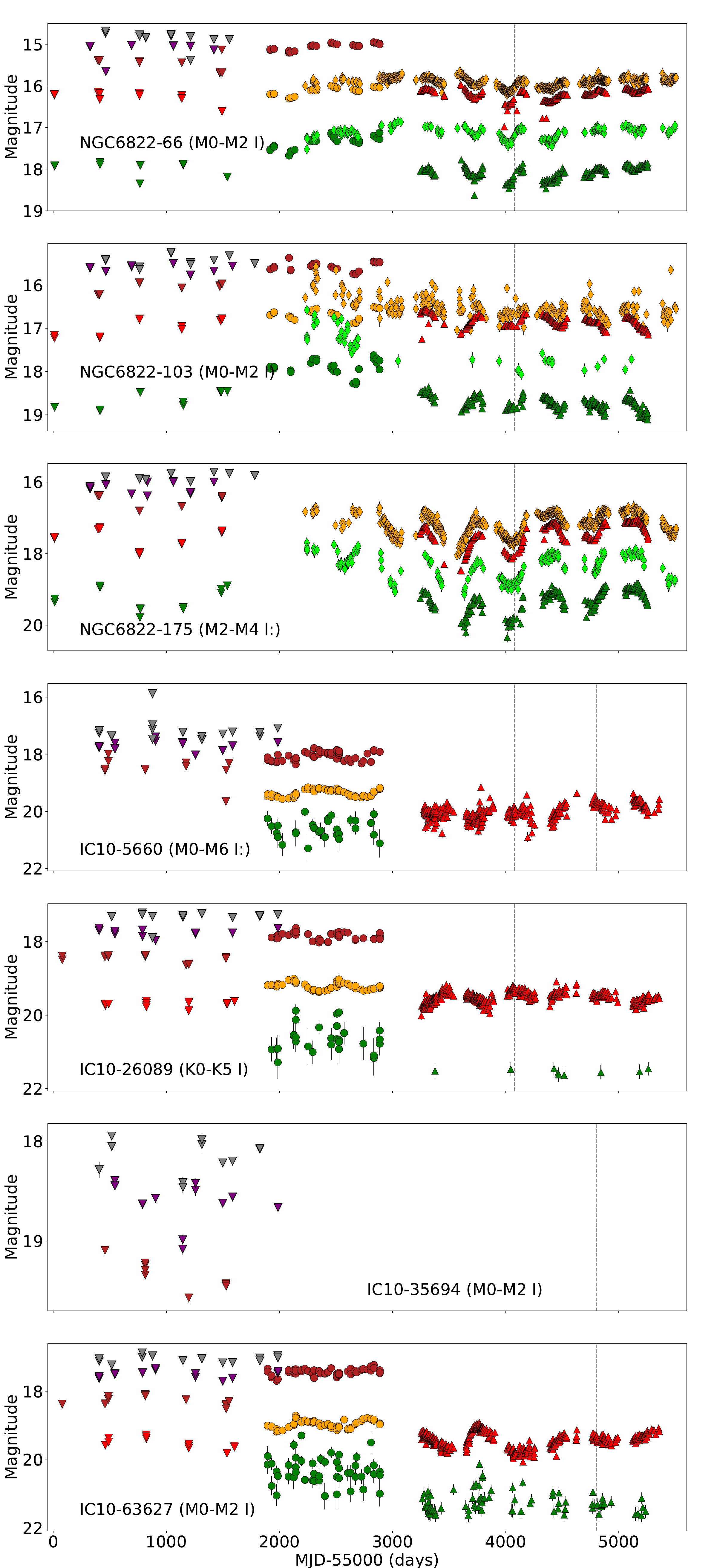}
\end{subfigure}
    \caption{Multi-survey optical light curve of the RSGs. Apparent magnitudes from Pan-STARRS1 DR2 (triangles: $gri$), \textit{Gaia}-DR3 (circles: $G, BP, RP$), ZTF (upper triangles: $g, r$), and ATLAS (diamonds: $c, o$) are shown. The vertical dashed line indicates the epoch of the OSIRIS spectroscopy.}
\end{figure*}
\clearpage
\begin{figure*}
\ContinuedFloat
\begin{subfigure}[t]{0.47\textwidth}
    \includegraphics[width=1\columnwidth]{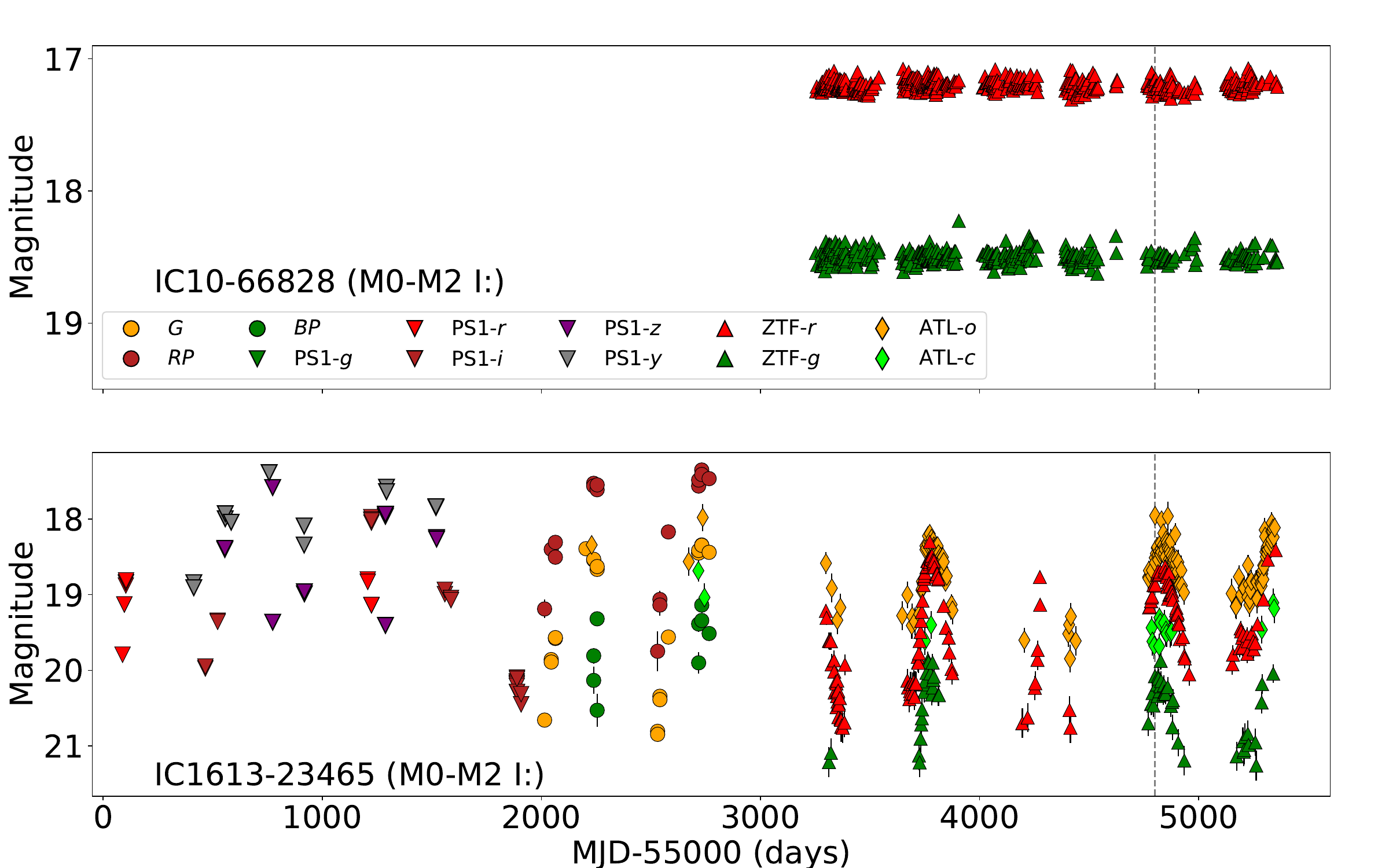}
\end{subfigure}
\begin{subfigure}[t]{0.47\textwidth}
    \includegraphics[width=1\columnwidth]{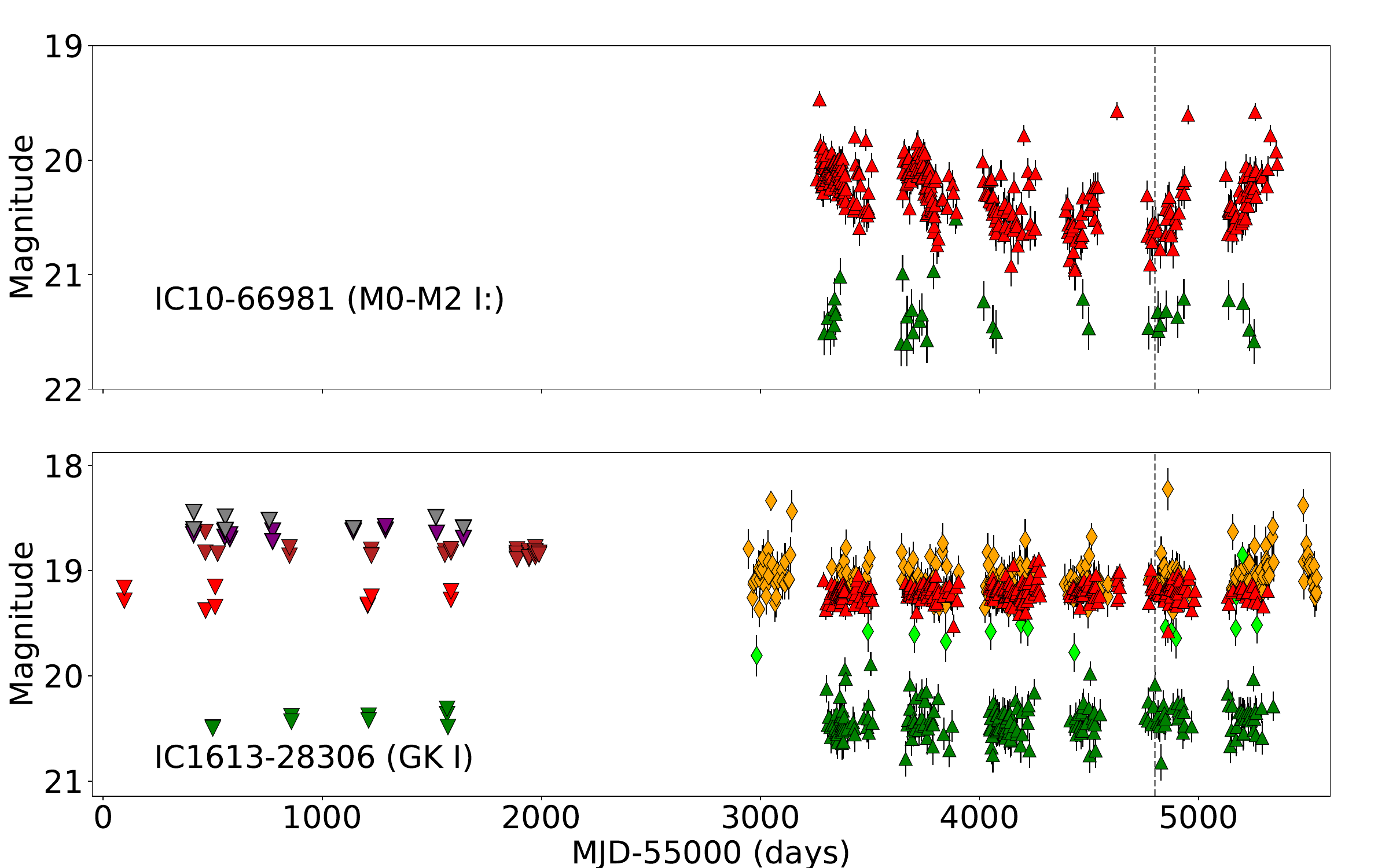}
\end{subfigure}
\caption{continued}
\label{fig:LC1}
\end{figure*}
\vspace{1.cm}

\begin{figure*}[h]
\begin{center}
\centerline{\includegraphics[width=0.8\columnwidth]{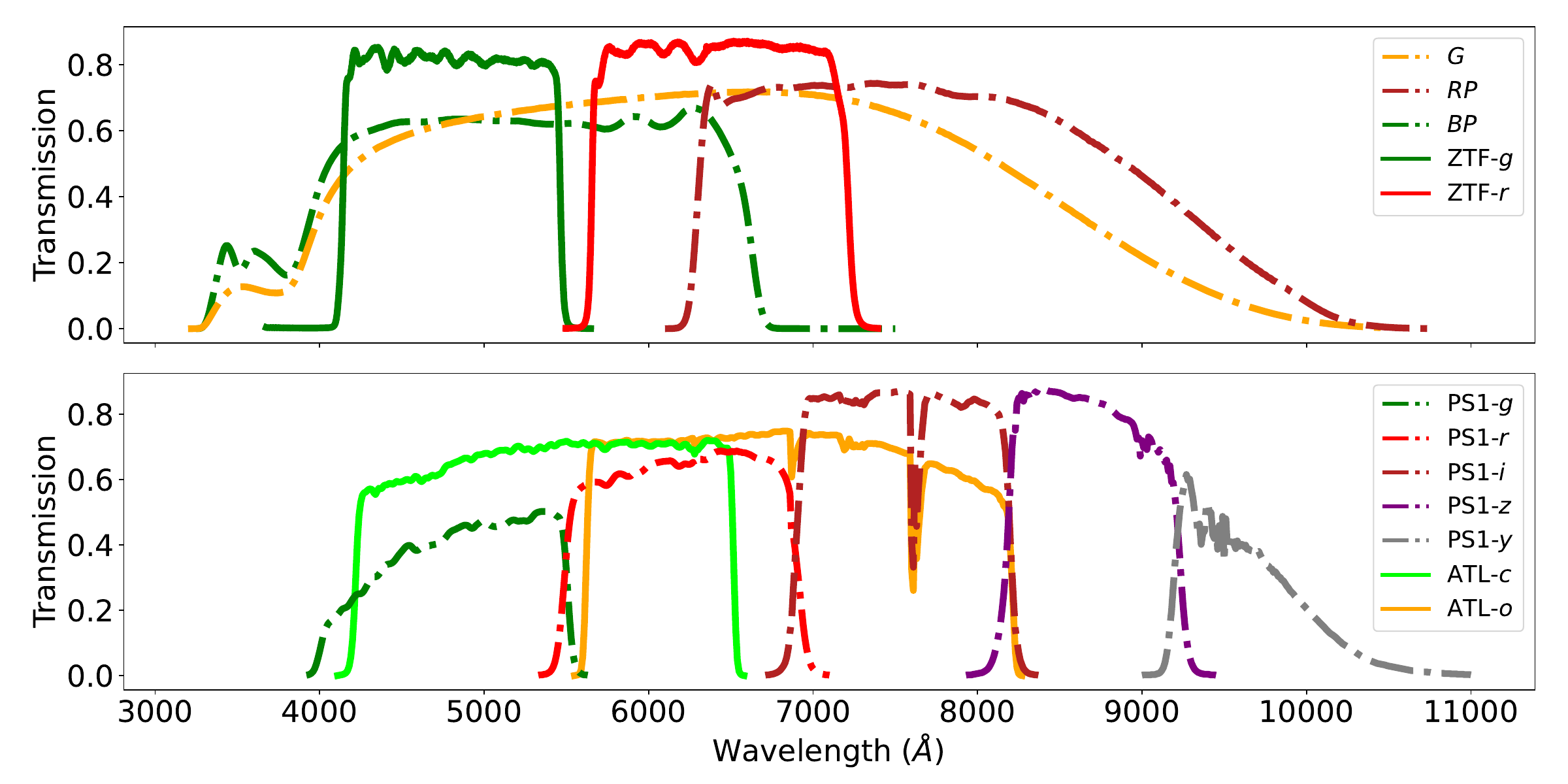}} %referee 1.1, aa 2 column width
    \caption{Transmission of the photometric filters used in the light curves}
    \label{fig:filters}
\end{center}
\end{figure*}

\clearpage

\section{Spectra and fitted models} \label{sec:fitsall}

\begin{figure*}[h]
\begin{subfigure}[t]{0.33\textwidth}
    \includegraphics[width=1\columnwidth]{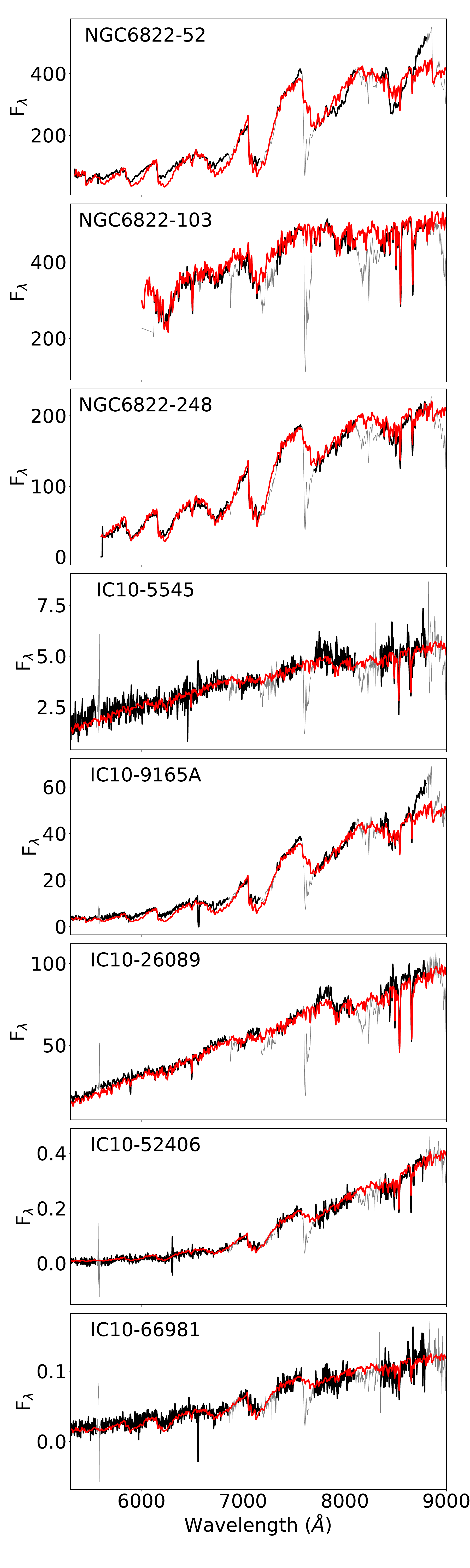}
\end{subfigure}
\begin{subfigure}[t]{0.33\textwidth}
    \includegraphics[width=1\columnwidth]{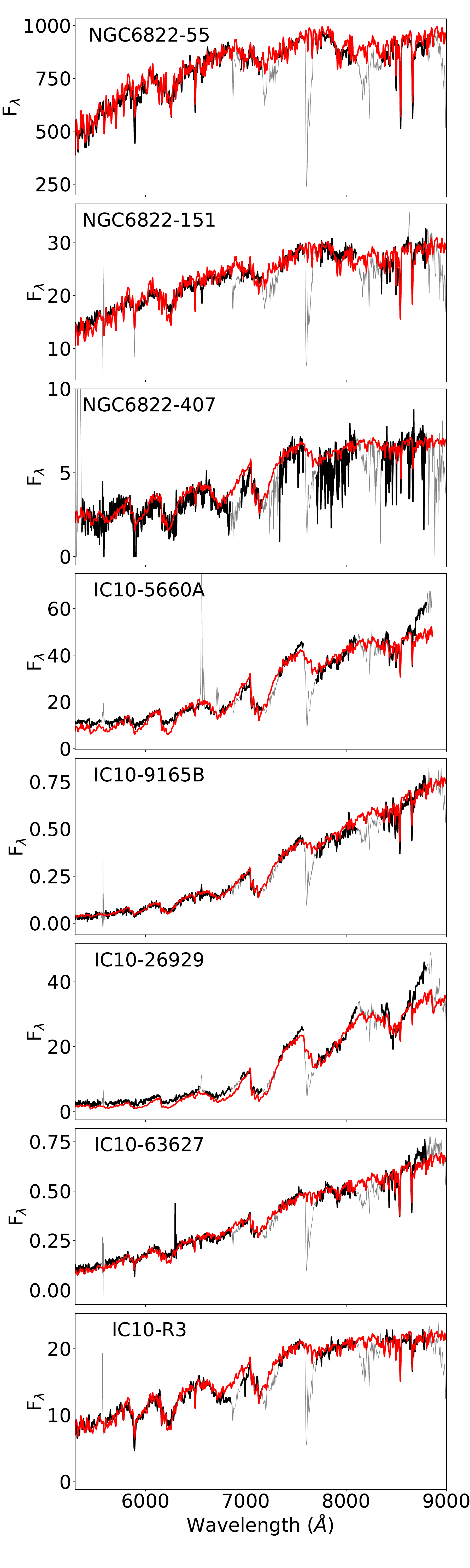}
\end{subfigure}
\begin{subfigure}[t]{0.33\textwidth}
    \includegraphics[width=1\columnwidth]{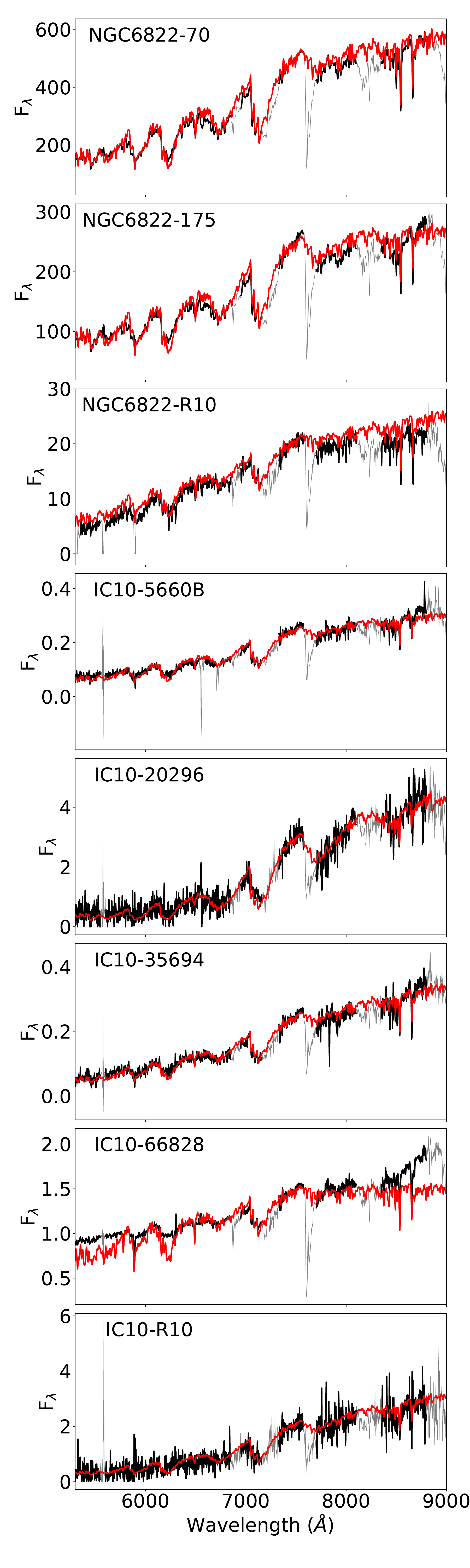}
\end{subfigure}
\caption{Best-fit \textsc{marcs} models (red) to all spectra of RSGs (black). Gray regions are excluded from the fit due to contamination by telluric absorption. F$_\lambda$ has units $\rm 10^{-18} erg \,\ cm^{-2}s^{-1}A^{-1} $.}
\label{fig:fitAll}
\end{figure*}

\clearpage

\begin{figure*}[t]
\ContinuedFloat
\begin{subfigure}[t]{0.33\textwidth}
    \includegraphics[width=1\columnwidth]{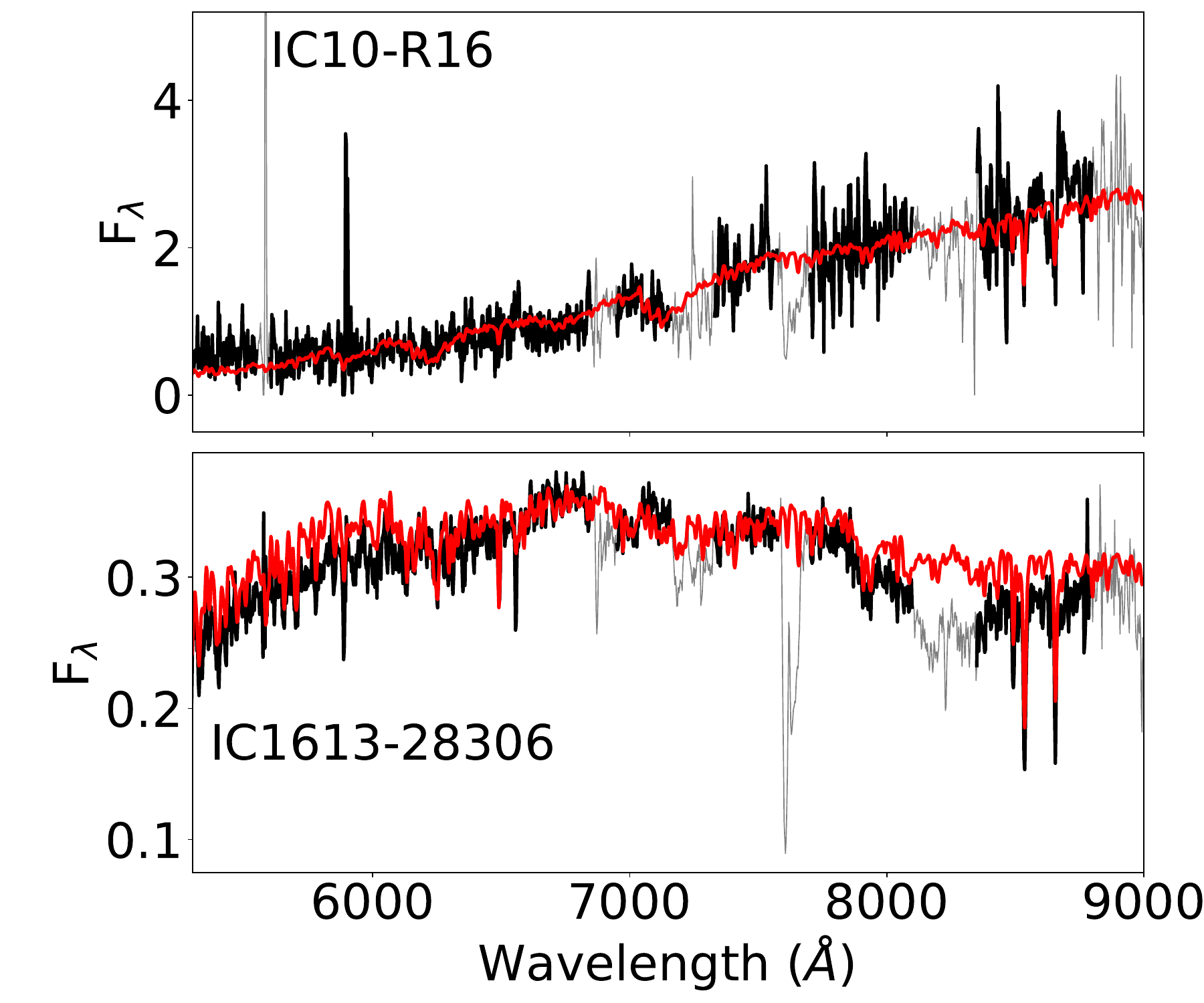}
\end{subfigure}
\begin{subfigure}[t]{0.33\textwidth}
    \includegraphics[width=1\columnwidth]{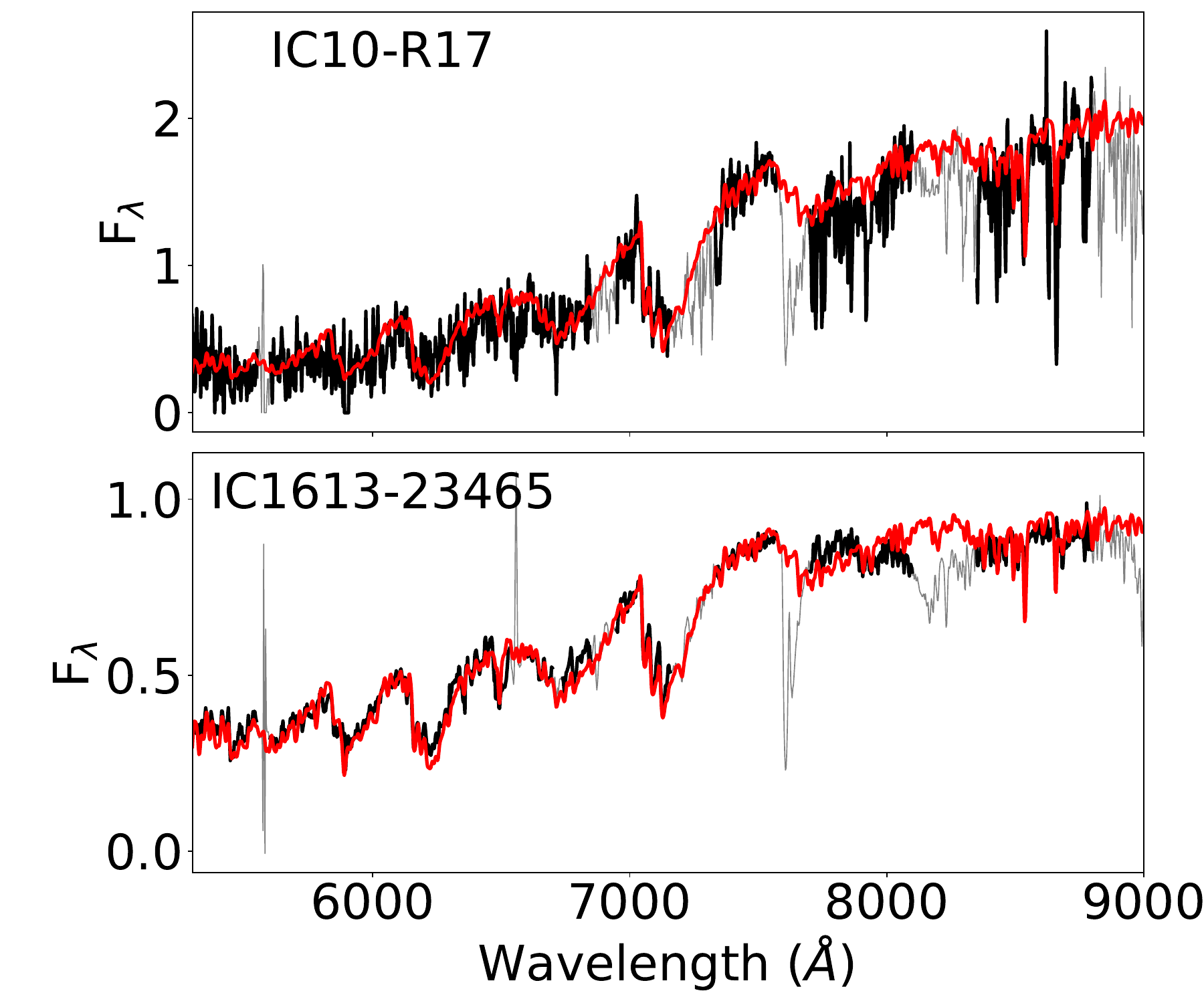}
\end{subfigure}
\begin{subfigure}[t]{0.33\textwidth}
    \includegraphics[width=1\columnwidth]{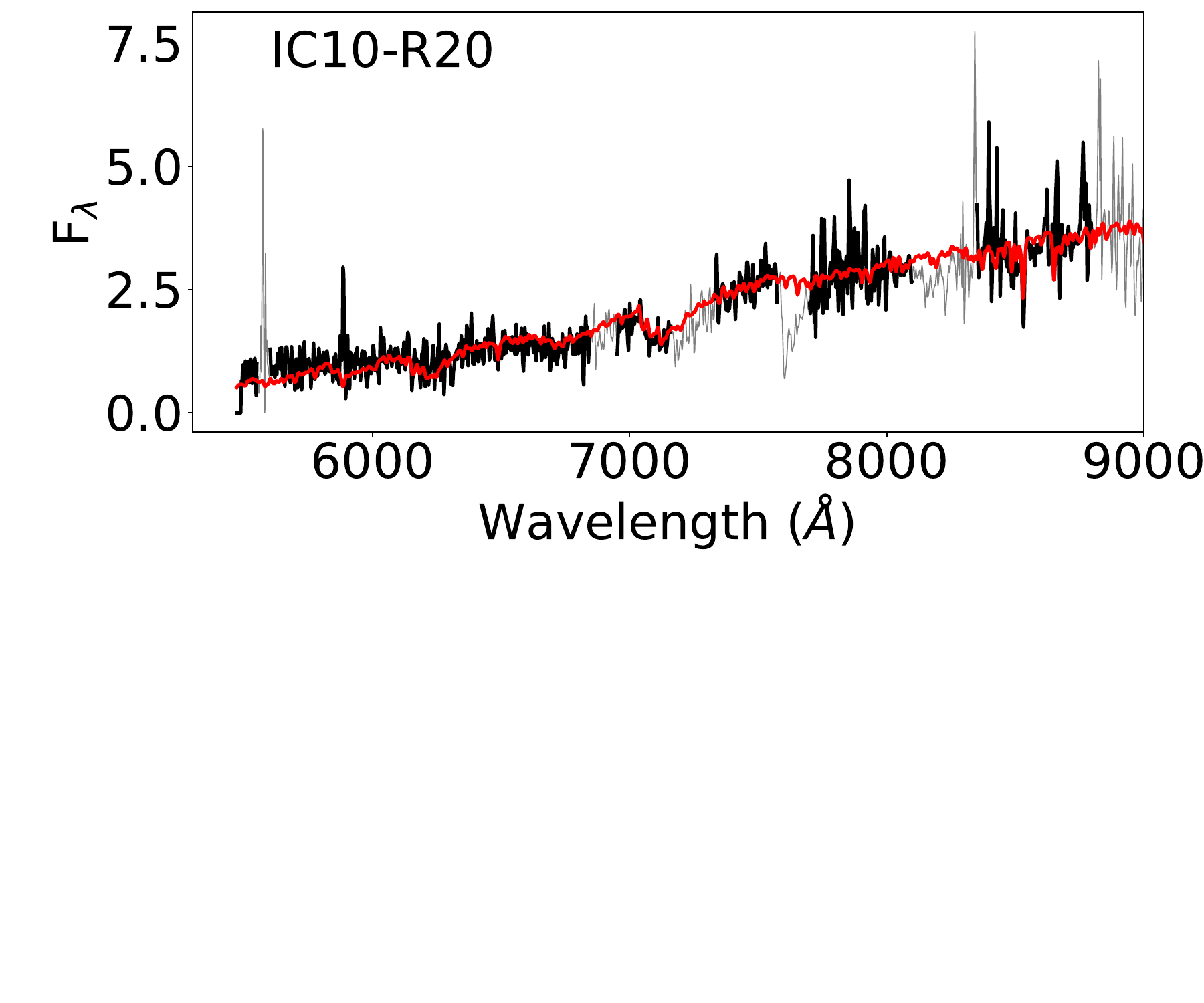}
\end{subfigure}
\caption{continued.}
\end{figure*}
\vspace{-1.cm}

\end{appendix}

\end{document}